%% file: article_main.tex
\documentclass[preprint,3p]{elsarticle}

\usepackage{natbib}
\usepackage{geometry}
\usepackage{empheq}
\usepackage[breaklinks=true]{hyperref}

\usepackage{amsmath,amsfonts,amsthm,amssymb,amsxtra,dsfont,mathtools,bm}
\usepackage{amsxtra, amssymb, mathrsfs, pstricks}
\usepackage[english]{babel}
\usepackage{amsmath,amsfonts,amstext,amsbsy,amsopn,amssymb,amsthm,amscd}
\usepackage{amsxtra,latexsym}
\usepackage{exscale}
\usepackage{graphicx,mathrsfs}
\usepackage{psfrag}
\usepackage{paralist,eucal,enumerate}
\usepackage{color,graphics}
\usepackage{blindtext}
\usepackage[ruled]{algorithm2e}
\usepackage{relsize}
\usepackage{subcaption}
\usepackage{comment}
\usepackage{tikz}

\include{definitions}

\journal{arXiv}

\begin{document}

\begin{frontmatter}



\title{A time-adaptive finite element phase-field model suitable for rate-independent fracture mechanics}


\author[1]{Felix Rörentrop}
\ead{felix.roerentrop@tu-dortmund.de}
\author[2]{Samira Boddin}
\ead{sboddin@mathematik.uni-kassel.de}
\author[2]{Dorothee Knees}
\ead{dknees@mathematik.uni-kassel.de}
\author[1]{Jörn Mosler\corref{cor1}}
\ead{joern.mosler@tu-dortmund.de}
\cortext[cor1]{Corresponding author}

\affiliation[1]{organization={Institute of Mechanics, TU Dortmund University},
addressline={Leonhard-Euler-Strasse 5},
postcode={44227},
city={Dortmund},
country={Germany}}

\affiliation[2]{organization={Institute of Mathematics, University of Kassel},
addressline={Heinrich-Plett Str.~40},
city={Kassel},
postcode={34132},
country={Germany}}

\begin{abstract}
The modeling of cracks is an important topic -- both in engineering as well as in mathematics. Since crack propagation is characterized by a free boundary value problem (the geometry of the crack is not known beforehand, but part of the solution), approximations of the underlying sharp-interface problem based on phase-field models are often considered. Focusing on a rate-independent setting, these models are defined by a unidirectional gradient-flow of an energy functional. Since this energy functional is non-convex, the evolution of the variables such as the displacement field and the phase-field variable might be discontinuous in time leading to so-called brutal crack growth. For this reason, solution concepts have to be carefully chosen in order to predict discontinuities that are physically reasonable. One such concept is that of Balanced Viscosity solutions (BV solutions). This concept predicts physically sound energy trajectories that do not jump across energy barriers. The paper deals with a time-adaptive finite element phase-field model for rate-independent fracture which converges to BV solutions. The model is motivated by constraining the pseudo-velocity of the crack tip. The resulting constrained minimization problem is solved by the augmented Lagrangian method. Numerical examples highlight the predictive capabilities of the model and furthermore show the efficiency and the robustness of the final algorithm.
\end{abstract}


\end{frontmatter}

\input{Chap_Introduction}
\input{Chap_Theory}
\input{Chap_NumImp}
\input{Chap_ConceptEx}

\input{Introduction_Example}
\input{Chap_NumExp}

\section{Conclusion}

A time-adaptive finite element phase-field model suitable for rate-independent fracture was elaborated in this paper. The model is based on the Efendiev \& Mielke scheme \cite{efendiev_rate-independent_2006} which is known to converge to BV solutions (Balanced Viscosity solutions). In contrast to global energy minimization, BV solutions predict brutal crack growth significantly more realistically. More precisely, the solution jumps usually too early in the case of global energy minimization. That was shown by means of two new examples.

Evidently, the Efendiev \& Mielke scheme is not the only scheme predicting local solutions. A frequently employed algorithm also leading to local energy minima is classic alternate minimization. However, although alternate minimization is indeed very robust, it usually does not converge to BV solutions. Unlike BV solutions, alternate minimization does not necessarily predict energetically optimal trajectories. To be more precise, physically unreasonable jumps across energy barriers might occur. Furthermore, the  Efendiev \& Mielke scheme is based on a time-adaptive discretization -- another unique selling property. This adaptivity automatically detects the onset of brutal crack growth and reduces the time steps. This, in turn, increases the robustness of the algorithm significantly.

The novel phase-field model is characterized by constrained energy minimization. While one constraint is associated with the standard irreversibility condition reflecting the monotonicity of crack growth, the second constraint is related to the Efendiev \& Mielke scheme. It was shown that the latter can be motivated by constraining the (pseudo) velocity of the crack tip. The resulting constrained non-linear minimization problem was implemented by means of the augmented Lagrangian method, combined with alternate minimization. Both sub-problems are solved using Newton's method. However, one of the involved stiffness matrices is dense. For this reason, a novel algorithm based on the Sherman-Morrison formula was elaborated. It leads to a linearized system of equations which is sparse. Thus, fast solvers can be applied. Numerical examples finally highlighted the predictive capabilities of the model and furthermore, showed the efficiency and the robustness of the final algorithm.


\section*{Acknowledgment}
Financial support from the German Research Foundation (DFG) via SPP 2256 (project number 441222077), project 13, is gratefully acknowledged. Furthermore, the authors gratefully acknowledge the computing time provided on the Linux HPC cluster at TU Dortmund University (LiDO3), partially funded in the course of the Large-Scale Equipment Initiative by the German Research Foundation (DFG) as project 271512359.
\section*{Data availability}
The numerical settings and material parameters used for the simulations are provided in the main text and respecting figures.
\section*{Declaration of competing interests}
The authors declare that they have no known competing financial interests or personal relationships that could
have appeared to influence the work reported in this paper.


  \bibliographystyle{elsarticle-num} 
  \bibliography{literature}


\end{document}

%% file: definitions.tex
\newcommand{\dd}{\mathrm{d}}

\newcommand{\field}[1]{\mathbb{#1}}

\newcommand{\R}{\field{R}}

  \def\calF{{\mathcal F}}
\def\calG{{\mathcal G}}  
  \def\calL{{\mathcal L}}
  
  \def\calR{{\mathcal R}}
\def\calS{{\mathcal S}}  
\def\calV{{\mathcal V}}  
 
\def\rmd{{\mathrm d}}

\def\BS{\boldsymbol} 

\def\bfepsilon{{\BS\varepsilon}}

\DeclareMathOperator{\argmin}{argmin}

\makeatletter%
\newcommand\xrupharpoonast[2][]{%
	\ext@arrow 0099{\rupharpoonfill@}{#1}{#2}\hspace{-4pt}{}^\ast\hspace{3pt}}
\def\rupharpoonfill@{%
	\arrowfill@\relbar\relbar\rightharpoonup}
\makeatother
\makeatletter%
\newcommand\xrupharpoon[2][]{%
	\ext@arrow 0099{\rupharpoonfill@}{#1}{#2}}
\def\rupharpoonfill@{%
	\arrowfill@\relbar\relbar\rightharpoonup}
\makeatother
\makeatletter%
\newcommand\xrarrow[2][]{%
	\ext@arrow 0099{\rarrowfill@}{#1}{#2}}
\def\rarrowfill@{%
	\arrowfill@\relbar\relbar\longrightarrow}
\makeatother

\newcommand{\abs}[1]{\left\lvert#1\right\rvert}      

\newcommand{\norm}[1]{\left\lVert#1\right\rVert}

\definecolor{darkgreen}{rgb}{0.0, 0.5, 0.0}

\usepackage{eucal}

\newcommand{\gradz}{\grad z}

\renewcommand{\dd}[1]{\ensuremath{\operatorname{d}\!{#1}}}

\newcommand{\uvec}{\boldsymbol{\mathrm{u}}}

\newcommand{\grad}{\nabla}
\renewcommand{\u}{\boldsymbol{u}}
\newcommand{\x}{\boldsymbol {x}}
\newcommand{\sig}{\boldsymbol{\sigma}}
\newcommand{\eps}{\boldsymbol{\varepsilon}}

\newcommand{\ulist}{\boldsymbol{\mathrm u}}
\newcommand{\zlist}{\boldsymbol{\mathrm z}}

%% file: Chap_Introduction.tex
\section{Introduction}

The initiation and propagation of cracks is one of the major sources for failure of engineering structures. Accordingly, the modeling of cracks enjoys a long tradition in the mechanics and physics community, cf. \cite{griffith_phenomena_1921,barenblatt_mathematical_1962}. At the same time, the equations describing crack propagation are mathematically very challenging and give rise to a complex temporal evolution of the mechanical system. An illustrative example is provided by so-called {\em brutal crack growth} for rate-independent systems. In this case, a finite crack extension occurs in a zero time-interval, i.e., the evolution of the displacement field is discontinuous in time, cf. \cite{francfort_revisiting_1998}. Due to the aforementioned mathematical challenges, the modeling of cracks also enjoys a long tradition in the mathematics community, cf. \cite{dal_maso_model_2002,mielke_rate-independent_2015,knees_inviscid_2008}. For this reason, a sound modeling of crack propagation requires a concerted collaboration between mechanics and mathematics. Within this paper, a novel mathematically and physically sound time-adaptive finite element phase-field model suitable for rate-independent fracture mechanics is presented. In particular, it is able to detect the correct loads at which discontinuities in crack propagation occur.

The focus is on rate-independent brittle fracture. Most sound models dealing with this subject go back to the pioneering work by Griffith \cite{griffith_phenomena_1921}. Within Griffith's model, a crack can propagate (infinitesimally), if the energy release rate reaches a critical material threshold -- the fracture energy. It bears emphasis that  Griffith's model only captures crack propagation, not crack initiation. More precisely, crack propagation requires a stress singularity -- such as that characterizing the already existing crack tip, cf. \cite{bourdin_variational_2008,kumar_revisiting_2020}. Griffith's model was modified in \cite{francfort_revisiting_1998}. Partly motivated by the incapability to predict crack initiation, Francfort and co-workers embedded Griffith's criterion into the framework of global energy minimization. As a consequence, the models \cite{griffith_phenomena_1921} and \cite{francfort_revisiting_1998} are not equivalent, see \cite{knees_fully_2022} for an illustrative example.

The models in \cite{griffith_phenomena_1921} and \cite{francfort_revisiting_1998} are based on so-called sharp-interface descriptions, i.e., the crack width is zero with respect to the reference configuration. Although the models  \cite{griffith_phenomena_1921} and \cite{francfort_revisiting_1998} are conceptually very simple, their numerical implementation is indeed challenging, since one deals with free boundary value problems (the geometry and the location of the crack is not known beforehand, but part of the solution). For this reason, phase-field approximations have been proposed, cf. \cite{bourdin_numerical_2000}. Within these models, the crack is represented by a diffuse interface having a finite thickness with respect to the reference configuration. Unlike the underlying sharp-interface problem, phase-field models do not lead to free boundary value problems and tracking as well as updating the crack's geometry becomes straightforward. The mathematical justification of phase-field models relies on the framework of $\Gamma$-convergence, cf. \cite{bourdin_variational_2008,bourdin_numerical_2007,ambrosio_approximation_1990}. As shown, e.g., in \cite{bourdin_numerical_2000,Giacomini-2005}, phase-field models often $\Gamma$-converge to an underlying sharp interface problem.

Phase-field models of fracture show the canonical variational principle of energy minimization. For instance, the evolution of the order-parameter defining the diffuse crack follows from a unidirectional gradient-flow. The variational structure of such approaches has been intensely studied  both from a mathematical point of view (see \cite{mielke_rate-independent_2015}), as well as from a mechanical point of view (see e.g. \cite{bourdin_variational_2008,miehe_thermodynamically_2010}).

The energy defining the phase-field model of rate-independent fracture is non-convex. This leads to two effects: (1) Brutal crack growth might occur. The mechanical systems might jump from a certain minimum to a different minimum and thus, the displacement field evolves discontinuously in time. (2) Local energy minimizers are often characterized by a different mechanical response compared to that associated with the respective global energy minimizer. Furthermore, to the best knowledge of the authors, no algorithm has been proposed yet which identifies the global energy minimum. An attempt to identify minima closer to the global one is represented by the backtracking algorithm, cf. \cite{bourdin_numerical_2007}. While for some mechanical problems the algorithm indeed finds the global minimum, counterexamples can be easily found as well. The non-convexity of the energy resulting in non-uniqueness of the solution animated the mathematical community to propose different solution concepts. For a recent overview on this subject, the interested reader is referred to \cite{mielke_rate-independent_2015}.

Among the different solution concepts, global energetic solutions are very popular. This is, at least, related to two reasons. First, it represents the natural solution concept for $\Gamma$-convergence -- the framework by which convergence of the phase-field model to a sharp interface problem was shown, cf. \cite{bourdin_numerical_2000,Giacomini-2005}, and also the approximation by means of the backtracking algorithm \cite{bourdin_numerical_2007}. Second and as mentioned before, global energy minimization also captures crack initiation. However, it is also well known that it usually predicts crack initiation and propagation too early. A physically unreasonable prediction of global energy minimization can be found, for instance, in \cite{knees_fully_2022}. The example analyzed shows that global energy minimization might predict crack initiation, if the driving force is not even close to the fracture energy and the crack-growth is over-predicted.

Within this article an algorithm is proposed that avoids such physically unreasonable predictions. From a mathematical point of view the interest is in approximating Balanced Viscosity solutions (BV solutions). Conceptually, such solutions appear as inviscid limit of a viscous damage regularization. Since from a computational point of view it is very delicate to relate the viscosity/damping parameter and the time-increment in a practically reasonable way (see \cite{knees_computational_2013} for a discussion/illustration of this issue), an alternative approach is proposed here. It relies on a time adaptive discretization first proposed in \cite{efendiev_rate-independent_2006} by Efendiev and Mielke (E\&M) and analyzed mathematically in \cite{boddin_approximation_2022}.

Particularly, the (physical) time is fixed in the case of brutal crack growth. This is implemented by prescribing the arc-length of the internal variable (such as the phase-field) and the time to be controlled.
In this context, the term ``pseudo-velocity'' will refer to the change of the internal variable in relation to its arc-length parametrization. In \cite{knees_fully_2022}, the E\&M-algorithm was applied to a discrete model of brittle fracture (sharp interface) and analyzed from a mathematical point of view. A phase-field approximation for damage was studied in \cite{sievers_convergence_2022}. In the present paper, a numerically efficient phase-field implementation is elaborated. It shows the following novel features:
\begin{itemize}
	\item Consistent approximation of BV solutions (Balanced Viscosity)
	\item The algorithm is well-motivated from a physics point of view:
	\begin{itemize}
		\item It predicts energetically optimal trajectories (no jump across energy barriers)
		\item The pseudo-velocity of the crack tip is bounded 
	\end{itemize}
	\item Control of pseudo-velocity of the crack by means of the augmented Lagrangian method
	\item Enforcement of thermodynamical consistency by means of augmented Lagrangian method
	\item Increasing the numerical efficiency by avoiding dense stiffness matrices
\end{itemize}
To illustrate different solution concepts a new  finite dimensional example that has a similar structure as the energies appearing in phase-field damage models is presented.  The example in particular shows that pure alternate minimization might develop jumps that do cross local minima and where every trajectory in the phase space that connects the starting point and the end point of the jump crosses an energy barrier at some point. The time-adaptive local minimization scheme \cite{efendiev_rate-independent_2006} does not have these problems.
Furthermore, in order to highlight the difference between local and global energy minima as well as in order to demonstrate the robustness of the proposed algorithm, a new benchmark is introduced: a simple snap-through problem.\\[1ex]
The paper is organized as follows: Section~\ref{sec:Theory} gives a concise review of the phase-field theory of brittle fracture. The section introduces the notations used in the paper and finishes with a discussion on the irreversibility condition intrinsic to cracking, i.e., the second-law of thermodynamics. The major novel contribution is associated with Section~\ref{sec:Implementation}. Here, an efficient algorithmic formulation of the scheme from \cite{boddin_approximation_2022} recapped in Section~\ref{sec:Mielke} is elaborated. The focus is on the implementation of the constraints related to the second-law of thermodynamics (monotonicity of cracking) and the Efendiev \& Mielke scheme. As mentioned above two conceptual examples with finite dimensional state spaces are presented in Section~\ref{sec:ConceptualExamples}. The predictive capabilities, the numerical robustness and the efficiency of the novel algorithm are finally highlighted in Section~\ref{sec:Experiments}.

%% file: Chap_Theory.tex
\section{Phase-field theory of brittle fracture in a nutshell} \label{sec:Theory}

This section gives a concise summary of the phase-field method for (quasi-)brittle fracture. It is based on the pioneering work \cite{bourdin_numerical_2000}. The thermodynamic fundamentals as well as the modeling of the tension-compression asymmetry of crack propagation (MCR effect; Microcrack-Closure-Reopening effect) follow \cite{miehe_thermodynamically_2010}.

\begin{figure}[htbp]
	\centering
	\begin{psfrags}%
		\psfrag{s1}[t][t][1.5]{\color[rgb]{0,0,0}\setlength{\tabcolsep}{0pt}\begin{tabular}{c}$\Omega$\end{tabular}}%
		\psfrag{s2}[t][t][1]{\color[rgb]{0,0,0}\setlength{\tabcolsep}{0pt}\begin{tabular}{c}$\partial\Omega_D$\end{tabular}}%
		\psfrag{s3}[b][b][1]{\color[rgb]{0,0,0}\setlength{\tabcolsep}{0pt}\begin{tabular}{c}$\partial\Omega_N$\end{tabular}}%
		\psfrag{s4}[b][b][1.5]{\color[rgb]{0,0,0}\setlength{\tabcolsep}{0pt}\begin{tabular}{c}$\Gamma$\end{tabular}}%
		\psfrag{s5}[t][t][1.5]{\color[rgb]{0,0,0}\setlength{\tabcolsep}{0pt}\begin{tabular}{c}$\Gamma_\epsilon$\end{tabular}}%
		\psfrag{s8}[t][t][1]{\color[rgb]{0,0,0}\setlength{\tabcolsep}{0pt}\begin{tabular}{c}$x_1$\end{tabular}}%
		\psfrag{s9}[tl][tl][1]{\color[rgb]{0,0,0}\setlength{\tabcolsep}{0pt}\begin{tabular}{c}$x_2$\end{tabular}}%
		\psfrag{s10}[t][t][1]{\color[rgb]{0,0,0}\setlength{\tabcolsep}{0pt}\begin{tabular}{c}$x_3$\end{tabular}}%
		\psfrag{b1}[t][t][1]{\color[rgb]{0,0,0}\setlength{\tabcolsep}{0pt}\begin{tabular}{c}$a$\end{tabular}}%
		\psfrag{b2}[t][t][1]{\color[rgb]{0,0,0}\setlength{\tabcolsep}{0pt}\begin{tabular}{c}\end{tabular}}%
		\psfrag{z0}[t][t][1]{\color[rgb]{0,0,0}\setlength{\tabcolsep}{0pt}\begin{tabular}{c}$z=0$\end{tabular}}%
		\psfrag{z1}[t][t][1]{\color[rgb]{0,0,0}\setlength{\tabcolsep}{0pt}\begin{tabular}{c}$z=1$\end{tabular}}%
		\includegraphics[scale=0.95]{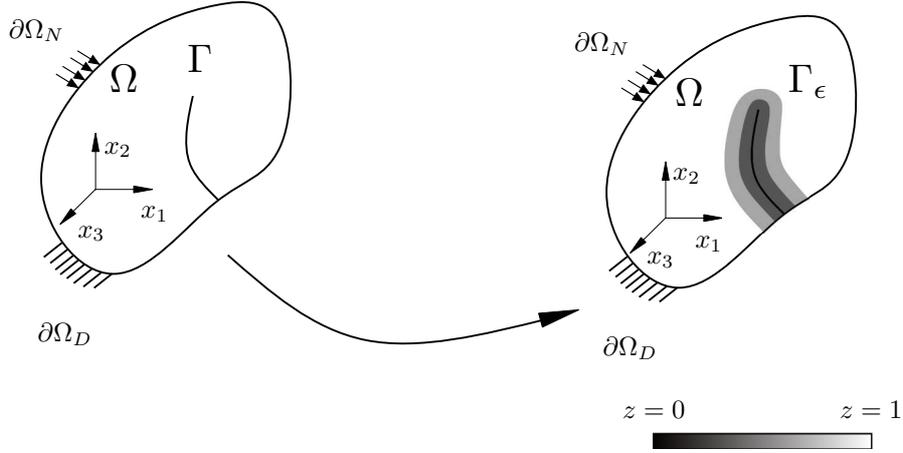}
	\end{psfrags}%
	\caption{Sketch of the phase field approximation of cracks: (left) sharp interface problem with a crack $\Gamma$; (right) approximation by means of diffuse interface $\Gamma_\epsilon$ showing a finite thickness}
	\label{pic:phase-field}
\end{figure}

In line with \cite{bourdin_numerical_2000}, the idea of the phase-field method for (quasi-)brittle fracture is the spatial approximation of sharp cracks by means of an order parameter $z$. While $z=1$ signals the initial, fully intact material, $z=0$ is associated with the fully damaged stated, i.e., a crack. However, and in contrast to the underlying sharp crack description, order parameter $z$ can also attain values between 0 and 1. This leads to a diffuse interface, i.e., the underlying sharp interface (crack) with a zero-thickness is approximated by a diffuse interface with a finite thickness, see Fig.~\ref{pic:phase-field}.

\subsection{Energy minimization -- variational structure}

The phase field approximation \cite{bourdin_numerical_2000} of the sharp interface problem characterizing fracture mechanics is based on the framework of $\Gamma$-convergence. Hence, an energetic variational formulation represents the canonical starting point. Such a formulation will be summarized here.

Cracking in brittle materials can be interpreted as a competition between surface and bulk energies. The latter are assumed to be of the type 
\begin{flalign}
\label{eq:bulk-energy-1}
\Psi(\nabla\u,z) &= \int_{\Omega}\psi(\nabla\u,z) \dd{V}= \int_{\Omega} [z^2+k]\psi_0^+(\nabla \u)+ \psi_0^-(\nabla\u) \dd{V}.
\end{flalign}
Here, $\psi$ is the volume-specific bulk energy (Helmholtz energy), $\Psi$ is its integrated global counterpart and $\u$ is the displacement field. According to Eq.~\eqref{eq:bulk-energy-1}, the volume-specific Helmholtz energy is decomposed into a positive and a negative part. While negative part $\psi_0^-$ corresponds to compression, positive part $\psi_0^+$ is related to tensile states. Only positive part $\psi_0^+$ is affected by cracking which can be seen by prefactor $[z^2+k]$ in Eq.~\eqref{eq:bulk-energy-1}. Small parameter $k>0$ is chosen for numerical reasons, cf. \cite{bourdin_numerical_2000}. Following \cite{miehe_thermodynamically_2010} and inspired by Hooke's model, the energies of the initial, fully intact material are chosen as
\begin{flalign}
	\label{eq:energies-plus-minus-1}
	\psi_0^\pm(\eps)=\frac{\lambda}{2} \left\langle \mathrm{tr}(\eps) \right \rangle _{\pm}^2 + \mu \mathrm{tr}(\eps_{\pm}^2),
\end{flalign}
where $\mu$ and $\lambda$ are the Lam\'e constants, $\bfepsilon=1/2[\nabla\u+(\nabla\u)^T]$ is the engineering strain tensor (geometrically linearized setting) and the bracket operator is defined by $\left \langle x \right \rangle_{\pm} := (x\pm \abs{x})/2$. The latter also defines the positive and the negative parts of the strain tensor $\eps_{\pm}$ in Eq.~\eqref{eq:energies-plus-minus-1} by means of the spectral decomposition
\begin{flalign}
	\eps_{\pm}:= \sum_{i=1}^{3}\left \langle \varepsilon_i \right \rangle_{\pm} \boldsymbol{n}_i \otimes \boldsymbol{n}_i
\end{flalign}
with $\boldsymbol{n}_i$ being the eigenvectors and $\varepsilon_i$ the eigenvalues of $\eps$. Since the decomposition of $\eps$ into these positive and negative parts is orthogonal ($\eps_-\cdot\eps_+=\mathbf{0}$ and $\eps_-:\eps_+=0$), $\psi_0^++\psi_0^-$ results in the standard Hooke's model.

The second important energy contribution is due to the generation of new interfaces (cracks). Denoting the area-specific fracture energy as  $g_{\text{c}}$, the total energy accounting for all new cracks is approximated by
\begin{flalign}
\label{eq:surface-energy-total-1}
\calG_{\text{f}}(z,\gradz) =\frac{g_{\text{c}}}{2l}\int\limits_\Omega [1-z]^2+l^2\,\grad z \cdot \grad z \dd{V}
=:g_{\text{c}}\,\int\limits_\Omega f_l(z,\nabla z)\,\dd{V}.
\end{flalign}
Term $f_l(z,\nabla z)$ can be interpreted as an approximation of the Dirac-delta distribution. More precisely and as shown in \cite{modica_gamma-convergence_1977}, functional $\int_\Omega f_l\dd{V}$ (under suitable boundary conditions) $\Gamma-$converges to the interface area occupied by the cracks. As consequence, $\calG_{\text{f}}$ indeed converges to the energy required for the generation of all cracks.

Finally, a third energy contribution is introduced. It captures the effect of external dead loads and reads
\begin{flalign}
\label{eq:energy-loads-1}
\Pi_{\text{ext}}(\u)=\int_\Omega \boldsymbol{b}\cdot \u \dd{V} + \int_{\partial \Omega_N}\boldsymbol t \cdot \u \dd{V},
\end{flalign}
where $\boldsymbol{b}$ are body forces and $\boldsymbol{t}$ are tractions prescribed at the Neumann boundary $\partial \Omega_N$.

By summarizing energy contributions~\eqref{eq:bulk-energy-1}, \eqref{eq:surface-energy-total-1} and \eqref{eq:energy-loads-1}, one obtains the total energy of the mechanical system
\begin{flalign}
\calF(\u,\nabla \u,z,\nabla z)=\Psi(\nabla\u,z)+\calG_{\text{f}}(z,\gradz)-\Pi_{\text{ext}}(\u). \label{eq:totalPotential}
\end{flalign}
The unknown fields $\u$ and $z$ follow from minimizing this functional under suitable boundary conditions and suitable irreversibility conditions (cracks cannot heal). However, due to the non-convexity of this energy, the solution is not unique in general and differs depending on the considered solution concept, e.g., global energetic solution or BV solution. The paper deals with the efficient algorithmic approximation of BV solutions.

\subsection{Balance equations}

The minimum of functional~\eqref{eq:totalPotential} is characterized by a non-negative first variation
\begin{flalign}
\delta \calF = \delta_{\u} \Psi \cdot \delta \u + \delta_z \Psi \delta z + \delta_z \calG_{\text{f}} \delta z - \delta_{\u} \Pi_{\text{ext}} \cdot \delta \u \geq  0\qquad\forall\delta z\le 0,\delta \u. 
\label{eq:first-variation-1}
\end{flalign}
While $\delta(\bullet)$ in Eq.~\eqref{eq:first-variation-1} denotes the variation of $(\bullet)$, $\delta_{a}(\bullet)$ is the variational derivative of $(\bullet)$ with respect to $a$. Furthermore, $\delta\u$ and $\delta z$ are the variations (test functions) of $\u$ and $z$. Since, $\u$ and $z$ are independent fields, Eq.~\eqref{eq:first-variation-1} yields on the one hand
\begin{flalign}
& \delta_{\u}\Psi\cdot \delta \u -\delta_{\u}\Pi_{ext}\cdot \delta \u = 0 \qquad\forall\delta \u \label{eq:totalvariation_u}.
\end{flalign}
By introducing the stress tensor $\sig=\partial_{\eps}\psi$, together with the variation of the strains $\delta\eps=1/2\,[\nabla\delta\u+(\nabla\delta\u)^T]$, Eq.~\eqref{eq:totalvariation_u} can be rewritten as
\begin{flalign}
0=\int_\Omega \sig:\delta \eps \dd{V}-\int_\Omega \boldsymbol b\cdot \delta \u \dd{V} - \int_{\partial \Omega_N}\boldsymbol t\cdot \delta \u \dd{A}\qquad\forall\delta\u, \label{eq:weakform_displacement}
\end{flalign}
which is the weak form of equilibrium. On the other hand, a variation of potential $\calF$ with respect to $z$ leads to
\begin{flalign}
\delta_z \Psi \delta z + \delta_z \calG_{\text{f}} \delta z \geq 0 \qquad\forall\delta z \leq 0. \label{eq:varzinequal}
\end{flalign}
By choosing $\delta z=\dot z\leq 0$ this variational inequality implies \cite{marengo_rigorous_2021}
\begin{flalign}
\left[\delta_z \Psi + \delta_z \calG_{\text{f}}\right]\dot z = 0, \qquad \delta_z \Psi + \delta_z \calG_{\text{f}} \leq 0.\label{eq:Griffithlaw}
\end{flalign}
Accordingly, $\delta_z \Psi + \delta_z \calG_{\text{f}}= 0$ defines a level set separating elastic from inelastic states. Eq.~\eqref{eq:Griffithlaw}$_1$ is thus a consistency condition, while $\dot z\leq 0$ enforces the irreversibility of the process.

To summarize, $z$ and $\u$ have to satisfy $\dot{z}\leq 0$, \eqref{eq:totalvariation_u} and \eqref{eq:Griffithlaw}.

\subsection{Remark on the mechanical dissipation}\label{subsec:Dissipation}

Although the irreversible nature of cracking was already considered in the pioneering work \cite{bourdin_numerical_2000}, thermodynamic aspects were not explicitly discussed. In particular, the mechanical dissipation was not introduced. This was only done in the works by Miehe and co-workers, e.g., \cite{miehe_thermodynamically_2010}. However, since an isothermal setting was adopted in \cite{bourdin_numerical_2000,miehe_thermodynamically_2010} -- and also in the present paper -- mechanical dissipation is not well-defined, cf. \cite{bartels_thermomechanical_2015}. In this subsection, a concise review of different ideas to incorporate the irreversibility associated with crack propagation is given.

In the model \cite{bourdin_numerical_2000}, the irreversible nature of cracking is only enforced for points that have already undergone complete failure, i.e., $z=0$. For these points, a respective Dirichlet boundary condition is set. By way of contrast, the generation of new cracks is considered to be a fully dissipative process in \cite{miehe_thermodynamically_2010}. Accordingly, and in line with the definition of the fracture energy~\eqref{eq:surface-energy-total-1}, the (volume-specific) dissipation reads
\begin{flalign}
	\label{eq:surface-energy-total-1-25}
	\mathcal{D}=g_{\text{c}}\,\dot{f}_l=g_{\text{c}}\,\delta_z f_l\,\dot{z}\ge 0.
\end{flalign}
It bears emphasis that $f_l=f_l(z,\nabla z)$. As a consequence, $\mathcal{D}$ can be interpreted as a state-dependent dissipation. As a matter of fact, this expression also depends on the gradient of the state. Clearly, $\dot{z}=0$ characterizes an elastic state and thus, $\mathcal{D}=0$. Furthermore, consistency condition~\eqref{eq:Griffithlaw}$_1$ shows that an inelastic step requires $\delta_z \calG_{\text{f}}=-\delta_z \Psi$. By localizing this equation one obtains (see Eq.~\eqref{eq:bulk-energy-1})
\begin{flalign}
	\label{eq:surface-energy-total-1-282}
	g_{\text{c}}\,\delta_z f_l=-2\, z\,\psi^+_0\le 0.
\end{flalign}
As a consequence, $\dot z\le 0$ is equivalent to $\mathcal{D}\ge 0$. A second, alternative dissipation functional reads
\begin{flalign}
	\label{eq:surface-energy-total-1-282322}
	\tilde{\mathcal{D}}=-\frac{g_{\text{c}}}{l} \dot z \ge 0
\end{flalign}
for admissible $\dot z$. It is given e.g. by \cite{boddin_approximation_2022}, where it is denoted as $\calR$ instead of $\tilde{\mathcal{D}}$.
Again, $\dot{z}\le 0$ is equivalent to $\tilde{\mathcal{D}}\ge 0$. However, and in contrast to Eq.~\eqref{eq:surface-energy-total-1-25}, dissipation functional~\eqref{eq:surface-energy-total-1-282322} is now state-independent. If the volume-specific bulk's energy is furthermore chosen as
\begin{flalign}
	\label{eq:surface-energy-total-1-28233222}
	\tilde{\psi}=\psi+\frac{g_{\text{c}}}{2l}\,[z^2+l^2 \nabla z\cdot\nabla z]
\end{flalign}
one observes that both models are equivalent (in an isothermal setting). More precisely,
\begin{flalign}
	\label{eq:surface-energy-total-1-28233cs222}
	\delta\calF=\delta\left[\int\limits_{\Omega}\left\{\tilde{\psi}+\int_{t_0}^t \tilde{\mathcal D}\dd{t} \right\}\dd{V}-\Pi_{\text{ext}}(\u,t)\right].
\end{flalign}
~
Within the numerical implementation, irreversibility is ensured by enforcing $\dot z\le 0$.

\section{An adaptive time-discretization of the phase-field model --- the Efendiev \& Mielke scheme}\label{sec:Mielke}

Since potential $\calF$ defining the phase-field model of fracture mechanics is non-convex, different solution concepts can be found in the literature, cf. \cite{knees_convergence_2017}. 
In general, they are based on a time-discretized setting with a discrete time series $\{t_1,\ldots,t_j,t_{j+1},\ldots,T\}$.  For a certain time step $t_j$  the aim is to find fields $\u_j:=\u(t_j)$ and $z_j:=z(t_j)$  that satisfy
\begin{flalign}
&z_{j}=\arg \underset{z}{\min}\left\{ \calF(t_j,\u_{j},z):z\leq z_{j-1}\right\},\label{eq:constminprob1} \\
&\u_{j}=\arg \underset{\u}{\min}\left\{\calF(t_j,\u,z_j)\right\}. \label{eq:constminprob12}
\end{flalign}
The corresponding variational equations/inequalities then are a discretized version of \eqref{eq:totalvariation_u} and \eqref{eq:Griffithlaw}.
It is again emphasized that due to the non-convexity of $\calF(t,\cdot,\cdot)$ the coupled problem~\eqref{eq:constminprob1}--\eqref{eq:constminprob12} does not have to have a unique solution. The approximated solution type depends on the chosen algorithm. Possible options to obtain functions $\u_j$ and $z_j$ satisfying \eqref{eq:constminprob1}--\eqref{eq:constminprob12} could be the following:

\begin{itemize}
	\item One option is to follow a time-incremental global minimization procedure with 
	\begin{align}
		(\u_j,z_j)\in\arg \underset{(\u,z)}{\min}\left\{ \calF(t_j,\u,z):z\leq z_{j-1}\right\}. \label{eq:glob-min}
	\end{align}
	This ultimately leads to global energetic solutions.
	\item Another option is to deploy the alternate minimization procedure, which will converge to a pair $(\u_j,z_j)$ with the desired property \eqref{eq:constminprob1}--\eqref{eq:constminprob12}.
\end{itemize}
However, in this paper, the sound concept of solutions with balanced viscosity is followed.  
In order to compute a Balanced Viscosity solution (BV), a more specific choice of appropriate iterates $\u_j$ and $z_j$ is made. It is based on the scheme proposed in \cite{efendiev_rate-independent_2006} and comprises a constrained minimization in combination with an adaptive time-discretization. Incorporating the selection process from \cite{efendiev_rate-independent_2006} in \eqref{eq:constminprob1}--\eqref{eq:constminprob12} the following time-adaptive incremental scheme for the phase-field model of fracture is obtained
\begin{subequations}\label{eq:EM_system}
	\begin{empheq}[left=\empheqlbrace]{align}
		&\u_j=\arg \underset{\u}{\min}\left\{\calF(t_j,\u,z_j)\right\}\label{eq:minprobdisp}\\
		&z_j \in \arg \underset{z}{\min}\left\{\calF(t_j,\u_j,z): z-z_{j-1}\leq 0, \norm{z-z_{j-1}}_\calV\leq \rho\right \}\label{eq:constminprob}\\
		&t_{j+1}=t_j+\rho - \norm{z_j-z_{j-1}}_\calV\label{eq:timestep}.
	\end{empheq}
\end{subequations}
Here $\rho$ is an arc-length parameter coupling the time-step size to the increment of the phase-field, i.e., $\rho=(t_{j+1}-t_j)+\norm{z_j-z_{j-1}}_\calV$, cf. Eq.~\eqref{eq:timestep}. If this parameter converges to zero, a BV solution is obtained. In order to understand the idea of this scheme, a constant (finite) $\rho$ is considered. Accordingly, phase-field $z$ is not allowed to evolve too fast, since $\norm{z_j-z_{j-1}}_\calV\leq \rho$. In the limiting case $\norm{z_j-z_{j-1}}_\calV= \rho$, the time $t_{j+1}=t_j$ does not evolve. This is precisely the case for brutal crack growth (finite crack extension within a zero time-interval). For brutal crack growth, the algorithm thus automatically sets the time step to zero and allows the crack to propagate in several increments (at the same point in time).

It bears emphasis that problem~\eqref{eq:minprobdisp}-\eqref{eq:timestep} does not completely define an algorithm, since it is not clarified how to  compute the points $(\u_j,z_j)$. For instance, the minimization with respect to $\u$ and $z$ can be implemented monolithically or in a staggered manner (alternate minimization). Furthermore, the inequality constraints can be resolved by means of different methods.
An excellent state-of-the-art overview on constrained optimization can be found in \cite{geiger_theorie_2002}. Among the different optimization schemes covered in this work, the augmented Lagrangian method (also known as method of multipliers) is very promising - both from a mathematical as well as from a numerical point of view. For instance, and unlike classic penalty or interior point methods, the augmented Lagrangian approach fulfills the constraints exactly. As a matter of fact, it can be shown under relatively mild assumptions that any Karush-Kuhn-Tucker point of the underlying constrained optimization problem is also a stationary point of the augmented Lagrangian function. Equally important, the well-established implementation of the augmented Lagrangian method by means of the Hestenes-Powell update leads to an unconstrained minimization problem. For this reason, the augmented Lagrangian method preserves the variational structure of the problem - one still seeks for minima of an (extended) energy functional with respect to $\u$ and $z$. For this reason, this method will be considered in what follows. Since the resulting sub-problems (minimization with respect to $\u$ and $z$) are also still separately convex, alternate minimization can be implemented by applying two Newton iterations. This is precisely the idea further elaborated in the next sections.

\subsection{Choice of the norm} \label{sec:Norm}

The core idea of the  Efendiev \& Mielke scheme \cite{efendiev_rate-independent_2006} is an adaptive parameterization of the real time. According to Eq.~\eqref{eq:timestep}, it depends on arc-length parameter $\rho$ (which has to be sent to zero from a mathematical point of view) as well as on the choice of the norm $||\bullet||_\calV$. Within the previous work \cite{boddin_approximation_2022}, the 
\begin{flalign}
	L_p\text{-norm:}\qquad\norm{z-z_{j-1}}_{L_p}=\bigg[\int_{\Omega}|z-z_{j-1}|^p\dd{V}\bigg]^{\frac{1}{p}}\label{eq:LP-Norm}
\end{flalign}
as well as the 
\begin{flalign}
H_1\text{-norm:}\qquad\norm{z-z_{j-1}}_{H_1}=\bigg[\int_{\Omega}[z-z_{j-1}]^2+|\grad (z-z_{j-1})|^2\dd{V}\bigg]^{\frac{1}{2}}
\label{eq:H1-Norm}
\end{flalign}
were considered. From a physics point of view, the $H_1$-norm is of particular interest, since it has a similar structure as the integrated fracture energy~\eqref{eq:surface-energy-total-1}. Indeed
\begin{flalign}
		\label{eq:surface-energy-total-121}
	\calG_{\text{f}}(z-z_{j-1}) =\frac{g_{\text{c}}}{2l}\int\limits_\Omega [1-(z-z_{j-1})]^2+l^2 |\grad (z-z_{j-1})|^2 \dd{V}.
\end{flalign}
By further elaborating this identity one could also replace norm $||\bullet||_\calV$ by
\begin{flalign}
	\label{eq:surface-energy-total-1221}
	\Delta\calG_{\text{f}}:=\calG_{\text{f}}(z)-\calG_{\text{f}}(z_{j-1}).
\end{flalign}
Evidently, $\Delta\calG_{\text{f}}$ is not a norm. However, it highlights the underlying physics idea of the Efendiev \& Mielke scheme very well: the pseudo crack tip velocity is constrained by the scheme. The adjective "pseudo" is used here for two reasons. First, rate-independent systems intrinsically do not depend on a physical time scale. Second and even more important, the Efendiev \& Mielke scheme restricts the pseudo crack tip velocity within an iteration -- not within a time step. Finally, it is noted that this physics analogy can be seen best by choosing $\Delta\calG_{\text{f}}$. However, both the $L_p$-norm as well as the $H_1$-norm also restrict the pseudo crack tip velocity -- certainly, in a more implicit manner. 

%% file: Chap_NumImp.tex
\newcommand{\Sint}{S_{\calV}}
\newcommand{\sint}{s}
\newcommand{\calFtild}{\tilde{\calF}}
\newcommand{\Deltaz}{\Delta z}
\newcommand{\AssOp}{\boldsymbol{\mathsf A}_{e=1}^{nel}}
\newcommand{\Kzstiff}{\boldsymbol{\mathrm{K}}_{zz}}
\newcommand{\Kzstiffsparse}{\boldsymbol{\mathrm{K}}_{sp}}
\newcommand{\KueAB}{\boldsymbol{\mathrm{K}}^{e,AB}_{\u \u}}
\newcommand{\epsfem}{\boldsymbol{\mathrm{\varepsilon}}}
\newcommand{\BAu}{\boldsymbol{B}^u_A}
\newcommand{\BBu}{\boldsymbol{B}^u_B}
\newcommand{\BA}{\boldsymbol{B}_A}
\newcommand{\BB}{\boldsymbol{B}_B}
\newcommand{\rsnu}{\boldsymbol{\mathrm{r}}_{\u}}
\newcommand{\rsnz}{\boldsymbol{\mathrm{r}}_z}
\newcommand{\fintzpd}{\underline{\mathrm{f}}^{int,AL}}
\newcommand{\flistz}{\boldsymbol{\mathrm{f}}_{z}}
\newcommand{\flistztransp}{\underline{\mathrm{f}}^{z2,T}}
\newcommand{\zlin}{\hat z}
\SetKwComment{Comment}{/* }{ */}

\section{Algorithmic formulation} \label{sec:Implementation}

An algorithm for realizing the Efendiev \& Mielke scheme (E\&M) applied to the phase-field model of fracture mechanics is elaborated in this section, cf. Eqs.~\eqref{eq:minprobdisp}-\eqref{eq:timestep}. Following the pioneering work \cite{bourdin_numerical_2000}, alternate minimization (AM) is used in order to compute stationary points. Since underlying potential $\calF$ is separately convex in $\u$ and $z$, this can be implemented in an efficient manner by applying two Newton schemes sequentially.\\
A straightforward combination of AM and E\&M yields the general structure of the algorithm shown in Fig.~\ref{pic:Algo_FEM}.
\begin{figure}[htbp]
  \parbox{\textwidth}{
  \begin{algorithm}[H]
  \caption{General Solution scheme combining AM with E\&M}\label{alg1}
\textbf{1. Input values:} $\rho$, $[0;T]$, \text{BC} \\
\textbf{2. Initialize:} $j=1$, $\u_0=\boldsymbol 0$, $z_0=1$, $t_1=0$\\
\While{$t_j<T$}{
 $(\u_j,z_j)=(\u_{j-1},z_{j-1})$\;
 \While{$\u_j\neq\arg \underset{\u}{\min}\left\{\calF(t_j,\u,z_j)\right\}$}{
 	$\u_j=\arg \underset{\u}{\min}\left\{\calF(t_j,\u,z_j)\right\}$\;
 	$z_j=\arg \underset{z}{\min}\left\{\calF(t_j,\u_j,z):z\leq z_{j-1},\norm{z-z_{j-1}}_\calV\leq \rho\right\}$\;
 }
 $t_{j+1}=t_j+\rho - \norm{z_j-z_{j-1}}_\calV$\;
 $j\leftarrow j+1$;
 }
\end{algorithm}}%
  \caption{General structure of an algorithm for applying the Efendiev \& Mielke scheme to the phase-field model of fracture mechanics}
  \label{pic:Algo_FEM}
\end{figure}
The convergence of such an algorithm to BV solutions was recently shown in \cite{boddin_approximation_2022}.
Although the general structure of the algorithm is indeed straightforward, the implementation is challenging and elaborated next.

\subsection{Implementation of the constraints associated with the phase-field -- Augmented Lagrangian approach}

Since field $\u$ is unconstrained, the minimization of separately convex potential $\calF$ with respect to $\u$ is straightforward, cf.  Eq.~\eqref{eq:minprobdisp}. As mentioned before, Newton's method is applied for that purpose. Due to convexity of $\calF$ in $\u$, no line search strategies are required. By way of contrast, the minimization of  $\calF$ with respect to $z$ is constrained. The respective constraints are the irreversibility condition $z-z_{j-1}\le 0$ as well as inequality
\begin{flalign}
	g(z):=\norm{z-z_{j-1}}_\calV-\rho\le 0\label{eq:ineq-2}
\end{flalign}
related to the E\&M scheme. While constraint $g(z)\le 0$ is defined over the whole body, i.e., by means of an integral, inequality $z-z_{j-1}\le 0$ has to be fulfilled pointwise. The final implementation is based on a spatial finite element discretization with bilinear quadrilateral elements, cf. Subsection~\ref{subsec:fe-implementation}. For such elements, the extrema of field $z$ occur at the nodal points. Accordingly, Ineq.~\eqref{eq:ineq-2} is precisely enforced at the nodes of the finite element triangulation.

An effective way to incorporate the nodal constraint $z-z_{j-1}\le 0$ and the global Ineq.~\eqref{eq:ineq-2} into the non-linear minimization problem is provided by the augmented Lagrangian approach, cf. \cite{geiger_theorie_2002}. It combines the advantages of penalty methods with those of Lagrange multipliers. Following \cite{geiger_theorie_2002}, the consideration of the two inequalities leads to the augmented Lagrangian
\begin{flalign}
\mathcal{L}_a\calF (t_j,\u_j,z,\lambda_1^{1},\ldots,\lambda_1^{n},\lambda_2,\alpha_1^{1},\ldots,\alpha_1^{n},\alpha_2):=\calF(t_j,\u_j,z) + \sum\limits_{A=1}^n\calL_1^{A}(z,\lambda_1^{A},\alpha_1^{A})+\calL_2(z,\lambda_2,\alpha_2),
\label{eq:augmented-potential}
\end{flalign}
where $\calL_1^{A}$ and $\calL_2$ are defined by
\begin{flalign}
&\calL_1^{A}(z,\lambda^{A}_1,\alpha^{A}_1):=\frac{1}{2\alpha^{A}_1}\left[\left[\max\{0,\lambda^{A}_1+\alpha^{A}_1 (z^{A}-z^{A}_{j-1})\}\right]^2-[\lambda^{A}_1]^2\right]\\
&\calL_2(z,\lambda_2,\alpha_2):=\frac{1}{2\alpha_2}\left[\left(\max\{0,\lambda_2+\alpha_2 g(z)\}\right)^2-\lambda_2^2\right].
\label{eq:augmented-constraint-2}
\end{flalign}
Constraint $z-z_{j-1}\le 0$ is enforced for all nodal points $A$ of the triangulation, i.e., $z^{A}-z^{A}_{j-1}\le 0$ for all $1\le A\le n$. Consequently, $n$ Lagrange multipliers $\lambda_1^{A}$ and penalty factors $\alpha_1^{A}$ are required. Further following the augmented Lagrangian approach, the penalty factors $\alpha_1^{A}$ and $\alpha_2$ are iteratively updated and increased, while Lagrange multipliers $\lambda_1^{A}$ and $\lambda_2$ follow a Hestenes-Powell update, cf. \cite{geiger_theorie_2002}. As a consequence, field $z_j$ follows from unconstrained optimization problem
\begin{flalign}
z_j=\arg \underset{z}{\min}\big\{\mathcal{L}_a\calF (\u_j,z,\lambda^{1}_1,\ldots,\lambda^{n}_1,\lambda_2,\alpha^{1}_1,\ldots,\alpha^{n}_1,\alpha_2)\big\}.
\label{eq:augemnted-lagrange-zj}
\end{flalign}
Again, the stationary point is computed by using Newton's method. Due to the convexity of $\mathcal{L}_a\calF$ with respect to $z$, this point is indeed well-defined and the algorithm is very robust.

\subsection{Finite element implementation}\label{subsec:fe-implementation}

Variational problems~\eqref{eq:minprobdisp} and \eqref{eq:augemnted-lagrange-zj} are discretized in space by means of finite elements. For that purpose, displacement-field $\u$ and phase-field $z$ are approximated within each finite element as
\begin{flalign}
&\u(\x)\approx \sum_{A=1}^{n_{e}}N_e^A(\x)\u_e^A && z(\x)\approx\sum_{A=1}^{n_{e}}N_e^A(\x)z_e^A \label{eq:shape-functions-1}&\\
&\grad \u(\x)\approx\sum_{A=1}^{n_{e}}\u_e^A \otimes \grad(N_e^A(\x)) && \gradz(\x) \approx \sum_{A=1}^{n_{e}} z_e^A \grad ( N_e^A(\x)).&
\label{eq:shape-functions-2}
\end{flalign}
Here, $\u_e^A$ and $z_e^A$ are the values at node $A$ and $N_e^A$ is the shape function of the respective node in element $e$. Clearly, continuity of fields $\u$ and $z$ has to be enforced at the boundaries of the finite elements by means of a standard assembly step. According to Eqs.~\eqref{eq:shape-functions-1} and \eqref{eq:shape-functions-2}, the same space is chosen for $\u$ and for $z$. Clearly, this is not mandatory.
First, minimization problem~\eqref{eq:minprobdisp} is considered. Due to the additive interval property, one obtains
\begin{flalign}
\left.\delta\calF\right|_{z_j=\text{const}}=\sum\limits_{e=1}^{\#\text{elements}}\left.\delta\calF_{e}\right|_{z_j=\text{const}}
\qquad\text{with}\qquad
\left.\delta\calF_{e}\right|_{z_j=\text{const}}=\sum\limits_{A=1}^{n_e}
\underbrace{\left.\partial_{\u_e^A}\calF_{e}\right|_{z_j=\text{const}}}_{\displaystyle =:\boldsymbol{\mathrm{r}}^{e,A}_{\u}}\cdot\delta\u^A_e
\label{eq:additive-integral}
\end{flalign}
where $\#\text{elements}$ is the number of finite elements within the triangulation and $n_e$ is the number of nodes per element, cf. Eqs.~\eqref{eq:shape-functions-1}-\eqref{eq:shape-functions-2}. The assembled counterpart of $\boldsymbol{\mathrm{r}}^{e,A}_{\u}=\left.\partial_{\u_e^A}\calF_{e}\right|_{z_j=\text{const}}$ defines the discrete equilibrium condition and thus, it has to vanish. A straightforward computation yields
\begin{flalign}
&\boldsymbol{\mathrm{r}}^{e,A}_{\u}=\int_{\Omega_e}\underbrace{\left[ [z^2+\eta]\partial_{\eps} \psi_0^+ + \partial_{\eps} \psi_0^- \right]}_{\displaystyle \sig=\partial_{\eps}\psi}\cdot\grad N_e^A \dd{V}-\int_{\Omega_e}\boldsymbol{b}N^A_e\dd{V} - \int_{\partial\Omega_e^N} \boldsymbol{t}N^A_e\dd{A}.
\label{eq:discrete-residuum-equilibrium}
\end{flalign}

While Eq.~\eqref{eq:discrete-residuum-equilibrium} is well-established (see, e.g., \cite{miehe_thermodynamically_2010}), the stationary condition of novel variational problem~\eqref{eq:augemnted-lagrange-zj} is not. As a matter of fact, special attention is required in order to derive an efficient finite element implementation. Precisely this point is addressed next. The starting point is Eq.~\eqref{eq:augmented-potential}. The standard first term of $\calF$ in Eq.~\eqref{eq:augmented-potential} leads to residual
\begin{flalign}
	\mathrm{r}^{e,A,I}_z=\int_{\Omega_e} 2z\psi_0^+\,N^A_e\dd{V}
	+
	\int_{\Omega_e}\left[\frac{g_{\text{c}}}{l} [z-1]\,N^A_e+ g_{\text{c}}\,l\,\grad(z) \cdot \grad N^A_e\right]\dd{V}\label{eq:resdual_F_z}
\end{flalign}
where the first term results from the variation of the Helmholtz energy and the second term from the variation of the total fracture energy -- both with respect to $z$. Again, the global counterpart of $\mathrm{r}^{e,A,I}_z$ is computed by a standard assembly step. The respective residual is denoted as $\mathrm{r}^{A,I}_z$. The second term in Eq.~\eqref{eq:augmented-potential}, does not require an assembly step. It yields residual
\begin{flalign}
\mathrm{r}^{A,II}_z=\frac{\partial\mathcal{L}_1^{A}}{\partial z^A_g}.\label{eq:resdual_L1_z}
\end{flalign}
The notation $\bullet_g^A$ refers to the value of $\bullet$ at global node $A$ (assembled triangulation). Finally, the third residual $\mathrm{r}^{A,III}_z$ is addressed. It is related to global constraint~\eqref{eq:ineq-2}. To be more precise, it is computed from the linearization of Eq.~\eqref{eq:augmented-constraint-2}. One observes that this energy cannot be additively split into elemental contributions. This is due to the definitions of the involved norms, cf. Eqs.~\eqref{eq:LP-Norm} and \eqref{eq:H1-Norm}. In order to highlight the numerical challenges associated with this energy constraint, $g\le 0$ is rewritten as a composition. For instance and focusing on the $L_p$-norm~\eqref{eq:LP-Norm}
\begin{flalign}
g=S^{1/p}-\rho
\quad\text{with}\quad
S=\int\limits_{\Omega}s(z)\,\dd{V}
\quad\text{and}\quad
s(z)=|z-z_{j-1}|^p.
\label{eq:parametrization-g}
\end{flalign}
Evidently, the same idea also applies to the $H_1$-norm. Based on composition~\eqref{eq:parametrization-g}, one obtains the variation
\begin{flalign}
\left.\delta\mathcal{L}_2\right|_{\lambda_2=\text{const},\alpha_2=\text{const}}=
\partial_g\mathcal{L}_2\,\partial_S g\,\int\limits_\Omega\delta s\,\dd{V}.
\label{eq:parametrization-g-variation}
\end{flalign}
This, in turn, defines the final residual
\begin{flalign}
\mathrm{r}^{A,III}_z=\partial_g\mathcal{L}_2\,\partial_S g\,\int\limits_\Omega\delta_z s\,N^A_g\,\dd{V}.
\label{eq:parametrization-g-variation-resi}
\end{flalign}
If $s$ does not depend on the gradient of $z$ -- as it is the case for the $L_p$-norm -- the variational derivative simplifies to the partial derivative, i.e., $\delta_z s=\partial_z s$. Based on $\boldsymbol{\mathrm{r}}^{e,A}_{\u}$ (Eq.~\eqref{eq:discrete-residuum-equilibrium}) and $\mathrm{r}^{A}_z=\mathrm{r}_z^{A,I}+\mathrm{r}_z^{A,II}+\mathrm{r}_z^{A,III}$ (see Eqs.~\eqref{eq:resdual_F_z}, \eqref{eq:resdual_L1_z} and \eqref{eq:parametrization-g-variation-resi}) the resulting nonlinear equation-system is given by 
\begin{flalign}
&\rsnu = \boldsymbol 0 \text{ and } \rsnz= \boldsymbol 0
\end{flalign}
which is solved for the vectors $\boldsymbol{\mathrm{u}},\boldsymbol{\mathrm{z}}$ (unknown nodal degrees of freedom) in a sequential manner. Since Newton's scheme is employed, the derivatives of the residuals are required. 
In fact, due to non-smoothness Newton-derivatives are used.
Straightforward computations yield
\begin{flalign}
\KueAB:=\frac{\partial \boldsymbol{\mathrm{r}}^{e,A}_{\u}}{\partial \u^B_e}=\int_{\Omega_e}\grad N^A_e\cdot \underbrace{
\partial^2_{\eps} \left[[z^2+\mu]\psi_0^+ + \psi_0^-\right]}_{\displaystyle =\partial_{\eps}\sig}\cdot \grad N^B_e\, \dd{V},
\label{eq:stiffness-u}
\end{flalign}
\begin{flalign}
\mathrm K^{e,AB,I}_{zz}=
\frac{\partial \mathrm{r}^{e,A,I}_z}{\partial z^B_e}
=\int_{\Omega_e}2N^A_e\,\psi^+_0\, N^B_e\,\dd{V}+
\int_{\Omega_e}\left[N^A_e \,\frac{g_{\text{c}}}{l}\,N^B_e+g_{\text{c}}\,l\grad N^A_e \cdot\grad N^B_e\right]\dd{V},
\label{eq:stiffness-z-I}
\end{flalign}
\begin{flalign}
\mathrm K^{AB,II}_{zz}=
\frac{\partial \mathrm{r}^{A,II}_z}{\partial z^B_g}
=\frac{\partial^2\mathcal{L}_1^{A}}{\partial z^A_g\,\partial z^B_g},
\label{eq:stiffness-z-II}
\end{flalign}
\begin{flalign}
	\nonumber
	&\mathrm K^{AB,III}_{zz}=
	\frac{\partial \mathrm{r}^{A,III}_z}{\partial z^B_g}
	=\hspace*{-2cm}&&\underbrace{\left[\partial_g\mathcal{L}_2\,\partial^2_S g+\partial^2_g\mathcal{L}_2\left(\partial^2_S g\right)^2\right]\,
	\underbrace{\left\{\int\limits_\Omega\delta_z s\,N^A_g\,\dd{V}\right\}}_{\displaystyle =:\mathrm{f}^A_z}
	\left\{\int\limits_\Omega\delta_z s\,N^B_g\,\dd{V}\right\}}_{\raisebox{-0.5pt}{\textcircled{1}}}&\\\label{eq:stiffness-z-III}
	&&+&\underbrace{\partial_g\mathcal{L}_2\,\partial_S g\,\int\limits_\Omega\delta^2_z s\,N^A_g\,N^B_g\,\dd{V}}_{\raisebox{-0.5pt}{\textcircled{2}}}.&
\end{flalign} 
Coupling terms $\boldsymbol{\mathrm K}^{AB}_{z\u}$ and $\boldsymbol{\mathrm K}^{AB}_{\u z}$ are not required, since a sequential algorithm is used. Furthermore, note that $\mathrm{K}^{AB,II}_{zz}=0$ $\forall A\not =B$. 

Due to the properties of the shape functions, matrices $\mathbf{K}_{\u\u}$ and $\mathbf{K}_{zz}^I$ are sparse and $\mathbf{K}_{zz}^{II}$ even shows a diagonal structure. By way of contrast, matrix $\boldsymbol{\mathrm K}^{III}_{zz}$ is not sparse, cf. Eq.~\eqref{eq:stiffness-z-III}. To be more precise, term \textcircled{2} in Eq.~\eqref{eq:stiffness-z-III} leads to a sparse matrix -- in contrast to term \textcircled{1}. The latter can be written as
\begin{flalign}
	\boldsymbol{\mathrm K}^{III,{\raisebox{-0.5pt}{\scriptsize \textcircled{1}}}}_{zz}=b\,\flistz\otimes\flistz
	\quad\text{with}\quad
	b=\left[\partial_g\mathcal{L}_2\,\partial^2_S g+\partial^2_g\mathcal{L}_2\,\left(\partial^2_S g\right)^2\right].
    \label{eq:stiffness-z-III-1}
\end{flalign} 
Based on notation~\eqref{eq:stiffness-z-III-1}, stiffness matrix $\Kzstiff$ is decomposed into a sparse part $\Kzstiff^{\text{sp}}$ and a dense part $\Kzstiff^{\text{den}}$, i.e.,
\begin{flalign}
\Kzstiff=\Kzstiff^{\text{sp}}+\Kzstiff^{\text{den}}\quad\text{with}\quad
\Kzstiff^{\text{sp}}=\boldsymbol{\mathrm K}^{I}_{zz}+\boldsymbol{\mathrm K}^{II}_{zz}+\boldsymbol{\mathrm K}^{III,{\raisebox{-0.5pt}{\scriptsize \textcircled{2}}}}_{zz},\qquad
\Kzstiff^{\text{den}}=\boldsymbol{\mathrm K}^{III,{\raisebox{-0.5pt}{\scriptsize \textcircled{1}}}}_{zz}=b\,\flistz\otimes\flistz.
\end{flalign}
Although the resulting matrix $\Kzstiff$ is thus dense, its inverse can be efficiently computed. This is crucial for the Newton-solver. The existence of the inverse is ensured by the strict convexity of $\calF$ and $\calL_a$ in $z$. In order to compute the inverse of $\Kzstiff$, the Sherman-Morrison formula is applied. This leads to the closed-form solution
\begin{flalign}
\left[\Kzstiff^{\text{sp}}+b\,\flistz \otimes \flistz \right]^{-1}= \left[ \Kzstiff^{\text{sp}}\right]^{-1}-b\,\frac{ \left[\left[ \Kzstiff^{\text{sp}} \right ]^{-1} \cdot \flistz \right] \otimes \left[\left[ \Kzstiff^{\text{sp}} \right ]^{-1} \cdot \flistz \right]}{1+b\,\flistz \cdot \left[ \Kzstiff^{\text{sp}} \right ]^{-1} \cdot \flistz}.
\end{flalign}
which, in turn, yields the Newton-update
\begin{flalign}
\zlist \leftarrow \zlist-\left[ \Kzstiff^{\text{sp}} \right ]^{-1}\cdot \rsnz +b\, \frac{ \left[\left[\Kzstiff^{\text{sp}} \right ]^{-1} \cdot \flistz \right]\left[\flistz \cdot \left[ \Kzstiff^{\text{sp}} \right ]^{-1}\cdot  \rsnz \right]}{1+b\,\flistz \cdot \left[ \left[ \Kzstiff^{\text{sp}} \right ]^{-1} \cdot \flistz\right]}.
\label{eq:newton-update-sherm}
\end{flalign} 
Thanks to the symmetric structure of the phase-field method (Schwarz's Theorem for smooth parts), matrix $\Kzstiff^{\text{sp}}$ is symmetric and only sparse problems
\begin{flalign}
	\Kzstiff^{\text{sp}}\cdot \mathbf{x}=\flistz,\qquad
	\Kzstiff^{\text{sp}}\cdot \mathbf{y} = \rsnz
	\label{eq:aux-problem}
\end{flalign} 
have to be solved which can be done very efficiently. Once Eqs.~\eqref{eq:aux-problem} have been solved, update~\eqref{eq:newton-update-sherm} can be performed. The resulting algorithm is very efficient and robust. It is summarized in Fig.~\ref{pic:Algo2_FEM}
\begin{figure}[htbp]
	\parbox{\textwidth}{
		\begin{algorithm}[H]
			\caption{Finite element solver combining AM with E\&M}\label{alg2}
			\textbf{1. Input values:} $\rho,[0;T]$, BC, \\
			\textbf{2. Initialize:} $j=1$, $\uvec_0=\boldsymbol 0$, $\zlist_0=1$, $t_1=0$\\
			\While{$t_j\leq T$}{
				set $(\ulist_j,\zlist_j):=(\ulist_{j-1},\zlist_{j-1})$\;
				\While{$Res_{stag}>\mathtt{TOL}_{stag}$}{
					solve $\ulist_j:=\left\{\ulist:\rsnu(t_j,\ulist,\zlist_j)=0\right\}$\;
					initialize $k=1$\;
					initialize $(\lambda_1^{A,k},\alpha_1^{A,k}):=(0,\alpha_1^{init})\;\forall A \in \{1,\dots,n\}$ \;
					initialize $(\lambda_2^k,\alpha_2^k):=(0,\alpha_2^{init})$\;
					\While{$(\zlist_j^{k},\lambda_1^{A,k})$ or $(\zlist_j^{k},\lambda_2^k)$ are not KKT-points of \eqref{eq:constminprob}}{
						solve $\zlist_j^k:=\big\{\zlist:\rsnz(t_j,\ulist_j,\zlist,\lambda_1^{1,k},\dots,\lambda_1^{n,k},\lambda^k_2,\alpha^{1,k}_1,\dots,\alpha^{n,k}_1,\alpha^k_2)=0\big\}$\;
						update $\lambda_1^{A,k}\text{ and } \alpha_1^{A,k}\;\forall A \in \{1,\dots,n\}$\Comment*[r]{Hestenes-Powell}
						update $\alpha_2^k \text{ and } \lambda_2^k$\;
						$k \leftarrow k+1$
					}
					set $\zlist_j:=\zlist_j^k$\;
					compute $Res_{stag}=\norm{\rsnu(t_j,\ulist_j,\zlist_j)}$\;
				}
				$t_{j+1}=t_j+\rho - \norm{\zlist_j-\zlist_{j-1}}_\calV$\;
				$j \leftarrow j+1$\;
			}
	\end{algorithm}}%
	\caption{Finite-element implementation of time-incremental AM-E\&M-algorithm based on augmented Lagrangian~\eqref{eq:augmented-potential}}
	\label{pic:Algo2_FEM}
\end{figure}

%% file: Chap_ConceptEx.tex
\section{Conceptual examples} 
\label{sec:ConceptualExamples}
In this section two examples that highlight the differences between global energetic solutions (derived with global minimization) and balanced viscosity (BV) solutions (derived with local minimization) are given. The snap-through problem in Section~\ref{subsec:snap-through} illustrates these mathematical concepts by means of a mechanical application. Moreover, it allows to demonstrate the numerical robustness and efficiency of the adaptive finite element model elaborated in Section~\ref{sec:Experiments}.
Beyond that, the example from Section~\ref{subsec:AMnotBV} demonstrates how the chosen scheme affects which solution concept is approximated. In particular, it shows that alternate minimization (AM) without any further constraints in general does not give a BV solution even in case of a separately convex energy.

\subsection{Example: Alternate minimization (AM) in general does not generate Balanced Viscosity (BV) solutions}\label{subsec:AMnotBV}

All the numerical experiments presented in Section~\ref{sec:Experiments} and also in \cite{boddin_approximation_2022} suggest that approximate solutions generated by a pure AM scheme without the E\&M constraint converge to  BV solutions, as well. In order to show that this in general is not the case, an example with a structure similar to the damage model but with finite dimensional  state spaces is given in this section. 

Let $\calF:[0,T]\times \R^2\times \R\to\R$ be given by
\begin{align}\notag
	\calF(t,u_1,u_2,z) \coloneqq 58(z+14)^2&+ u_1^2\left(60(z+2)^2+76\right)+u_1\left(-4(z-16)^2-8\right)\\
	&+u_2^2\left(15(z-33)^2+100\right)+25 u_2(z-20)^2\underbrace{-2000t u_2}_{\eqqcolon \ell(t,u_2)}\quad\text{for } t\geq 0, u_1,u_2,z\in\R. \label{eq:energy}
\end{align}
Observe that this function is 
structurally similar to the energy functional of the damage model considered in \cite{boddin_approximation_2022}: it is separately convex and even separately quadratic in the variables $\u=(u_1,u_2)$ and $z$. An external loading in $u_2$ direction is included with the term $\ell(t,u_2)$. 

\begin{figure}[http]
	\centering
	{\begin{psfrags}%
			\psfrag{z}[mc][tc][1][270]{\color[rgb]{0,0,0}\setlength{\tabcolsep}{0pt}\begin{tabular}{c}\Large$z$\end{tabular}}%
			\psfrag{t}[tc][tc][1]{\color[rgb]{0,0,0}\setlength{\tabcolsep}{0pt}\begin{tabular}{c}\Large$t$\end{tabular}}%
			\begin{tikzpicture}
				\node at (0,0) {\includegraphics[scale=0.65]{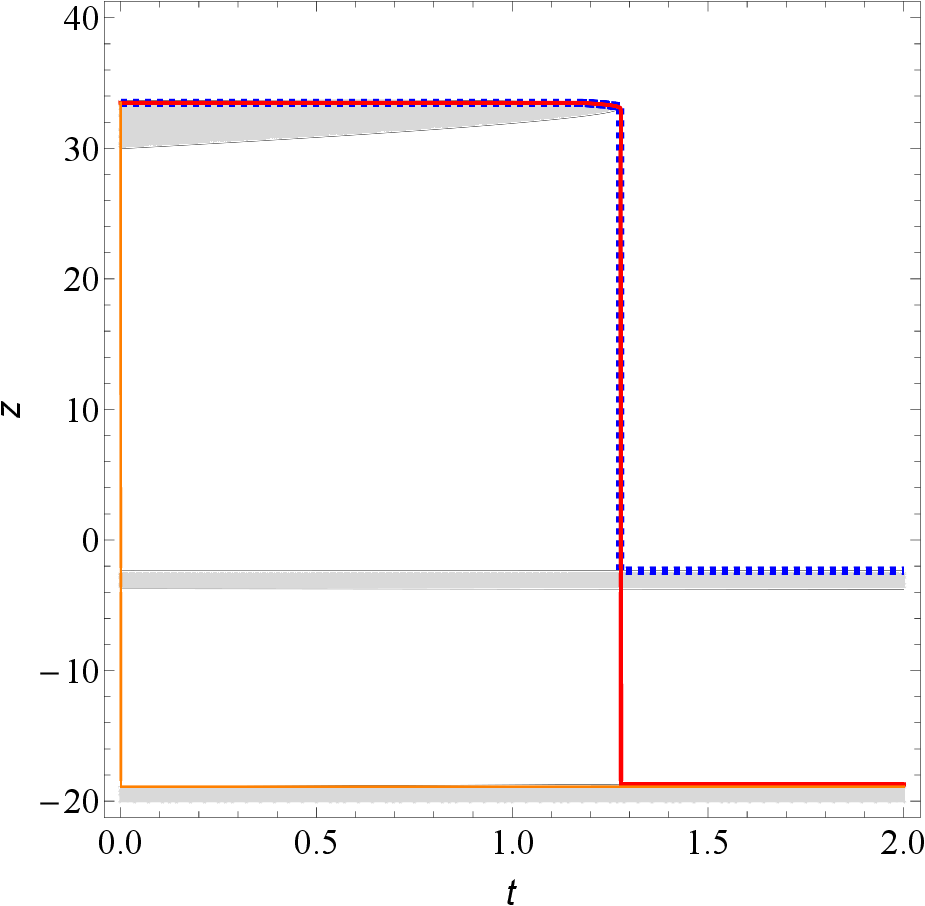}};
				\node at (6,2) {\includegraphics[scale=0.6]{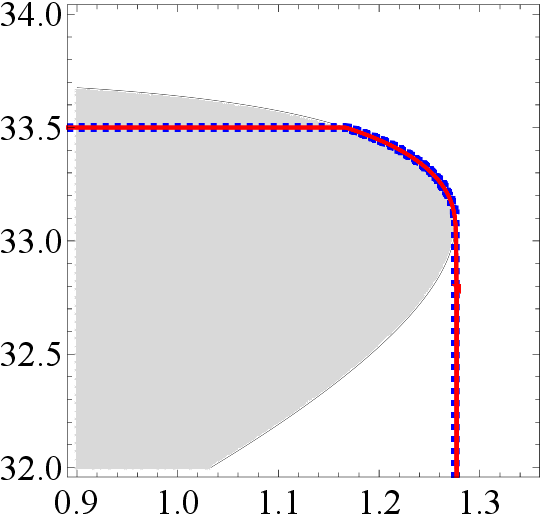}};
				\node (rect) at (1.15,3.77) [draw,black,thick,minimum width=2cm,minimum height=0.3cm] {};
				\draw[black,thick] (0.15,3.92)--(3.94,4.555);
				\draw[black,thick] (0.15,3.615)--(3.938,-0.24);
				\node (rect) at (8.1,-2.35) [draw,black,thick,minimum width=4.8cm,minimum height=3.3cm] {};
				\draw[red,line width=0.5mm] (6,-2)--(6.5,-2);
				\node at (8.2,-2) [black] {pure AM scheme};
				\draw[blue,line width=0.8mm,dash pattern=on 0.7mm off 0.7mm] (6,-1.3)--(6.5,-1.3);
				\node at (8.22,-1.3) [black] {combined scheme};
				\draw[orange,line width=0.5mm] (6,-2.7)--(6.5,-2.7);
				\node at (8.42,-2.7) [black] {global minimization};
				\node (rect) at (6.25,-3.4) [draw,gray!75,fill=lightgray!60,minimum width=0.5cm,minimum height=0.5cm] {};
				\node at (8.41,-3.4) [black] {locally stable states};
			\end{tikzpicture}
	\end{psfrags}}
	\caption{Outcome of schemes for approximating solutions of the rate-independent system driven by the energy given in \eqref{eq:energy} and satisfying irreversibility $\dot{z}\leq 0$: Set of locally stable states $\calS_\text{loc}$
	(gray area); the combined E\&M scheme (dashed blue) for $\rho=0.001$, the pure alternate minimization scheme (red) and the global minimization scheme (orange) with equidistant time steps $\Delta t=\rho$, all for initial value $z_0=33.5$ and the former two with 20 alternate minimization iterations per time step}\label{fig:localStability}
\end{figure}

In line with the phase-field fracture model presented in Section \ref{sec:Theory} and in particular with \eqref{eq:totalvariation_u} and \eqref{eq:Griffithlaw}, functions $\u:[0,T]\to\R^2$ and $z:[0,T]\to\R$ have to satisfy 
\begin{gather}
 \delta_{\u}\calF(t,\u,z)=0,
 \label{eq:ex-u}
\\
 \delta_z\calF(t,\u,z)\dot z=0,\quad \dot z\leq 0,\quad \delta_z\calF(t,\u,z)\leq 0\,.
 \label{eq:ex-z}
\end{gather}
Due to the non-convexity of $\calF(t,\cdot,\cdot)$ in the pair $(\u,z)$ solutions satisfying \eqref{eq:ex-u}--\eqref{eq:ex-z} might be discontinuous in time. This can be seen as follows:
Since for fixed $t$ and $z$ the mapping $\u\mapsto\calF(t,\u,z)$ is strictly convex, for each pair $(t,z)$ there exists a unique $\u=\u(t,z)$ satisfying \eqref{eq:ex-u}. Consequently, a reduced energy $\calF_\text{red}(t,z):=\calF(t,\u(t,z),z)$ can be defined.
By doing so, the  system \eqref{eq:ex-u}--\eqref{eq:ex-z} can equivalently be rewritten as follows: To find $z:[0,T]\to\R$ with
\begin{align}
 \delta_z\calF_\text{red}(t,z)\dot z=0,\quad \dot z\leq 0,\quad \delta_z\calF_\text{red}(t,z)\leq 0\,.
 \label{eq:ex-z-red}
\end{align}
Due to the last condition in \eqref{eq:ex-z-red} the trajectory  $t\mapsto(t,z(t))$ of a  continuous solution necessarily belongs to the set of \textit{locally stable states} that is given by
\[
 \calS_\text{loc}=\left\{(t,z)\in \R^2\,;\, \delta_z\calF_\text{red}(t,z)\leq 0\right\}\subset\R^2\,.
\]
For the function $\calF$ from \eqref{eq:energy} the set of locally stable states $\calS_\text{loc}$ is visualized as the gray area in Figure \ref{fig:localStability}. Starting at $t=0$ with an initial value in the (upper) gray area, e.g., $z_0=33.5$, as time evolves, a solution necessarily has to jump across the white area (where the last condition in  \eqref{eq:ex-z-red} is not satisfied) to reach the (lower) gray area for times larger than $t\approx 1.28$.

Different solution concepts such as the  concept of global energetic solutions and the concept of Balanced Viscosity solutions enrich the conditions in  \eqref{eq:ex-z-red} with jump criteria. While the jump criteria that are included in the global energetic framework origin from (incremental) global minimizations (see \eqref{eq:glob-min}), the jump criteria in BV solutions are derived from a vanishing viscosity procedure. Here, one starts with a viscously regularized version of \eqref{eq:ex-z-red}. For the considered example it reads (for $\varepsilon>0$):
\begin{align}
 (\delta_z\calF_\text{red}(t,z_\varepsilon)+\varepsilon\dot z_\varepsilon)\dot z_\varepsilon=0,\quad \dot z_\varepsilon\leq 0,\quad \delta_z\calF_\text{red}(t,z_\varepsilon)+\varepsilon\dot z_\varepsilon\leq 0\,.
 \label{eq:visc-reg}
\end{align}
The applied regularization is similar to viscoplastic regularizations. For $\varepsilon\to 0$ solutions of \eqref{eq:visc-reg} converge to BV solutions. In our example, a Lipschitz continuous curve $s\mapsto (\hat t(s),\hat z(s))$ is a (parameterized, normalized) Balanced Viscosity solution if it satisfies:
\begin{itemize}
 \item the initial conditions: $\hat t(0)=0,\, \hat z(0)=z_0$,
 \item the normalization and complementarity conditions:
 \[
\hat t'(s)\geq 0,\quad \hat z'(s)\leq 0,\quad  \hat t'(s) + \abs{\hat z'(s)}=1,
 \quad 
 \hat t'(s)\min\{-\delta_z\calF_\text{red}(\hat t(s),\hat z(s)),0\}=0,
\]
 \item the energy balance: For all $0\leq s_1<s_2$
  \begin{multline}
  	\calF_\text{red}(\hat t(s_2),\hat z(s_2)) + \int_{s_1}^{s_2}\abs{\hat z'(s)}
  	\abs{\min\{-\delta_z\calF_\text{red}(\hat t(s),\hat z(s)),0\}}\,\rmd s = \\
  	 \calF(\hat t(s_1),\hat z(s_1)) + \int_{s_1}^{s_1}\delta_t\calF_\text{red}(\hat t(s),\hat z(s))\hat t'(s)\,\rmd s\,.
  \end{multline}
\end{itemize}
Next, from the energy balance a condition is derived that necessarily has to be satisfied along jump trajectories of BV solutions. Let $s_-<s_+$ be a parameter interval with $\hat t(s)=const.=t_*$ for all $s$ from this interval and let $z_\pm:=\hat z(s_\pm)$. Thanks to the normalization condition $\abs{\hat z'(s)}=1$ holds on this interval. Accordingly, function $z$ jumps from $z_-$ to $z_+$ at the (true physical) time $t_*$. For $s_1<s_2\in [s_-,s_+]$ the energy balance reduces to 
\[
 \calF_\text{red}(t_*,\hat z(s_2)) + \int_{s_1}^{s_2}
  \abs{\min\{-\delta_z\calF_\text{red}(t_*,\hat z(s)),0\}}\,\rmd s 
  = \calF(t_*,\hat z(s_1))\,.
\]
Taking the derivative  with respect to the parameter $s$ one ultimately finds that for (almost all) $s\in [s_-,s_+]$ 
\begin{align}
 \delta_z\calF_\text{red}(t_*,\hat z(s))=\abs{\min\{-\delta_z\calF_\text{red}(t_*,\hat z(s)),0\}}\,.
 \label{eq:necessary}
\end{align}
Hence, one necessarily has $\delta_z\calF_\text{red}(t_*,\hat z(s))\geq 0$ for (a.a.) $s\in [s_-,s_+]$ along a jump trajectory. This implies in particular that jump trajectories of BV solutions do not cross the set of locally stable states $\calS_\text{loc}$. 

In Figure \ref{fig:localStability}, a solution generated with the combined E\&M scheme (cf.\ Alg.~\ref{alg1})) and a solution generated with a pure alternate minimization scheme are depicted alongside a solution obtained from global minimization. 
For the combined E\&M scheme arc-length parameter $\rho=0.001$ is chosen, whereas the pure alternate minimization scheme and the global minimization scheme are realized with equidistant time steps $\Delta t=\rho$. All schemes are initialized with $z_0=33.5$ (20 alternate minimization iterations per increment).

It can be observed that global minimization leads to an immediate jump, whereas the other two approximated solutions jump at $t\approx 1.28$. However, they jump into two differing states. The combined E\&M scheme produces a solution that jumps just as far as necessary to reach again a locally stable state (gray area). This is in perfect coincidence with the necessary condition \eqref{eq:necessary} for jumps of BV solutions. By way of contrast, the pure alternate minimization scheme jumps much further, in particular the solution jumps across some locally stable states. In summary, the solution generated by the alternate minimization scheme is not a BV solution since it violates \eqref{eq:necessary} along its jump path.

\begin{figure}[http]
	\centering
	\subcaptionbox{$t_{1}=0.001$\label{fig:redEner_Plots1}}
	{\begin{psfrags}%
			\psfrag{z}[mc][tc][1]{\color[rgb]{0,0,0}\setlength{\tabcolsep}{0pt}\begin{tabular}{c}\small$z$\end{tabular}}%
			\psfrag{Fred}[mc][tc][1]{\color[rgb]{0,0,0}\setlength{\tabcolsep}{0pt}\begin{tabular}{c}\small$\calF_{\text{red}}(t_1,z)$\end{tabular}}%
			\includegraphics[scale=0.35]{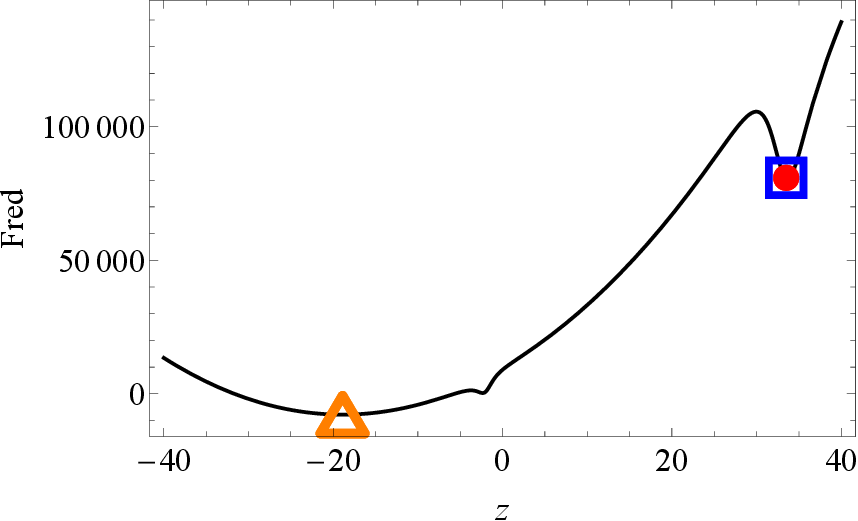}
	\end{psfrags}}
	\subcaptionbox{$t_{1680}=1.27558$\label{fig:redEner_Plots3}}
	{\begin{psfrags}%
			\psfrag{z}[mc][tc][1]{\color[rgb]{0,0,0}\setlength{\tabcolsep}{0pt}\begin{tabular}{c}\small$z$\end{tabular}}%
			\psfrag{Fred}[mc][tc][1]{\color[rgb]{0,0,0}\setlength{\tabcolsep}{0pt}\begin{tabular}{c}\small$\calF_{\text{red}}(t_{1680},z)$\end{tabular}}%
			\includegraphics[scale=0.35]{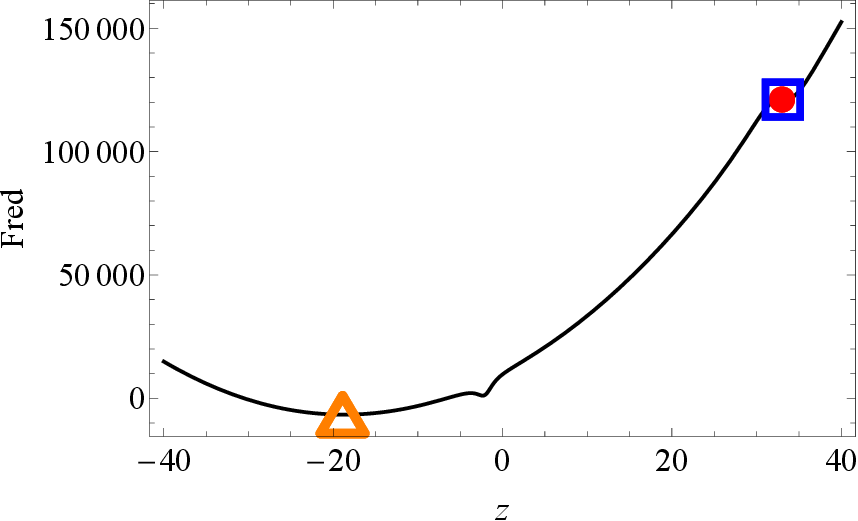}
	\end{psfrags}}
	\subcaptionbox{$t_{1701}=1.27562$\label{fig:redEner_Plots4}}
	{\begin{psfrags}%
			\psfrag{z}[mc][tc][1]{\color[rgb]{0,0,0}\setlength{\tabcolsep}{0pt}\begin{tabular}{c}\small$z$\end{tabular}}%
			\psfrag{Fred}[mc][tc][1]{\color[rgb]{0,0,0}\setlength{\tabcolsep}{0pt}\begin{tabular}{c}\small$\calF_{\text{red}}(t_{1701},z)$\end{tabular}}%
			\includegraphics[scale=0.35]{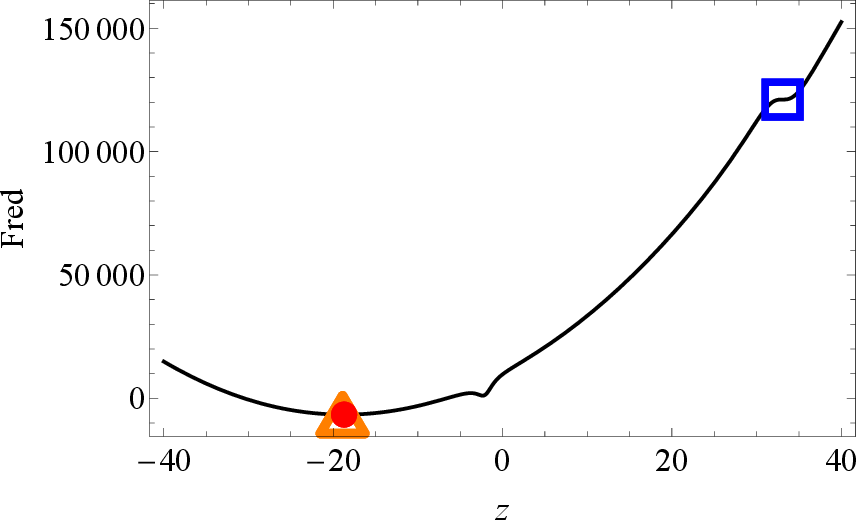}
	\end{psfrags}}
	\\\vspace{10pt}
	\subcaptionbox{$t_{11001}=1.27562$\label{fig:redEner_Plots5}}
	{\begin{psfrags}%
			\psfrag{z}[mc][tc][1]{\color[rgb]{0,0,0}\setlength{\tabcolsep}{0pt}\begin{tabular}{c}\small$z$\end{tabular}}%
			\psfrag{Fred}[mc][tc][1]{\color[rgb]{0,0,0}\setlength{\tabcolsep}{0pt}\begin{tabular}{c}\small$\calF_{\text{red}}(t_{11001},z)$\end{tabular}}%
			\includegraphics[scale=0.35]{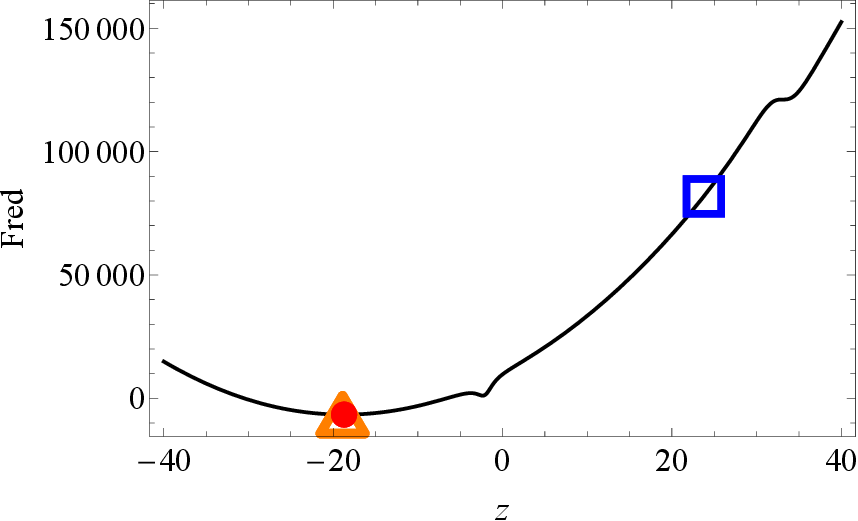}
	\end{psfrags}}
	\subcaptionbox{$t_{37124}=1.27562$\label{fig:redEner_Plots7}}
	{\begin{psfrags}%
			\psfrag{z}[mc][tc][1]{\color[rgb]{0,0,0}\setlength{\tabcolsep}{0pt}\begin{tabular}{c}\small$z$\end{tabular}}%
			\psfrag{Fred}[mc][tc][1]{\color[rgb]{0,0,0}\setlength{\tabcolsep}{0pt}\begin{tabular}{c}\small$\calF_{\text{red}}(t_{37124},z)$\end{tabular}}%
			\includegraphics[scale=0.35]{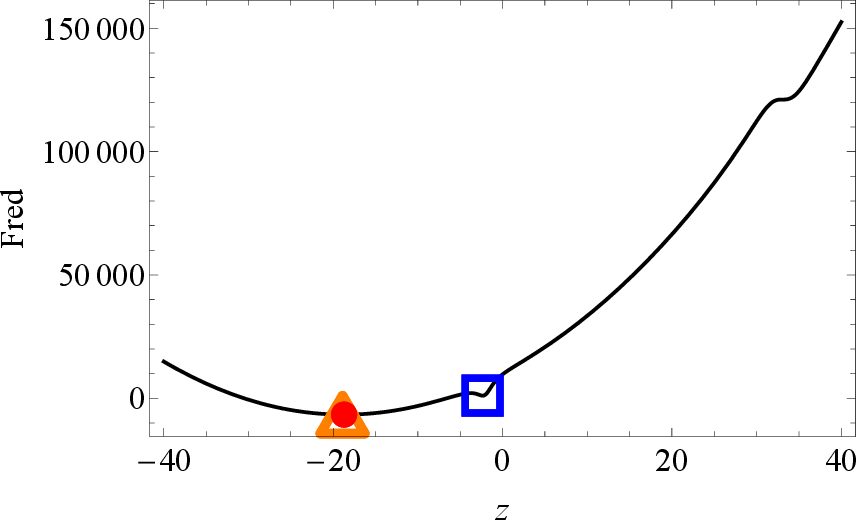}
	\end{psfrags}}
	\subcaptionbox{$t_{37849}=2.00052$\label{fig:redEner_Plots8}}
	{\begin{psfrags}%
			\psfrag{z}[mc][tc][1]{\color[rgb]{0,0,0}\setlength{\tabcolsep}{0pt}\begin{tabular}{c}\small$z$\end{tabular}}%
			\psfrag{Fred}[mc][tc][1]{\color[rgb]{0,0,0}\setlength{\tabcolsep}{0pt}\begin{tabular}{c}\small$\calF_{\text{red}}(t_{37849},z)$\end{tabular}}%
			\includegraphics[scale=0.35]{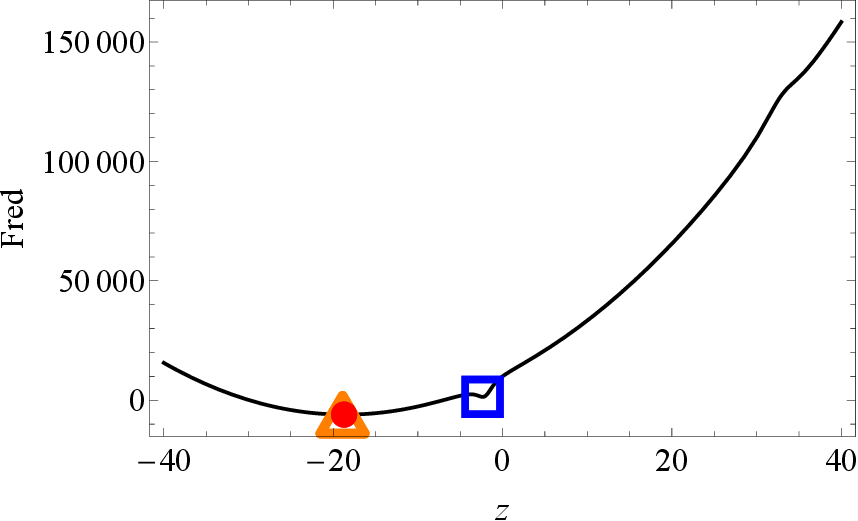}
	\end{psfrags}}
	\caption{Reduced energy $\calF_{\text{red}}(t,z)$ at several time steps $t$: Result of the time-adaptive scheme combined with alternate minimization (blue square) for $\rho=0.001$, the pure alternate minimization scheme (red circle) and the global minimization scheme (orange triangle)}\label{fig:reducedEnergy}
\end{figure}

In Figure~\ref{fig:reducedEnergy} the reduced energy $F_{\text{red}}(t,z)$ is plotted for several times $t$. Moreover, its values for the solutions generated by the different schemes are highlighted to illustrate what type of minima (global/local) is attained.

In summary, all three schemes predict different solutions. Furthermore, while alternate minimization jumps across energy barriers, this is not the case for the algorithm combined with the Efendiev \& Mielke scheme.

%% file: Introduction_example.tex
\subsection{Snap-through problem}
\label{subsec:snap-through}

\begin{figure}[htbp]
	\centering
	\subcaptionbox{Mechanical system}{\label{fig:snap_through_sketch}
	\begin{psfrags}%
		\psfrag{l1}[cr[cr][1]{\color[rgb]{0,0,0}\setlength{\tabcolsep}{0pt}\begin{tabular}{c}$1$\end{tabular}}%
		\psfrag{l2}[cm][cm][1]{\color[rgb]{0,0,0}\setlength{\tabcolsep}{0pt}\begin{tabular}{c}$1$\end{tabular}}%
		\psfrag{F}[cm][cm][1]{\color[rgb]{0,0,0}\setlength{\tabcolsep}{0pt}\begin{tabular}{c}$F$\end{tabular}}%
		\psfrag{u}[cm][cm][1]{\color[rgb]{0,0,0}\setlength{\tabcolsep}{0pt}\begin{tabular}{c}$u$\end{tabular}}%
		\includegraphics[scale=1]{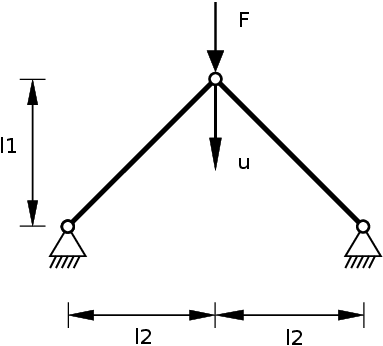}
	\end{psfrags}}
	\subcaptionbox{Force-displacement diagrams}{\label{fig:snap-through-system_b}
	\begin{psfrags}
		\psfrag{local}[tm][tm][1]{\color[rgb]{0,0,0}\setlength{\tabcolsep}{0pt}\begin{tabular}{c}local\end{tabular}}%
		\psfrag{global}[tm][tm][1]{\color[rgb]{0,0,0}\setlength{\tabcolsep}{0pt}\begin{tabular}{c}global\end{tabular}}%
		\psfrag{s1}[tm][tm][1]{\color[rgb]{0,0,0}\setlength{\tabcolsep}{0pt}\begin{tabular}{c}Displacement $u$\end{tabular}}%
		\psfrag{s2}[bm][bm][1]{\color[rgb]{0,0,0}\setlength{\tabcolsep}{0pt}\begin{tabular}{c}Force $F$\end{tabular}}%
		\psfrag{x1}[tm][tm][1]{\color[rgb]{0,0,0}\setlength{\tabcolsep}{0pt}\begin{tabular}{c}$0.0$\end{tabular}}%
		\psfrag{x2}[tm][tm][1]{\color[rgb]{0,0,0}\setlength{\tabcolsep}{0pt}\begin{tabular}{c}$0.5$\end{tabular}}%
		\psfrag{x3}[tm][tm][1]{\color[rgb]{0,0,0}\setlength{\tabcolsep}{0pt}\begin{tabular}{c}$1.0$\end{tabular}}%
		\psfrag{x4}[tm][tm][1]{\color[rgb]{0,0,0}\setlength{\tabcolsep}{0pt}\begin{tabular}{c}$1.5$\end{tabular}}%
		\psfrag{x5}[tm][tm][1]{\color[rgb]{0,0,0}\setlength{\tabcolsep}{0pt}\begin{tabular}{c}$2.0$\end{tabular}}%
		\psfrag{y1}[cr][cr][1]{\color[rgb]{0,0,0}\setlength{\tabcolsep}{0pt}\begin{tabular}{c}$0.0$\end{tabular}}%
		\psfrag{y2}[cr][cr][1]{\color[rgb]{0,0,0}\setlength{\tabcolsep}{0pt}\begin{tabular}{c}$0.1$\end{tabular}}%
		\psfrag{y3}[cr][cr][1]{\color[rgb]{0,0,0}\setlength{\tabcolsep}{0pt}\begin{tabular}{c}$0.2$\end{tabular}}%
		\psfrag{y4}[cr][cr][1]{\color[rgb]{0,0,0}\setlength{\tabcolsep}{0pt}\begin{tabular}{c}$0.3$\end{tabular}}%
		\psfrag{y5}[cr][cr][1]{\color[rgb]{0,0,0}\setlength{\tabcolsep}{0pt}\begin{tabular}{c}$0.4$\end{tabular}}%
		\includegraphics[scale=1]{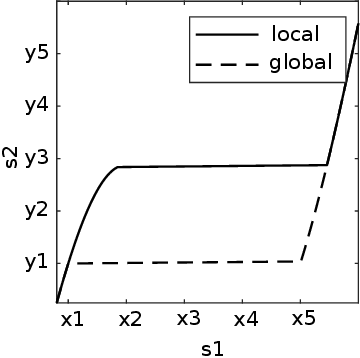}
	\end{psfrags}}
	\caption{Snap-through problem: (a) sketch of the mechanical system. Hooke's model with Young's modulus of $E=1$ and a cross-sectional area $A=1$ are assumed. Finite displacements are considered (geometrically exact description). (b) computed structural response local vs. global energy minimization\label{fig:snap-through-system}}
\end{figure}

To illustrate the difference between global and local energy minimization on the one hand and the robustness of the novel time adaptive method on the other hand, a classic snap-through problem is considered next, see Fig.~\ref{fig:snap-through-system}. Effects due to inertia are neglected. Since Hooke's model is adopted (linear elasticity), the system is rate-independent.
In order to visualize the difference between local and global energy minimization, the structure is loaded in a force-driven manner. The load at $t=0$ is set to $F=-0.1$. Accordingly, the trusses are initially under tension and thus, stability problems do not occur (geometrically exact setting; finite displacements). Consequently, global and local energy minimization predict the same structural response, cf. Fig.~\ref{fig:snap-through-system}(b).
However, if $F$ becomes positive -- an infinitesimal perturbation from zero is sufficient -- global energy minimization leads immediately to a snap-through. Compressive stresses in the trusses are therefore never predicted. Clearly, this observation is not in line with physics. By way of contrast, local energy minimization -- Efendiev \& Mielke scheme or any other method -- indeed captures the real physics response: A positive force $F$ can be applied first without a snap-through. Only if a threshold is reached (critical positive force $F$), the system snaps through, see Fig.~\ref{fig:snap-through-system}(b). 

\begin{figure}[htbp]
	\centering
	\subcaptionbox{Potential energy for $F=-0.1$}{
		\begin{psfrags}%
			\psfrag{s3}[tm][tm][1]{\color[rgb]{0,0,0}\setlength{\tabcolsep}{0pt}\begin{tabular}{c}$u$\end{tabular}}%
			\psfrag{s4}[bm][bm][1]{\color[rgb]{0,0,0}\setlength{\tabcolsep}{0pt}\begin{tabular}{c}$E$\end{tabular}}%
			\psfrag{x1}[tm][tm][1]{\color[rgb]{0,0,0}\setlength{\tabcolsep}{0pt}\begin{tabular}{c}$0$\end{tabular}}%
			\psfrag{x2}[tm][tm][1]{\color[rgb]{0,0,0}\setlength{\tabcolsep}{0pt}\begin{tabular}{c}$1$\end{tabular}}%
			\psfrag{x3}[tm][tm][1]{\color[rgb]{0,0,0}\setlength{\tabcolsep}{0pt}\begin{tabular}{c}$2$\end{tabular}}%
			\psfrag{x4}[tm][tm][1]{\color[rgb]{0,0,0}\setlength{\tabcolsep}{0pt}\begin{tabular}{c}$3$\end{tabular}}%
			\psfrag{y1}[cr][cr][1]{\color[rgb]{0,0,0}\setlength{\tabcolsep}{0pt}\begin{tabular}{c}$0$\end{tabular}}%
			\psfrag{y2}[cr][cr][1]{\color[rgb]{0,0,0}\setlength{\tabcolsep}{0pt}\begin{tabular}{c}$0.2$\end{tabular}}%
			\psfrag{y3}[cr][cr][1]{\color[rgb]{0,0,0}\setlength{\tabcolsep}{0pt}\begin{tabular}{c}$0.4$\end{tabular}}%
			\psfrag{y4}[cr][cr][1]{\color[rgb]{0,0,0}\setlength{\tabcolsep}{0pt}\begin{tabular}{c}$0.6$\end{tabular}}%
			\psfrag{y5}[cr][cr][1]{\color[rgb]{0,0,0}\setlength{\tabcolsep}{0pt}\begin{tabular}{c}$0.8$\end{tabular}}%
			\includegraphics[scale=1.1]{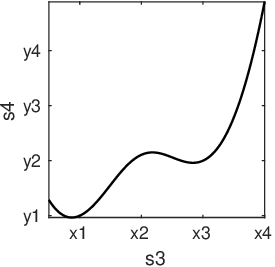}
		\end{psfrags}
	}
	\subcaptionbox{Potential energy for $F=0$}{
		\begin{psfrags}%
			\psfrag{s3}[tm][tm][1]{\color[rgb]{0,0,0}\setlength{\tabcolsep}{0pt}\begin{tabular}{c}$u$\end{tabular}}%
			\psfrag{s4}[bm][bm][1]{\color[rgb]{0,0,0}\setlength{\tabcolsep}{0pt}\begin{tabular}{c}$E$\end{tabular}}%
			\psfrag{x1}[tm][tm][1]{\color[rgb]{0,0,0}\setlength{\tabcolsep}{0pt}\begin{tabular}{c}$0$\end{tabular}}%
			\psfrag{x2}[tm][tm][1]{\color[rgb]{0,0,0}\setlength{\tabcolsep}{0pt}\begin{tabular}{c}$1$\end{tabular}}%
			\psfrag{x3}[tm][tm][1]{\color[rgb]{0,0,0}\setlength{\tabcolsep}{0pt}\begin{tabular}{c}$2$\end{tabular}}%
			\psfrag{x4}[tm][tm][1]{\color[rgb]{0,0,0}\setlength{\tabcolsep}{0pt}\begin{tabular}{c}$3$\end{tabular}}%
			\psfrag{y1}[cr][cr][1]{\color[rgb]{0,0,0}\setlength{\tabcolsep}{0pt}\begin{tabular}{c}$0$\end{tabular}}%
			\psfrag{y2}[cr][cr][1]{\color[rgb]{0,0,0}\setlength{\tabcolsep}{0pt}\begin{tabular}{c}$0.1$\end{tabular}}%
			\psfrag{y3}[cr][cr][1]{\color[rgb]{0,0,0}\setlength{\tabcolsep}{0pt}\begin{tabular}{c}$0.2$\end{tabular}}%
			\psfrag{y4}[cr][cr][1]{\color[rgb]{0,0,0}\setlength{\tabcolsep}{0pt}\begin{tabular}{c}$0.3$\end{tabular}}%
			\psfrag{y5}[cr][cr][1]{\color[rgb]{0,0,0}\setlength{\tabcolsep}{0pt}\begin{tabular}{c}$0.4$\end{tabular}}%
			\includegraphics[scale=1.1]{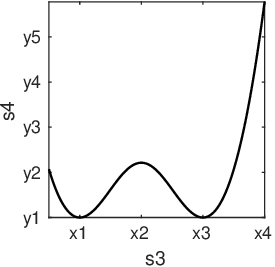}
		\end{psfrags}
	}
	\subcaptionbox{Potential energy for $F=0.1$}{
		\begin{psfrags}%
			\psfrag{s3}[tm][tm][1]{\color[rgb]{0,0,0}\setlength{\tabcolsep}{0pt}\begin{tabular}{c}$u$\end{tabular}}%
			\psfrag{s4}[bm][bm][1]{\color[rgb]{0,0,0}\setlength{\tabcolsep}{0pt}\begin{tabular}{c}$E$\end{tabular}}%
			\psfrag{x1}[tm][tm][1]{\color[rgb]{0,0,0}\setlength{\tabcolsep}{0pt}\begin{tabular}{c}$0$\end{tabular}}%
			\psfrag{x2}[tm][tm][1]{\color[rgb]{0,0,0}\setlength{\tabcolsep}{0pt}\begin{tabular}{c}$1$\end{tabular}}%
			\psfrag{x3}[tm][tm][1]{\color[rgb]{0,0,0}\setlength{\tabcolsep}{0pt}\begin{tabular}{c}$2$\end{tabular}}%
			\psfrag{x4}[tm][tm][1]{\color[rgb]{0,0,0}\setlength{\tabcolsep}{0pt}\begin{tabular}{c}$3$\end{tabular}}%
			\psfrag{y1}[cr][cr][1]{\color[rgb]{0,0,0}\setlength{\tabcolsep}{0pt}\begin{tabular}{c}$-0.2$\end{tabular}}%
			\psfrag{y2}[cr][cr][1]{\color[rgb]{0,0,0}\setlength{\tabcolsep}{0pt}\begin{tabular}{c}$-0.1$\end{tabular}}%
			\psfrag{y3}[cr][cr][1]{\color[rgb]{0,0,0}\setlength{\tabcolsep}{0pt}\begin{tabular}{c}$0.0$\end{tabular}}%
			\psfrag{y4}[cr][cr][1]{\color[rgb]{0,0,0}\setlength{\tabcolsep}{0pt}\begin{tabular}{c}$0.1$\end{tabular}}%
			\includegraphics[scale=1.1]{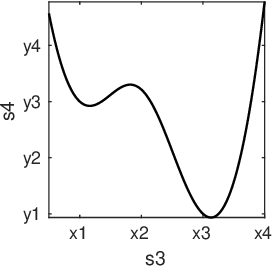}
		\end{psfrags}
	}
	\caption{Snap-through problem: Total energy of the system depending on displacement field $u$ for different prescribed forces $F$}
	\label{fig:Energies_Snap_through}
\end{figure}

The system in Fig.~\ref{fig:snap-through-system}(a) is rate-independent and conservative. Hence, it can be defined by an energy. This is precisely the reason why it has been chosen. The energy associated with the system is given by
\begin{align}
	\calF(u,F)=\frac{1}{\sqrt{2}}(\sqrt{(1-u)^2+1}-\sqrt{2})^2-Fu
\end{align}
and visualized in Fig.~\ref{fig:Energies_Snap_through}. It depends on displacement field $u$ and the prescribed force $F$ und furthermore, its double-well form shows a striking analogy to the energy characterizing phase-field models.
As evident from Fig.~\ref{fig:Energies_Snap_through}(b), the stress free reference configuration is not uniquely defined. Either $u=0$ or $u=2$ (the mirrored configuration) are possible. If the initial system corresponds to $u=0$, the minimum at $u=2$ is of course not reached, since an energy barrier must be overcome. However, both solutions are equally possible in the case of global energy minimization. For a negative force $F$ (Fig.~\ref{fig:Energies_Snap_through}(a)), tension occurs within the trusses. As already mentioned before, stability problems such as a snap-through cannot occur in this case. Accordingly, the local minimum of the energy which is closer to the initial reference configuration is also the global minimum. Therefore, local and global energy minimization predict the same solution. By way of contrast, the global minimum is further away from the initial configuration for positive forces, see Fig.~\ref{fig:Energies_Snap_through}(c). However, this global minimum is not reached by local energy minimization and thus, both minimization strategies lead to different results.

\begin{figure}[htbp]
\centering
\begin{psfrags}%
	\psfrag{s3}[tm][tm][1]{\color[rgb]{0,0,0}\setlength{\tabcolsep}{0pt}\begin{tabular}{c}time-step\end{tabular}}%
	\psfrag{s4}[bm][bm][1]{\color[rgb]{0,0,0}\setlength{\tabcolsep}{0pt}\begin{tabular}{c}iterations\end{tabular}}%
	\psfrag{x1}[tm][tm][1]{\color[rgb]{0,0,0}\setlength{\tabcolsep}{0pt}\begin{tabular}{c}$0$\end{tabular}}%
	\psfrag{x2}[tm][tm][1]{\color[rgb]{0,0,0}\setlength{\tabcolsep}{0pt}\begin{tabular}{c}$50$\end{tabular}}%
	\psfrag{x3}[tm][tm][1]{\color[rgb]{0,0,0}\setlength{\tabcolsep}{0pt}\begin{tabular}{c}$100$\end{tabular}}%
	\psfrag{x4}[tm][tm][1]{\color[rgb]{0,0,0}\setlength{\tabcolsep}{0pt}\begin{tabular}{c}$150$\end{tabular}}%
	\psfrag{y1}[cr][cr][1]{\color[rgb]{0,0,0}\setlength{\tabcolsep}{0pt}\begin{tabular}{c}$0$\end{tabular}}%
	\psfrag{y2}[cr][cr][1]{\color[rgb]{0,0,0}\setlength{\tabcolsep}{0pt}\begin{tabular}{c}$60$\end{tabular}}%
	\psfrag{y3}[cr][cr][1]{\color[rgb]{0,0,0}\setlength{\tabcolsep}{0pt}\begin{tabular}{c}$120$\end{tabular}}%
	\psfrag{y4}[cr][cr][1]{\color[rgb]{0,0,0}\setlength{\tabcolsep}{0pt}\begin{tabular}{c}$180$\end{tabular}}%
	\psfrag{y5}[cr][cr][1]{\color[rgb]{0,0,0}\setlength{\tabcolsep}{0pt}\begin{tabular}{c}$240$\end{tabular}}%
	\includegraphics[scale=1]{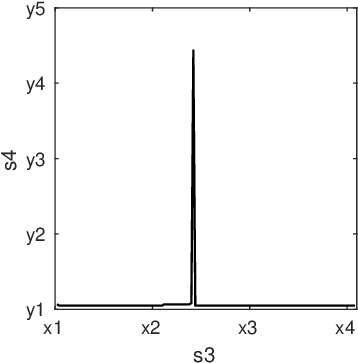}
\end{psfrags}
\begin{psfrags}%
	\psfrag{s3}[tm][tm][1]{\color[rgb]{0,0,0}\setlength{\tabcolsep}{0pt}\begin{tabular}{c}time-step\end{tabular}}%
	\psfrag{s4}[bm][bm][1]{\color[rgb]{0,0,0}\setlength{\tabcolsep}{0pt}\begin{tabular}{c}iterations\end{tabular}}%
	\psfrag{x1}[tm][tm][1]{\color[rgb]{0,0,0}\setlength{\tabcolsep}{0pt}\begin{tabular}{c}$0$\end{tabular}}%
	\psfrag{x2}[tm][tm][1]{\color[rgb]{0,0,0}\setlength{\tabcolsep}{0pt}\begin{tabular}{c}$100$\end{tabular}}%
	\psfrag{x3}[tm][tm][1]{\color[rgb]{0,0,0}\setlength{\tabcolsep}{0pt}\begin{tabular}{c}$200$\end{tabular}}%
	\psfrag{x4}[tm][tm][1]{\color[rgb]{0,0,0}\setlength{\tabcolsep}{0pt}\begin{tabular}{c}$300$\end{tabular}}%
	\psfrag{x5}[tm][tm][1]{\color[rgb]{0,0,0}\setlength{\tabcolsep}{0pt}\begin{tabular}{c}$400$\end{tabular}}%
	\psfrag{y1}[cr][cr][1]{\color[rgb]{0,0,0}\setlength{\tabcolsep}{0pt}\begin{tabular}{c}$0$\end{tabular}}%
	\psfrag{y2}[cr][cr][1]{\color[rgb]{0,0,0}\setlength{\tabcolsep}{0pt}\begin{tabular}{c}$4$\end{tabular}}%
	\psfrag{y3}[cr][cr][1]{\color[rgb]{0,0,0}\setlength{\tabcolsep}{0pt}\begin{tabular}{c}$8$\end{tabular}}%
	\psfrag{y4}[cr][cr][1]{\color[rgb]{0,0,0}\setlength{\tabcolsep}{0pt}\begin{tabular}{c}$12$\end{tabular}}%
	\includegraphics[scale=1]{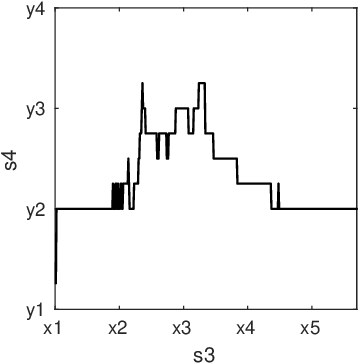}
\end{psfrags}
\caption{Snap-through problem: Number of Newton iterations for each time step: (left) standard local energy minimization; (right) local energy minimization combined with E\&M}
\label{fig:snap-through-Newton_iterations}
\end{figure}

Let the force applied at time $t$ be given by $F(t)$, then the Efendiev \& Mielke scheme for this problem reads as follows: Given $t_{j-1}, u_{j-1}$ calculate $t_j, u_j$ via
\begin{align*}
	u_j&=\underset{u}{\argmin}\left\{\calF(u,F(t_{j-1})):\abs{u-u_{j-1}}\leq\rho\right\},\\
	t_j&=t_{j-1}+\abs{u_j-u_{j-1}}.
\end{align*}
While the difference between local and global energy minimization is evident, standard local energy minimization and the Efendiev \& Mielke scheme yield the same results. However, only the algorithm based on the Efendiev \& Mielke scheme automatically detects the onset of the snap-through and refines the time-discretization. For this reason, it is expected that this adaptive algorithm is numerically more robust. This hypothesis is confirmed by Fig.~\ref{fig:snap-through-Newton_iterations}. Within this figure, the number of Newton iterations for each time step is summarized.
According to Fig.~\ref{fig:snap-through-Newton_iterations}(left), the standard algorithm requires about $200$ Newton iterations during the snap-through -- such a high value would not have been reached in a realistic application, since the algorithm had been terminated beforehand. By way of contrast, the algorithm based on the Efendiev \& Mielke scheme does not require more than 9 Newton iterations and it is significantly more robust, cf. Fig.~\ref{fig:snap-through-Newton_iterations}(right). The robustness, in turn, is related to the time-adaptivity. By comparing the left and the right diagram in Fig.~\ref{fig:snap-through-Newton_iterations}, one can see that the Efendiev \& Mielke scheme requires only 4 Newton iterations before and after the snap-back -- in line with the standard algorithm -- and it indeed refines the time step. This can be seen by comparing the number of time steps. Finally it should be mentioned that Fig.~\ref{fig:snap-through-Newton_iterations}(right) summarizes the maximum number of Newton iterations. Since the novel algorithm is based on an augmented Lagrangian approach, the Newton algorithm has to be performed more than once for a certain time step.

%% file: Chap_NumExp.tex
\section{Numerical examples} 
\label{sec:Experiments}

The predictive capabilities of the novel algorithmic formulation are numerically investigated in this section. For this purpose more complex boundary value problems are analyzed. While a CT-specimen is considered in Subsection~\ref{sec:ct}, an L-shaped plate is numerically investigated in Subsection~\ref{sec:L-shape}. Within both subsections, the influence of arc-length parameter $\rho$ and the choice of the norm is addressed.

\subsection{Compact-Tension specimen (CT)}
\label{sec:ct}

\begin{figure}[htbp]
	\centering
	\subcaptionbox{}{
		\begin{psfrags}%
			\psfrag{s1}[cl][cl][1]{\color[rgb]{0,0,0}\setlength{\tabcolsep}{0pt}\begin{tabular}{c}$\bar u$\end{tabular}}%
			\psfrag{l1}[cr][cr][1]{\color[rgb]{0,0,0}\setlength{\tabcolsep}{0pt}\begin{tabular}{c}$1\,\text{mm}$\end{tabular}}%
			\psfrag{l2}[cm][cm][1]{\color[rgb]{0,0,0}\setlength{\tabcolsep}{0pt}\begin{tabular}{c}$1\,\text{mm}$\end{tabular}}%
			\psfrag{s2}[cl][cl][1]{\color[rgb]{0,0,0}\setlength{\tabcolsep}{0pt}\begin{tabular}{c}$E = 100$ MPa\end{tabular}}%
			\psfrag{s3}[cl][cl][1]{\color[rgb]{0,0,0}\setlength{\tabcolsep}{0pt}\begin{tabular}{c}$\nu = 0.3$\end{tabular}}%
			\psfrag{s4}[cl][cl][1]{\color[rgb]{0,0,0}\setlength{\tabcolsep}{0pt}\begin{tabular}{c}$g_c = 1\frac{\text{N}}{\text{mm}} $\end{tabular}}%
			\includegraphics[scale=1]{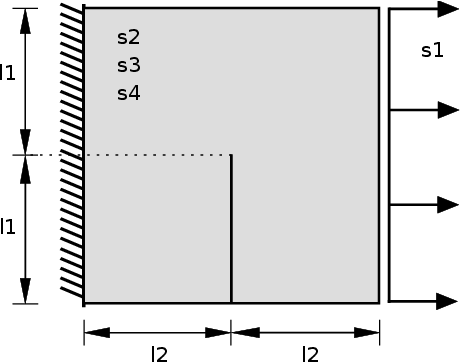}
	\end{psfrags}}
	\subcaptionbox{}{\includegraphics[scale=0.23]{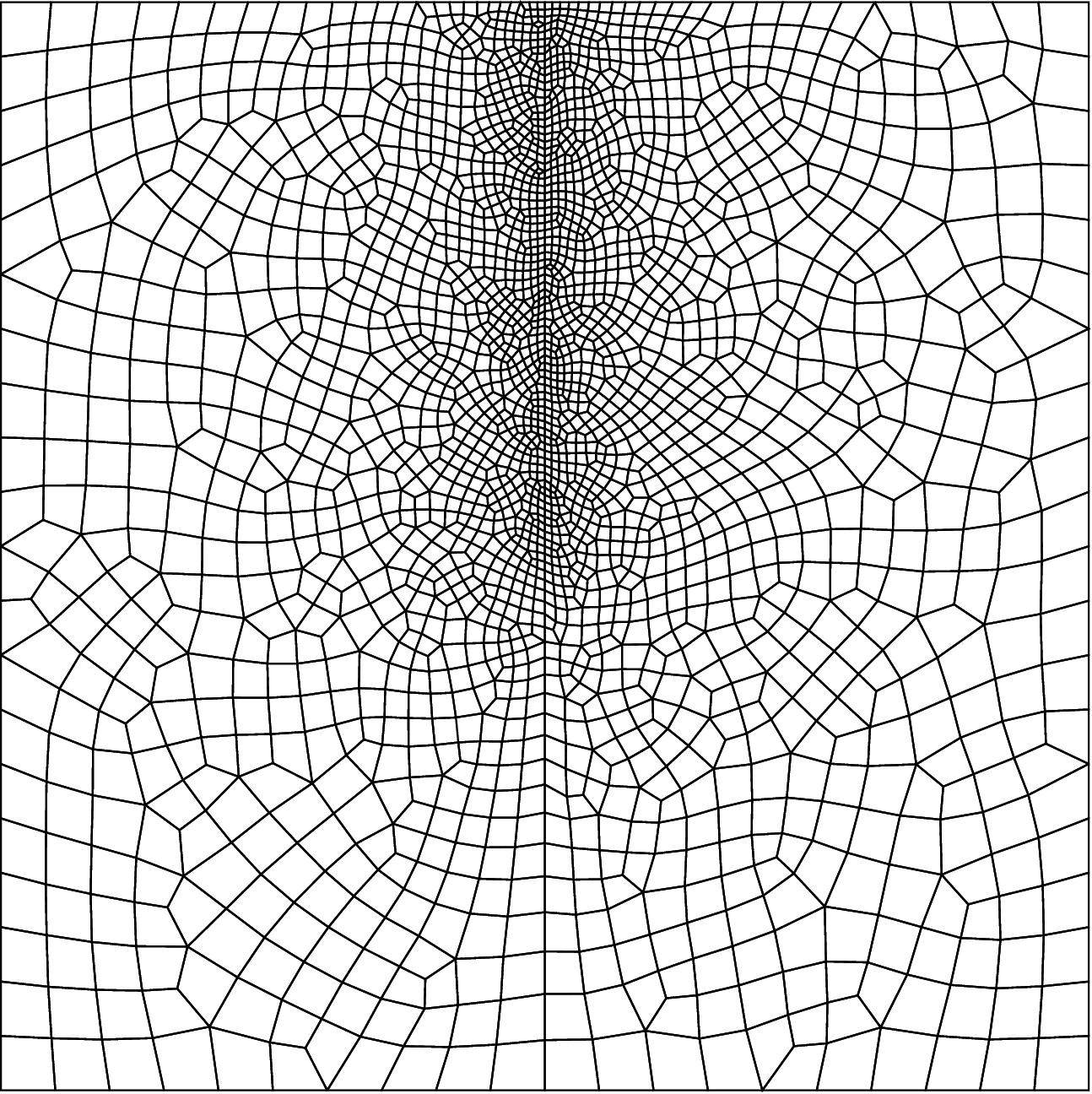}\vspace{2.7em}}
	\caption{Numerical analysis of a Compact-Tension specimen (CT). The specimen shows an initial notch of length $1$~mm: (a) mechanical system; (b) finite element triangulation}
	\label{fig:CT_Specimen_Sketch}
\end{figure}

The Compact-Tension specimen, shown in Fig.~\ref{fig:CT_Specimen_Sketch}, is a standard benchmark for brittle fracture. 
Loading is applied by prescribing a horizontal displacement $\bar u$ at the right hand side. The finite element mesh consists of $2361$ quadrilateral elements. According to Fig.~\ref{fig:CT_Specimen_Sketch}(b), it is refined in the region where the crack is expected to propagate. Plane strain conditions are assumed and the length-scale parameter of the phase-field approximation is set to $l=0.05\,\text{mm}$ throughout all numerical experiments.

\subsubsection{Influence of the arc-length parameter}

As shown in \cite{boddin_approximation_2022}, the solution predicted by the proposed algorithm converges to a BV solution for $\rho\to 0$. In this section, the sensitivity of the numerical results with respect to a finite $\rho$ is analyzed. The norm is chosen as $\calV=L_4(\Omega)$ within this study, while $\rho$ is varied in the range of $[0.005,\,0.5]$. Loading is prescribed by linear function $\bar u (t_j)=t_j \frac{u_{\text{max}}}{100\rho}$ with $T=100\rho$ and $u_{\text{max}}=0.3\,\text{mm}$. This leads to a lower bound of $100$ time-steps for each numerical experiment.
\begin{figure}[htbp]
	\centering
	\subcaptionbox{\label{fig:CT_FU_curves_rho}}
	{\begin{psfrags}%
			\psfrag{s1}[tc][tc][1]{\color[rgb]{0,0,0}\setlength{\tabcolsep}{0pt}\begin{tabular}{c}$\bar u$ [mm]\end{tabular}}%
			\psfrag{s2}[bc][bc][1]{\color[rgb]{0,0,0}\setlength{\tabcolsep}{0pt}\begin{tabular}{c}$F$ [N] \end{tabular}}%
			\psfrag{s000001}[cl][cl][0.8]{\color[rgb]{0,0,0}\setlength{\tabcolsep}{0pt}\begin{tabular}{c}AM \end{tabular}}%
			\psfrag{s000002}[cl][cl][0.8]{\color[rgb]{0,0,0}\setlength{\tabcolsep}{0pt}\begin{tabular}{c}$\rho=0.5$ \end{tabular}}%
			\psfrag{s000003}[cl][cl][0.8]{\color[rgb]{0,0,0}\setlength{\tabcolsep}{0pt}\begin{tabular}{c}$\rho=0.05$ \end{tabular}}%
			\psfrag{s000004}[cl][cl][0.8]{\color[rgb]{0,0,0}\setlength{\tabcolsep}{0pt}\begin{tabular}{c}$\rho=0.01$ \end{tabular}}%
			\psfrag{s000005}[cl][cl][0.8]{\color[rgb]{0,0,0}\setlength{\tabcolsep}{0pt}\begin{tabular}{c}$\rho=0.007$ \end{tabular}}%
			\psfrag{s000006}[cl][cl][0.8]{\color[rgb]{0,0,0}\setlength{\tabcolsep}{0pt}\begin{tabular}{c}$\rho=0.005$ \end{tabular}}%
			\psfrag{x1}[tm][tm][1]{\color[rgb]{0,0,0}\setlength{\tabcolsep}{0pt}\begin{tabular}{c}$0.0$\end{tabular}}%
			\psfrag{x2}[tm][tm][1]{\color[rgb]{0,0,0}\setlength{\tabcolsep}{0pt}\begin{tabular}{c}$0.08$\end{tabular}}%
			\psfrag{x3}[tm][tm][1]{\color[rgb]{0,0,0}\setlength{\tabcolsep}{0pt}\begin{tabular}{c}$0.16$\end{tabular}}%
			\psfrag{x4}[tm][tm][1]{\color[rgb]{0,0,0}\setlength{\tabcolsep}{0pt}\begin{tabular}{c}$0.24$\end{tabular}}%
			\psfrag{x5}[tm][tm][1]{\color[rgb]{0,0,0}\setlength{\tabcolsep}{0pt}\begin{tabular}{c}$0.32$\end{tabular}}%
			\psfrag{y1}[cr][cr][1]{\color[rgb]{0,0,0}\setlength{\tabcolsep}{0pt}\begin{tabular}{c}$0$\end{tabular}}%
			\psfrag{y2}[cr][cr][1]{\color[rgb]{0,0,0}\setlength{\tabcolsep}{0pt}\begin{tabular}{c}$3$\end{tabular}}%
			\psfrag{y3}[cr][cr][1]{\color[rgb]{0,0,0}\setlength{\tabcolsep}{0pt}\begin{tabular}{c}$6$\end{tabular}}%
			\psfrag{y4}[cr][cr][1]{\color[rgb]{0,0,0}\setlength{\tabcolsep}{0pt}\begin{tabular}{c}$9$\end{tabular}}%
			\psfrag{y5}[cr][cr][1]{\color[rgb]{0,0,0}\setlength{\tabcolsep}{0pt}\begin{tabular}{c}$12$\end{tabular}}%
			\includegraphics[scale=1]{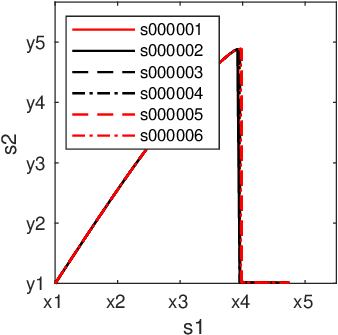}
	\end{psfrags}}
	\subcaptionbox{\label{fig:CT_timesteps_rho}}
	{\begin{psfrags}%
			\psfrag{s1}[tc][tc][1]{\color[rgb]{0,0,0}\setlength{\tabcolsep}{0pt}\begin{tabular}{c}iteration number $j$\end{tabular}}%
			\psfrag{s2}[bc][bc][1]{\color[rgb]{0,0,0}\setlength{\tabcolsep}{0pt}\begin{tabular}{c}relative time $t_j$/$T$\end{tabular}}%
			\psfrag{s000001}[cl][cl][0.8]{\color[rgb]{0,0,0}\setlength{\tabcolsep}{0pt}\begin{tabular}{c}AM \end{tabular}}%
			\psfrag{s000002}[cl][cl][0.8]{\color[rgb]{0,0,0}\setlength{\tabcolsep}{0pt}\begin{tabular}{c}$\rho=0.5$ \end{tabular}}%
			\psfrag{s000003}[cl][cl][0.8]{\color[rgb]{0,0,0}\setlength{\tabcolsep}{0pt}\begin{tabular}{c}$\rho=0.05$ \end{tabular}}%
			\psfrag{s000004}[cl][cl][0.8]{\color[rgb]{0,0,0}\setlength{\tabcolsep}{0pt}\begin{tabular}{c}$\rho=0.01$ \end{tabular}}%
			\psfrag{s000005}[cl][cl][0.8]{\color[rgb]{0,0,0}\setlength{\tabcolsep}{0pt}\begin{tabular}{c}$\rho=0.007$ \end{tabular}}%
			\psfrag{s000006}[cl][cl][0.8]{\color[rgb]{0,0,0}\setlength{\tabcolsep}{0pt}\begin{tabular}{c}$\rho=0.005$ \end{tabular}}%
			\psfrag{x1}[tm][tm][1]{\color[rgb]{0,0,0}\setlength{\tabcolsep}{0pt}\begin{tabular}{c}$1$\end{tabular}}%
			\psfrag{x2}[tm][tm][1]{\color[rgb]{0,0,0}\setlength{\tabcolsep}{0pt}\begin{tabular}{c}$100$\end{tabular}}%
			\psfrag{x3}[tm][tm][1]{\color[rgb]{0,0,0}\setlength{\tabcolsep}{0pt}\begin{tabular}{c}$200$\end{tabular}}%
			\psfrag{x4}[tm][tm][1]{\color[rgb]{0,0,0}\setlength{\tabcolsep}{0pt}\begin{tabular}{c}$300$\end{tabular}}%
			\psfrag{x5}[tm][tm][1]{\color[rgb]{0,0,0}\setlength{\tabcolsep}{0pt}\begin{tabular}{c}$400$\end{tabular}}%
			\psfrag{y01}[cr][cr][1]{\color[rgb]{0,0,0}\setlength{\tabcolsep}{0pt}\begin{tabular}{c}$0$\end{tabular}}%
			\psfrag{y02}[cr][cr][1]{\color[rgb]{0,0,0}\setlength{\tabcolsep}{0pt}\begin{tabular}{c}$0.25$\end{tabular}}%
			\psfrag{y03}[cr][cr][1]{\color[rgb]{0,0,0}\setlength{\tabcolsep}{0pt}\begin{tabular}{c}$0.50$\end{tabular}}%
			\psfrag{y04}[cr][cr][1]{\color[rgb]{0,0,0}\setlength{\tabcolsep}{0pt}\begin{tabular}{c}$0.75$\end{tabular}}%
			\psfrag{y05}[cr][cr][1]{\color[rgb]{0,0,0}\setlength{\tabcolsep}{0pt}\begin{tabular}{c}$1.00$\end{tabular}}%
			\includegraphics[scale=1]{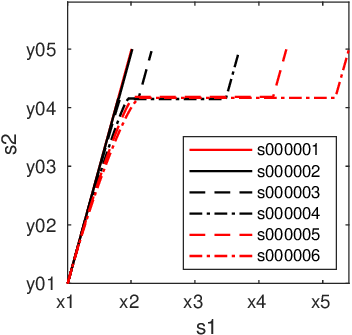}
	\end{psfrags}}
	\caption{Numerical analysis of a Compact-Tension specimen (CT). Sensitivity analysis with respect to arc-length parameter $\rho$: (a) Force-displacement diagram and (b) evolution of the time as a function in terms of the increment number. Results obtained from the standard alternate-minimization algorithm are also shown for the sake of comparison (AM)}\label{fig:CT_rhoexperiment}
\end{figure}
The computed structural responses are summarized in Fig.~\ref{fig:CT_rhoexperiment}. For the sake of comparison, the results associated with standard alternate-minimization are also included in the diagrams. As evident from Fig.~\ref{fig:CT_rhoexperiment}(a), the predicted response is invariant with respect to $\rho$. To be more precise, the proposed E\&M-scheme leads to the same load-displacement response as alternate minimization -- independent of $\rho$. Clearly, this invariance is not a general property of the algorithm, but strongly depends on the underlying mechanical system (see the example given in Section~\ref{subsec:AMnotBV}).

The evolution of the (pseudo-)time depending on the iteration number is depicted in Fig.~\ref{fig:CT_rhoexperiment}(b). In the case of alternate minimization, the evolution is a straight line, i.e., each increment corresponds to the same (pseudo-)time increment. By way of contrast, the proposed E\&M-scheme automatically detects the onset of brutal crack growth and refines the temporal discretization slightly before, during and after brutal crack growth. In the limiting case -- during brutal growth -- the time is kept fixed resulting in a horizontal line in Fig.~\ref{fig:CT_rhoexperiment}(b). Furthermore and as expected, the smaller $\rho$, the larger the number of (additional) increments.

\begin{figure}[htbp]
	\centering
	\subcaptionbox{$\rho=0.03$\label{fig:deltat_rho_large}}
	{\begin{psfrags}%
			\psfrag{s1}[tc][tc][1]{\color[rgb]{0,0,0}\setlength{\tabcolsep}{0pt}\begin{tabular}{c}time-step $j$\end{tabular}}%
			\psfrag{s2}[bc][bc][1]{\color[rgb]{0,0,0}\setlength{\tabcolsep}{0pt}\begin{tabular}{c}$\Delta t_{j+1}$ \end{tabular}}%
			\psfrag{x1}[tm][tm][1]{\color[rgb]{0,0,0}\setlength{\tabcolsep}{0pt}\begin{tabular}{c}$0$\end{tabular}}%
			\psfrag{x2}[tm][tm][1]{\color[rgb]{0,0,0}\setlength{\tabcolsep}{0pt}\begin{tabular}{c}$50$\end{tabular}}%
			\psfrag{x3}[tm][tm][1]{\color[rgb]{0,0,0}\setlength{\tabcolsep}{0pt}\begin{tabular}{c}$100$\end{tabular}}%
			\psfrag{x4}[tm][tm][1]{\color[rgb]{0,0,0}\setlength{\tabcolsep}{0pt}\begin{tabular}{c}$150$\end{tabular}}%
			\psfrag{x5}[tm][tm][1]{\color[rgb]{0,0,0}\setlength{\tabcolsep}{0pt}\begin{tabular}{c}$200$\end{tabular}}%
			\psfrag{y01}[cr][cr][1]{\color[rgb]{0,0,0}\setlength{\tabcolsep}{0pt}\begin{tabular}{c}$0.00$\end{tabular}}%
			\psfrag{y02}[cr][cr][1]{\color[rgb]{0,0,0}\setlength{\tabcolsep}{0pt}\begin{tabular}{c}$0.01$\end{tabular}}%
			\psfrag{y03}[cr][cr][1]{\color[rgb]{0,0,0}\setlength{\tabcolsep}{0pt}\begin{tabular}{c}$0.02$\end{tabular}}%
			\psfrag{y04}[cr][cr][1]{\color[rgb]{0,0,0}\setlength{\tabcolsep}{0pt}\begin{tabular}{c}$0.03$\end{tabular}}%
			\psfrag{y05}[cr][cr][1]{\color[rgb]{0,0,0}\setlength{\tabcolsep}{0pt}\begin{tabular}{c}$0.04$\end{tabular}}%
			\includegraphics[scale=1]{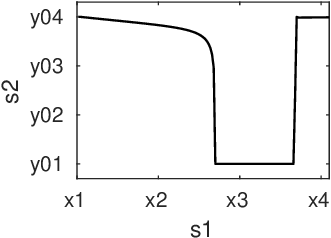}
			
	\end{psfrags}}
	\subcaptionbox{$\rho=0.003$\label{fig:CT_deltat_rho_small}}
	{\begin{psfrags}%
			\psfrag{s1}[tc][tc][1]{\color[rgb]{0,0,0}\setlength{\tabcolsep}{0pt}\begin{tabular}{c}time-step $j$\end{tabular}}%
			\psfrag{s2}[bc][bc][1]{\color[rgb]{0,0,0}\setlength{\tabcolsep}{0pt}\begin{tabular}{c}$\Delta t_{j+1}$ \end{tabular}}%
			\psfrag{x1}[tm][tm][1]{\color[rgb]{0,0,0}\setlength{\tabcolsep}{0pt}\begin{tabular}{c}$0$\end{tabular}}%
			\psfrag{x2}[tm][tm][1]{\color[rgb]{0,0,0}\setlength{\tabcolsep}{0pt}\begin{tabular}{c}$200$\end{tabular}}%
			\psfrag{x3}[tm][tm][1]{\color[rgb]{0,0,0}\setlength{\tabcolsep}{0pt}\begin{tabular}{c}$400$\end{tabular}}%
			\psfrag{x4}[tm][tm][1]{\color[rgb]{0,0,0}\setlength{\tabcolsep}{0pt}\begin{tabular}{c}$600$\end{tabular}}%
			\psfrag{x5}[tm][tm][1]{\color[rgb]{0,0,0}\setlength{\tabcolsep}{0pt}\begin{tabular}{c}$800$\end{tabular}}%
			\psfrag{y001}[cr][cr][1]{\color[rgb]{0,0,0}\setlength{\tabcolsep}{0pt}\begin{tabular}{c}$0.000$\end{tabular}}%
			\psfrag{y002}[cr][cr][1]{\color[rgb]{0,0,0}\setlength{\tabcolsep}{0pt}\begin{tabular}{c}$0.001$\end{tabular}}%
			\psfrag{y003}[cr][cr][1]{\color[rgb]{0,0,0}\setlength{\tabcolsep}{0pt}\begin{tabular}{c}$0.002$\end{tabular}}%
			\psfrag{y004}[cr][cr][1]{\color[rgb]{0,0,0}\setlength{\tabcolsep}{0pt}\begin{tabular}{c}$0.003$\end{tabular}}%
			\psfrag{y005}[cr][cr][1]{\color[rgb]{0,0,0}\setlength{\tabcolsep}{0pt}\begin{tabular}{c}$0.004$\end{tabular}}%
			\includegraphics[scale=1]{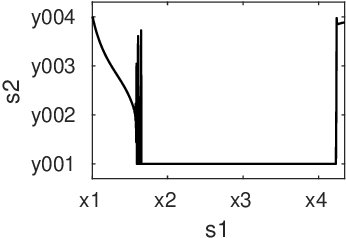}
	\end{psfrags}}
	\caption{Numerical analysis of a Compact-Tension test (CT). Sensitivity analysis with respect to arc-length parameter $\rho$: Evolution of the time step size as a function in terms of the increment number for two different arc-length parameters}\label{fig:CT_rhoexperiment_2}
\end{figure}

In order to show the time-adaptivity of the algorithm more explicitly, the time increment size computed from the novel E\&M-scheme is shown in Fig.~\ref{fig:CT_rhoexperiment_2} for two different arc-length parameters $\rho$.
It can be seen that the time is indeed kept fixed when brutal crack growth occurs. Furthermore, while the algorithm starts with relatively large time steps, these are more and more refined up to brutal crack growth. Finally, it is noted that for smaller $\rho$, some oscillations might occur immediately before brutal crack growth. These are related to the implementation of the arc-length constraint which can be understood as an explicit time-integration, cf. \cite{wriggers_nonlinear_2008}. It bears emphasis that such oscillations never led to any numerical problems.
\begin{figure}[htbp!]
	\centering
	\subcaptionbox{$\bar u(t_{1})=0$\label{fig:CT_Plots1}}
	{\includegraphics[scale=0.13]{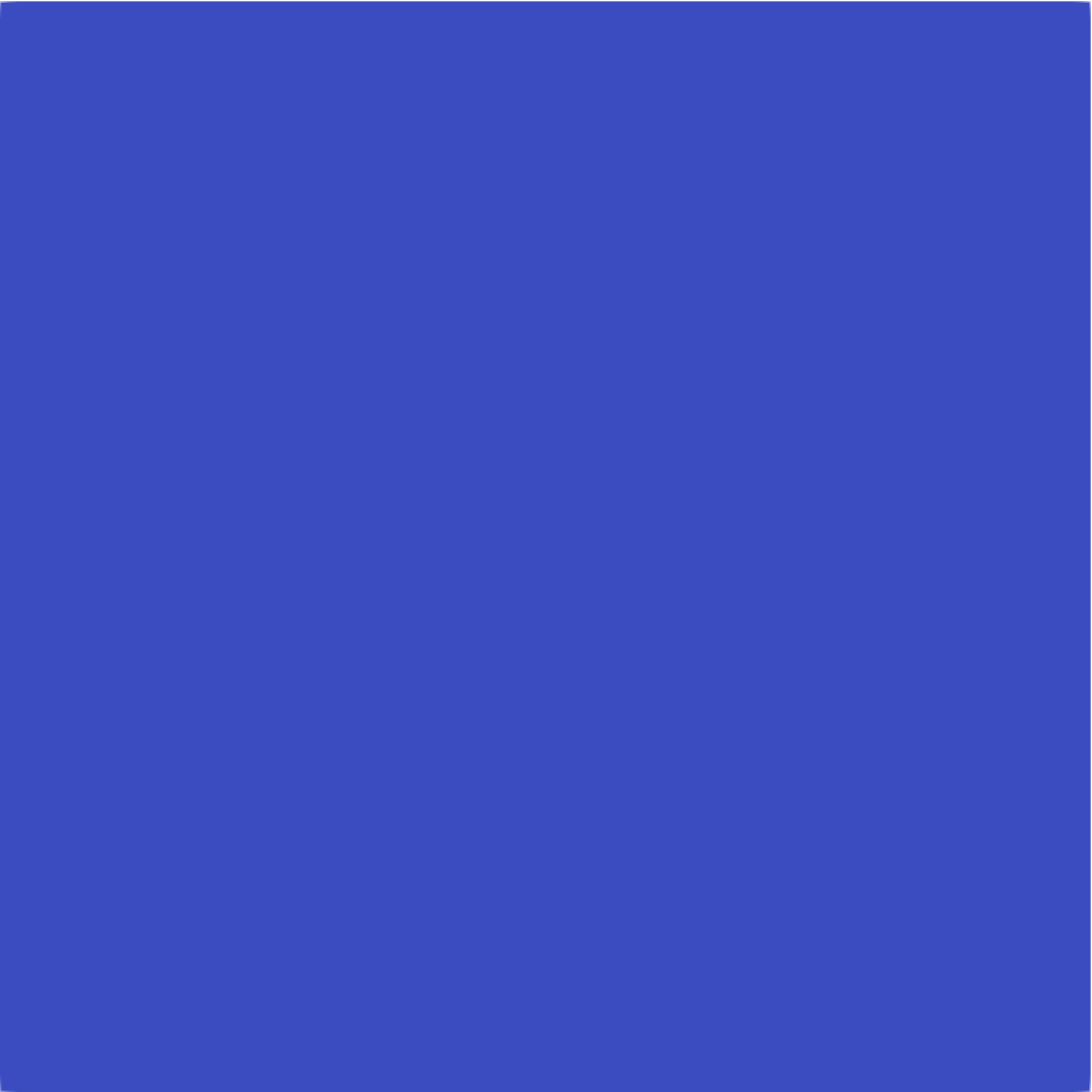}}
	\subcaptionbox{$\bar u(t_{96})=0.236\,\text{mm}$\label{fig:CT_Plots2}}
	{\includegraphics[scale=0.13]{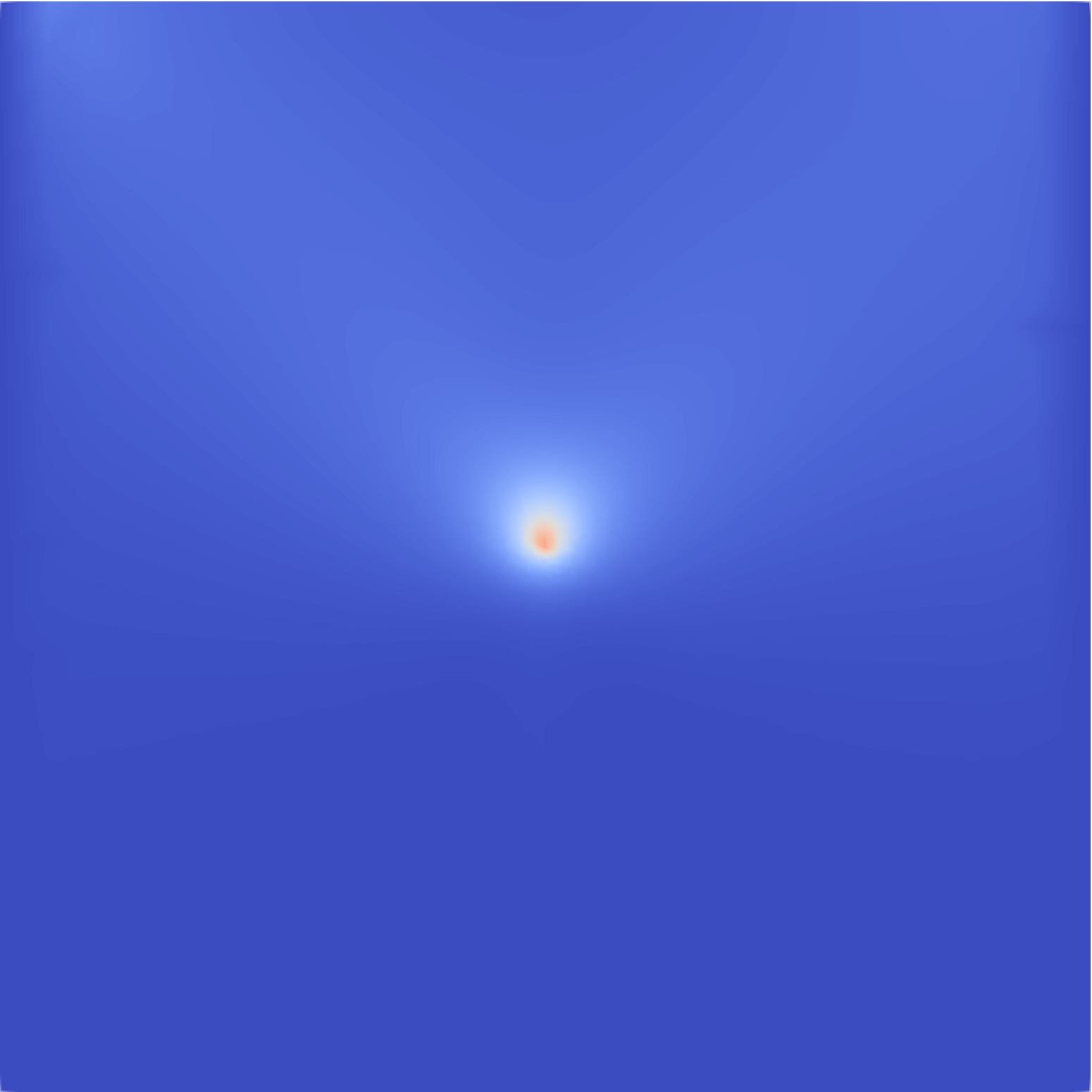}}
	\subcaptionbox{$\bar u(t_{122})=0.236\,\text{mm}$\label{fig:CT_Plots3}}
	{\includegraphics[scale=0.13]{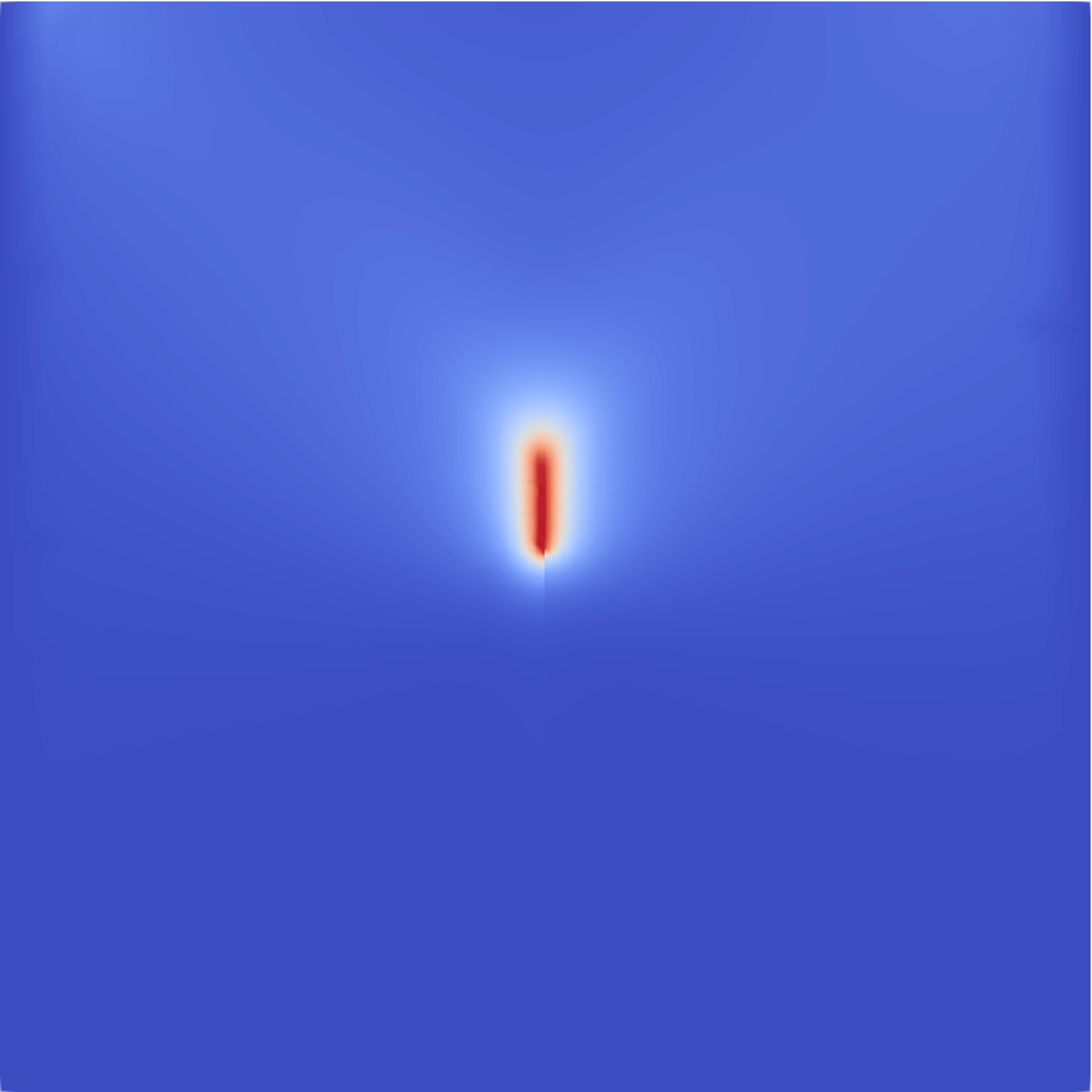}}
	\subcaptionbox{$\bar u(t_{148})=0.236\,\text{mm}$\label{fig:CT_Plots4}}
	{\includegraphics[scale=0.13]{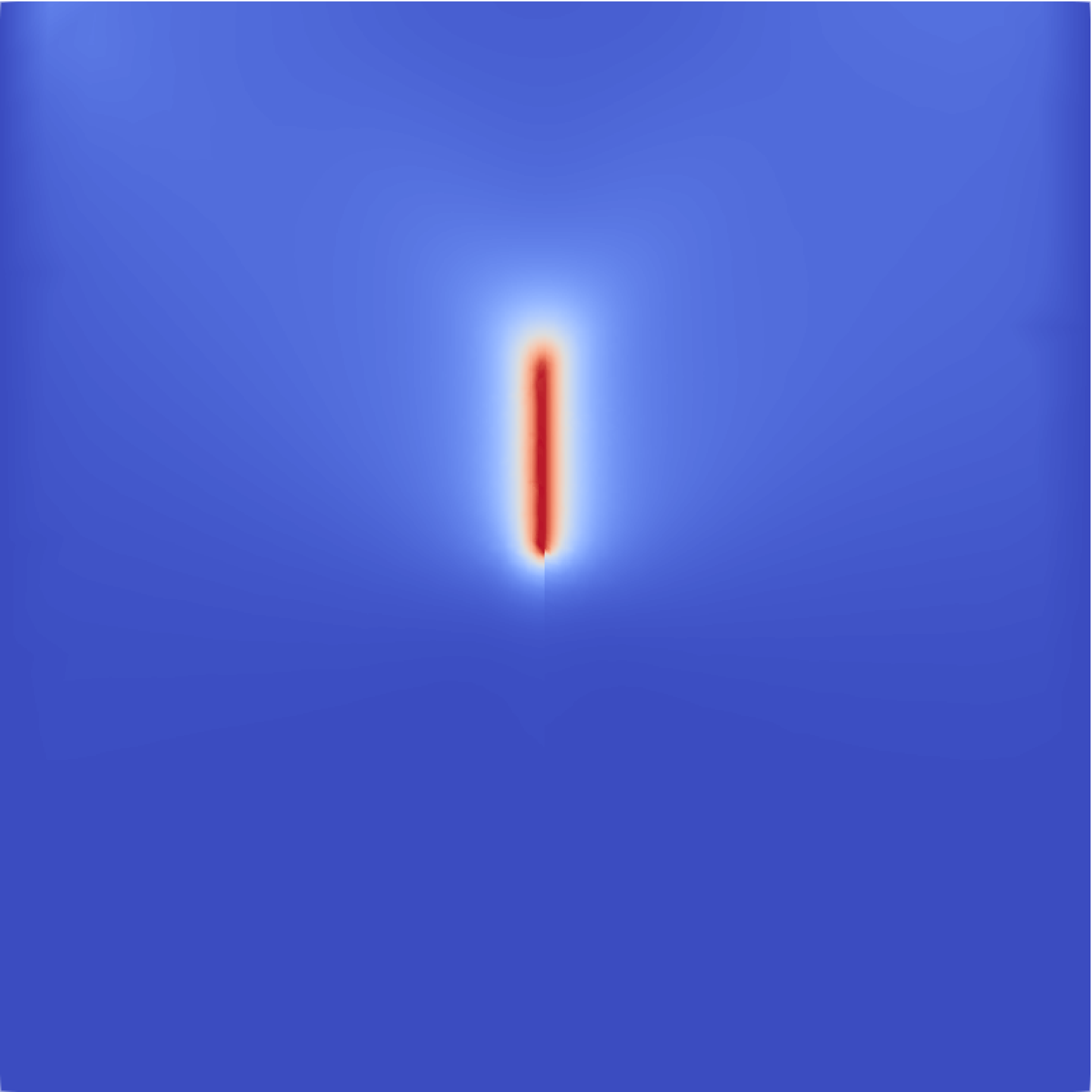}}
	\subcaptionbox{$\bar u(t_{174})=0.236\,\text{mm}$\label{fig:CT_Plots5}}
	{\includegraphics[scale=0.13]{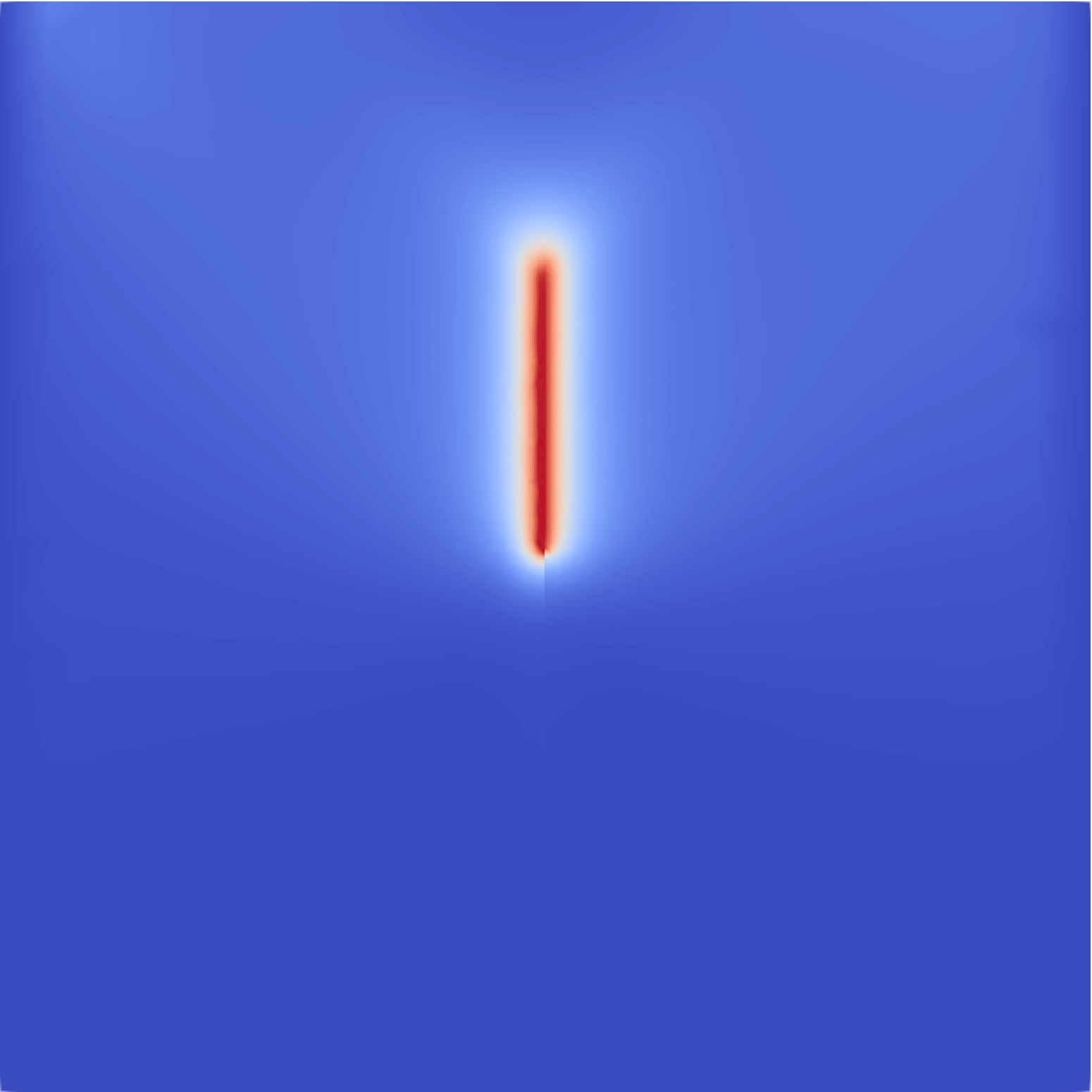}}
	\subcaptionbox{$\bar u(t_{200})=0.236\,\text{mm}$\label{fig:CT_Plots6}}
	{\includegraphics[scale=0.13]{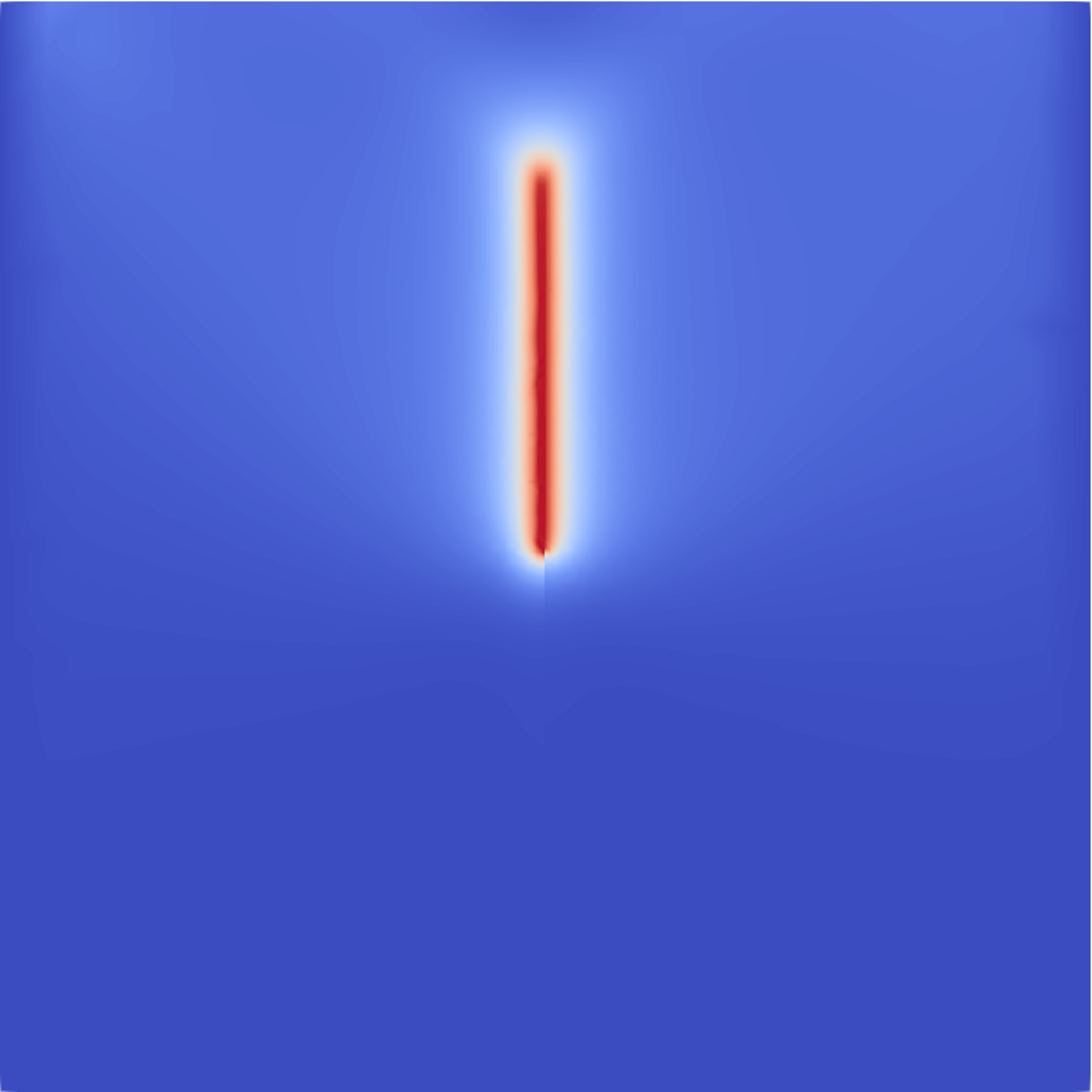}}
	\subcaptionbox{$\bar u(t_{226})=0.236\,\text{mm}$\label{fig:CT_Plots7}}
	{\includegraphics[scale=0.13]{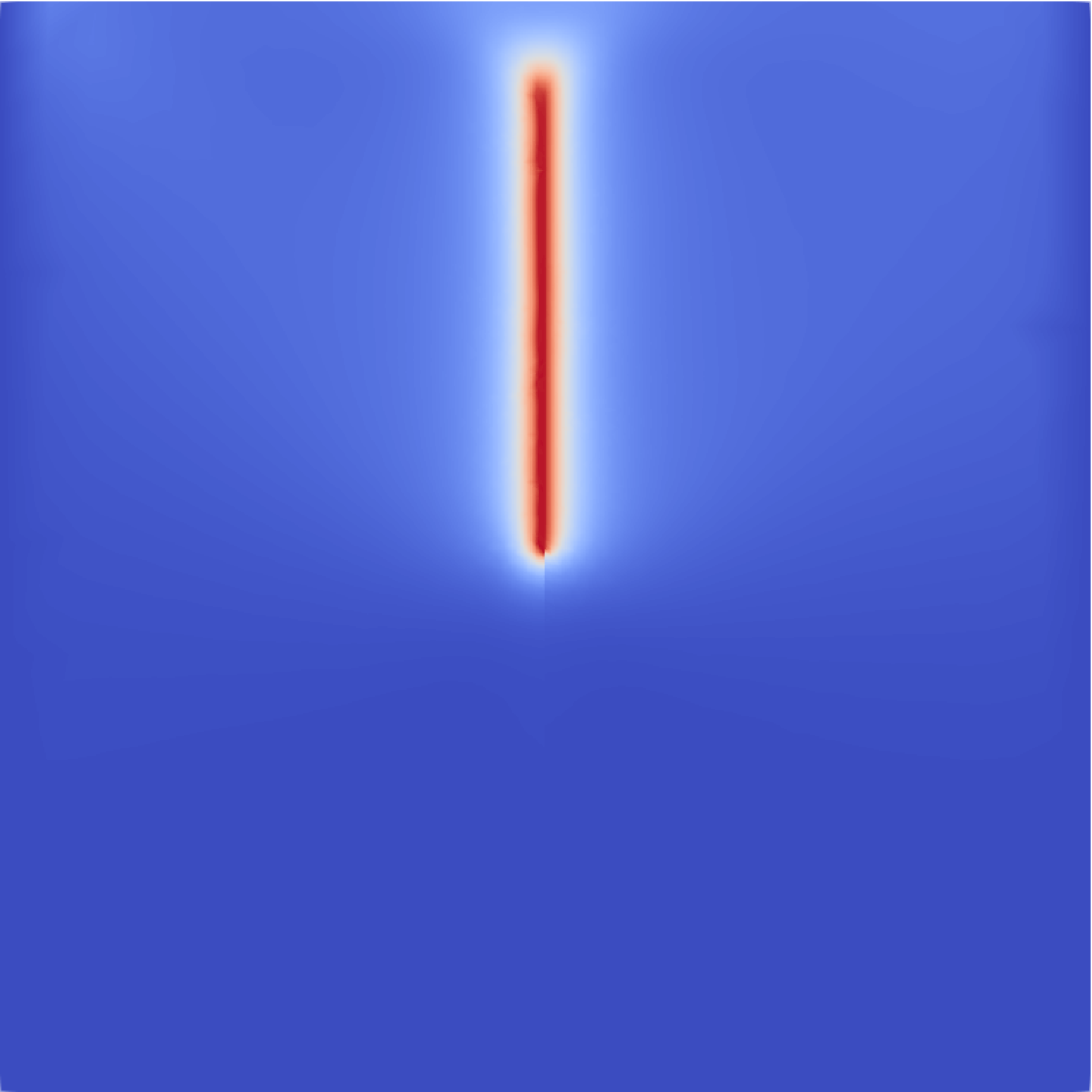}}
	\subcaptionbox{$\bar u(t_{248})=0.236\,\text{mm}$\label{fig:CT_Plots8}}
	{\includegraphics[scale=0.13]{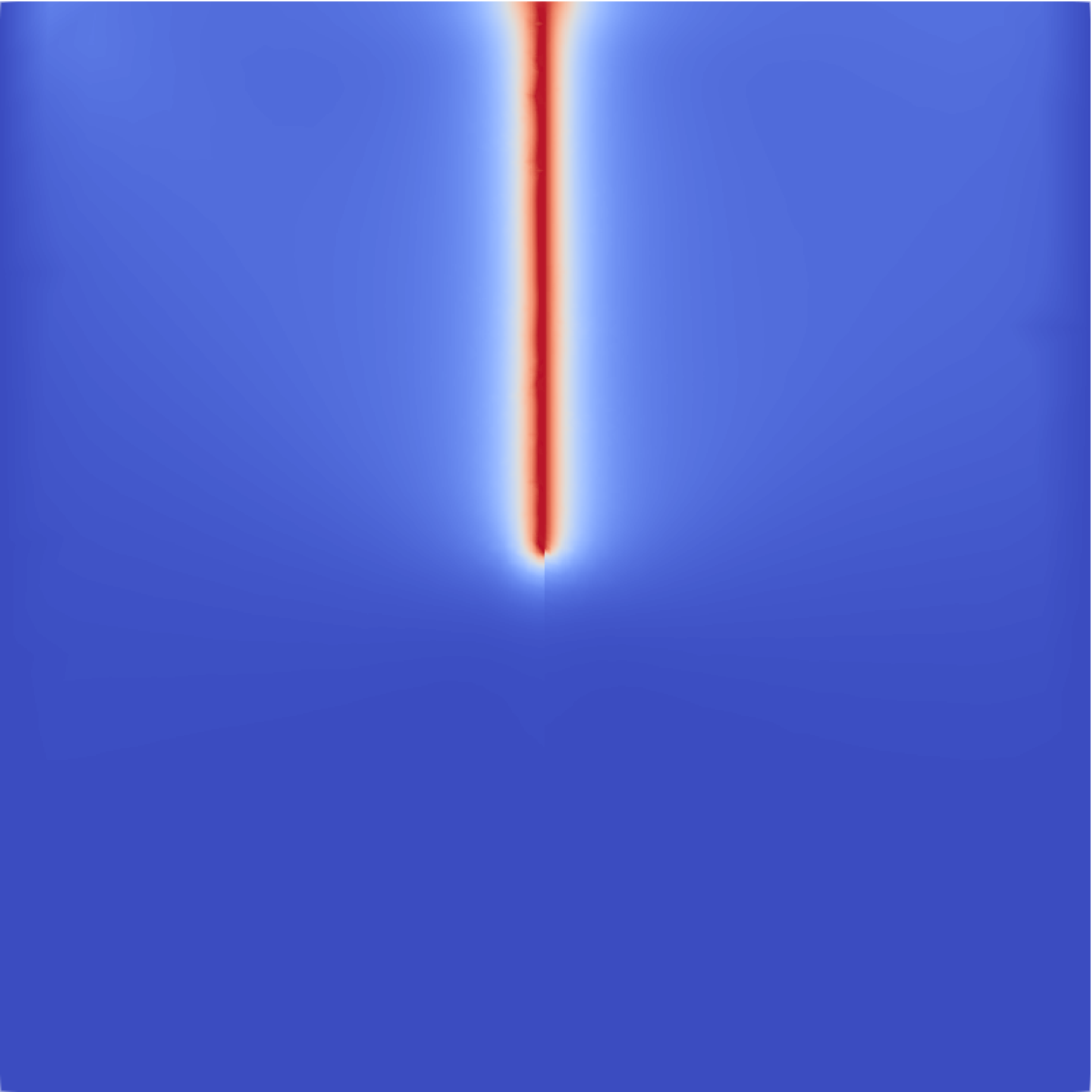}}
	\caption{Numerical analysis of a Compact-Tension test (CT): Distribution of the phase-field variable $z$ computed from the time-adaptive scheme combined with alternate minimization for $\rho=0.01$ and $p=4$. Brutal crack-growth occurs at $t=0.787$}\label{fig:CT_Plots}
\end{figure}

The propagation of the crack during the brutal growth is given in Fig.~\ref{fig:CT_Plots}. The respective computation is based on an arc-length parameter $\rho=0.01$.
It can be seen that the spatial distribution of the phase-field parameter which approximates the geometry of the crack indeed evolves discontinuously in time. The algorithm predicts brutal crack growth for a prescribed displacement of $\bar{u}=0.236$~mm and resolves the trajectory within the stage of  brutal crack growth by means of 152 increments (brutal crack growth occurs from increment $t_{96}$ up to $t_{248}$).

\subsubsection{Influence of the norm}

\begin{figure}[htbp]
	\centering
	\subcaptionbox{\label{fig:CT_timesteps_H1}}
	{\begin{psfrags}%
			\psfrag{s1}[tc][tc][1]{\color[rgb]{0,0,0}\setlength{\tabcolsep}{0pt}\begin{tabular}{c}iteration number $j$\end{tabular}}%
			\psfrag{s2}[bc][bc][1]{\color[rgb]{0,0,0}\setlength{\tabcolsep}{0pt}\begin{tabular}{c}relative time $t_j$/$T$ \end{tabular}}%
			\psfrag{s0000001}[cl][cl][0.8]{\color[rgb]{0,0,0}\setlength{\tabcolsep}{0pt}\begin{tabular}{c}AM \end{tabular}}%
			\psfrag{s0000002}[cl][cl][0.8]{\color[rgb]{0,0,0}\setlength{\tabcolsep}{0pt}\begin{tabular}{c}$H_1,\rho=0.1$ \end{tabular}}%
			\psfrag{s0000003}[cl][cl][0.8]{\color[rgb]{0,0,0}\setlength{\tabcolsep}{0pt}\begin{tabular}{c}$H_1,\rho=0.2$ \end{tabular}}%
			\psfrag{s0000004}[cl][cl][0.8]{\color[rgb]{0,0,0}\setlength{\tabcolsep}{0pt}\begin{tabular}{c}$H_1,\rho=0.5$ \end{tabular}}%
			\psfrag{x1}[tm][tm][1]{\color[rgb]{0,0,0}\setlength{\tabcolsep}{0pt}\begin{tabular}{c}$1$\end{tabular}}%
			\psfrag{x2}[tm][tm][1]{\color[rgb]{0,0,0}\setlength{\tabcolsep}{0pt}\begin{tabular}{c}$50$\end{tabular}}%
			\psfrag{x3}[tm][tm][1]{\color[rgb]{0,0,0}\setlength{\tabcolsep}{0pt}\begin{tabular}{c}$100$\end{tabular}}%
			\psfrag{x4}[tm][tm][1]{\color[rgb]{0,0,0}\setlength{\tabcolsep}{0pt}\begin{tabular}{c}$150$\end{tabular}}%
			\psfrag{x5}[tm][tm][1]{\color[rgb]{0,0,0}\setlength{\tabcolsep}{0pt}\begin{tabular}{c}$200$\end{tabular}}%
			\psfrag{y01}[cr][cr][1]{\color[rgb]{0,0,0}\setlength{\tabcolsep}{0pt}\begin{tabular}{c}$0$\end{tabular}}%
			\psfrag{y02}[cr][cr][1]{\color[rgb]{0,0,0}\setlength{\tabcolsep}{0pt}\begin{tabular}{c}$0.25$\end{tabular}}%
			\psfrag{y03}[cr][cr][1]{\color[rgb]{0,0,0}\setlength{\tabcolsep}{0pt}\begin{tabular}{c}$0.50$\end{tabular}}%
			\psfrag{y04}[cr][cr][1]{\color[rgb]{0,0,0}\setlength{\tabcolsep}{0pt}\begin{tabular}{c}$0.75$\end{tabular}}%
			\psfrag{y05}[cr][cr][1]{\color[rgb]{0,0,0}\setlength{\tabcolsep}{0pt}\begin{tabular}{c}$1.00$\end{tabular}}%
			\includegraphics[scale=1]{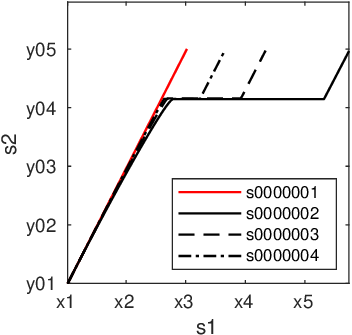}
	\end{psfrags}}
	\subcaptionbox{\label{fig:CT_timesteps_lp}}
	{\begin{psfrags}%
			\psfrag{s1}[tc][tc][1]{\color[rgb]{0,0,0}\setlength{\tabcolsep}{0pt}\begin{tabular}{c}iteration number $j$\end{tabular}}%
			\psfrag{s2}[bc][bc][1]{\color[rgb]{0,0,0}\setlength{\tabcolsep}{0pt}\begin{tabular}{c}relative time $t_j$/$T$ \end{tabular}}%
			\psfrag{s0001}[cl][cl][0.8]{\color[rgb]{0,0,0}\setlength{\tabcolsep}{0pt}\begin{tabular}{c}AM \end{tabular}}%
			\psfrag{s0002}[cl][cl][0.8]{\color[rgb]{0,0,0}\setlength{\tabcolsep}{0pt}\begin{tabular}{c}$p=2$ \end{tabular}}%
			\psfrag{s0003}[cl][cl][0.8]{\color[rgb]{0,0,0}\setlength{\tabcolsep}{0pt}\begin{tabular}{c}$p=4$ \end{tabular}}%
			\psfrag{s0004}[cl][cl][0.8]{\color[rgb]{0,0,0}\setlength{\tabcolsep}{0pt}\begin{tabular}{c}$p=6$ \end{tabular}}%
			\psfrag{s0005}[cl][cl][0.8]{\color[rgb]{0,0,0}\setlength{\tabcolsep}{0pt}\begin{tabular}{c}$p=8$ \end{tabular}}%
			\psfrag{x1}[tm][tm][1]{\color[rgb]{0,0,0}\setlength{\tabcolsep}{0pt}\begin{tabular}{c}$1$\end{tabular}}%
			\psfrag{x2}[tm][tm][1]{\color[rgb]{0,0,0}\setlength{\tabcolsep}{0pt}\begin{tabular}{c}$100$\end{tabular}}%
			\psfrag{x3}[tm][tm][1]{\color[rgb]{0,0,0}\setlength{\tabcolsep}{0pt}\begin{tabular}{c}$200$\end{tabular}}%
			\psfrag{x4}[tm][tm][1]{\color[rgb]{0,0,0}\setlength{\tabcolsep}{0pt}\begin{tabular}{c}$300$\end{tabular}}%
			\psfrag{x5}[tm][tm][1]{\color[rgb]{0,0,0}\setlength{\tabcolsep}{0pt}\begin{tabular}{c}$400$\end{tabular}}%
			\psfrag{y01}[cr][cr][1]{\color[rgb]{0,0,0}\setlength{\tabcolsep}{0pt}\begin{tabular}{c}$0$\end{tabular}}%
			\psfrag{y02}[cr][cr][1]{\color[rgb]{0,0,0}\setlength{\tabcolsep}{0pt}\begin{tabular}{c}$0.25$\end{tabular}}%
			\psfrag{y03}[cr][cr][1]{\color[rgb]{0,0,0}\setlength{\tabcolsep}{0pt}\begin{tabular}{c}$0.50$\end{tabular}}%
			\psfrag{y04}[cr][cr][1]{\color[rgb]{0,0,0}\setlength{\tabcolsep}{0pt}\begin{tabular}{c}$0.75$\end{tabular}}%
			\psfrag{y05}[cr][cr][1]{\color[rgb]{0,0,0}\setlength{\tabcolsep}{0pt}\begin{tabular}{c}$1.00$\end{tabular}}%
			\includegraphics[scale=1]{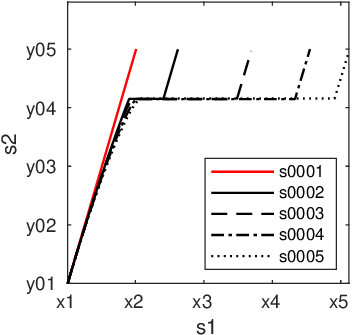}
	\end{psfrags}}
	\caption{Numerical analysis of a Compact-Tension test (CT): Sensitivity of the novel E\&M-algorithm with respect to the chosen norm. Evolution of the time as a function in terms of the increment number: (a) $H_1$-norm for different arc-length parameters $\rho$; (b) $L_p$-norm for a fixed arc-length parameter $\rho$ \label{fig:CT_lpexperiment_1}}
\end{figure}

Finally, the influence of the norm within Algorithm~\ref{alg1} is discussed. The predicted evolution of the physical time is depicted in Fig.~\ref{fig:CT_lpexperiment_1}.
First, it is mentioned that all computations lead to the same structural response. For this reason, the respective load-displacement diagrams are not shown here. Instead, only the evolution of the time as a function in terms of the increment number is analyzed. By comparing Fig.~\ref{fig:CT_lpexperiment_1}(a) and Fig.~\ref{fig:CT_lpexperiment_1}(b) one does not observe any qualitative difference between the predictions associated with the $H_1$-norm and those corresponding to the $L_p$-norm. Furthermore, it is evident that changing the size or the arc-length parameter has a similar effect. Essentially, it defines the admissible maximum growth of phase-field variable $z$ during each increment and thus, the number of increments. In the case of the CT-specimen, the trend "the larger the $p$, the larger the number of increments" holds. However, it will be shown that this is not a general property of the algorithm.

\subsection{L-shaped plate} \label{sec:L-shape}

\begin{figure}[htbp]
	\centering
	\subcaptionbox{}{
		\begin{psfrags}%
			\psfrag{s1}[cl][cl][1]{\color[rgb]{0,0,0}\setlength{\tabcolsep}{0pt}\begin{tabular}{c}$\bar u$\end{tabular}}%
			\psfrag{l1}[cr][cr][1]{\color[rgb]{0,0,0}\setlength{\tabcolsep}{0pt}\begin{tabular}{c}$250\,\text{mm}$\end{tabular}}%
			\psfrag{l2}[cm][cm][1]{\color[rgb]{0,0,0}\setlength{\tabcolsep}{0pt}\begin{tabular}{c}$250\,\text{mm}$\end{tabular}}%
			\psfrag{s2}[cl][cl][1]{\color[rgb]{0,0,0}\setlength{\tabcolsep}{0pt}\begin{tabular}{c}$E = 25840$ MPa\end{tabular}}%
			\psfrag{s3}[cl][cl][1]{\color[rgb]{0,0,0}\setlength{\tabcolsep}{0pt}\begin{tabular}{c}$\nu = 0.18$\end{tabular}}%
			\psfrag{s4}[cl][cl][1]{\color[rgb]{0,0,0}\setlength{\tabcolsep}{0pt}\begin{tabular}{c}$g_c = 0.65\frac{\text{N}}{\text{mm}} $\end{tabular}}%
			\includegraphics[scale=1]{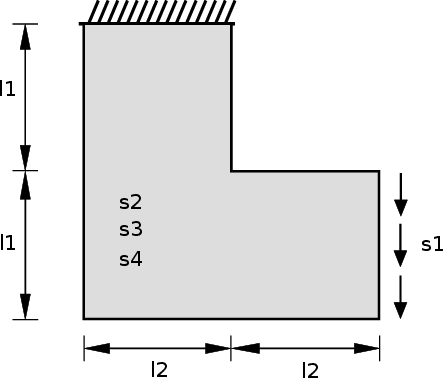}
	\end{psfrags}}
	\subcaptionbox{}{\includegraphics[scale=0.23]{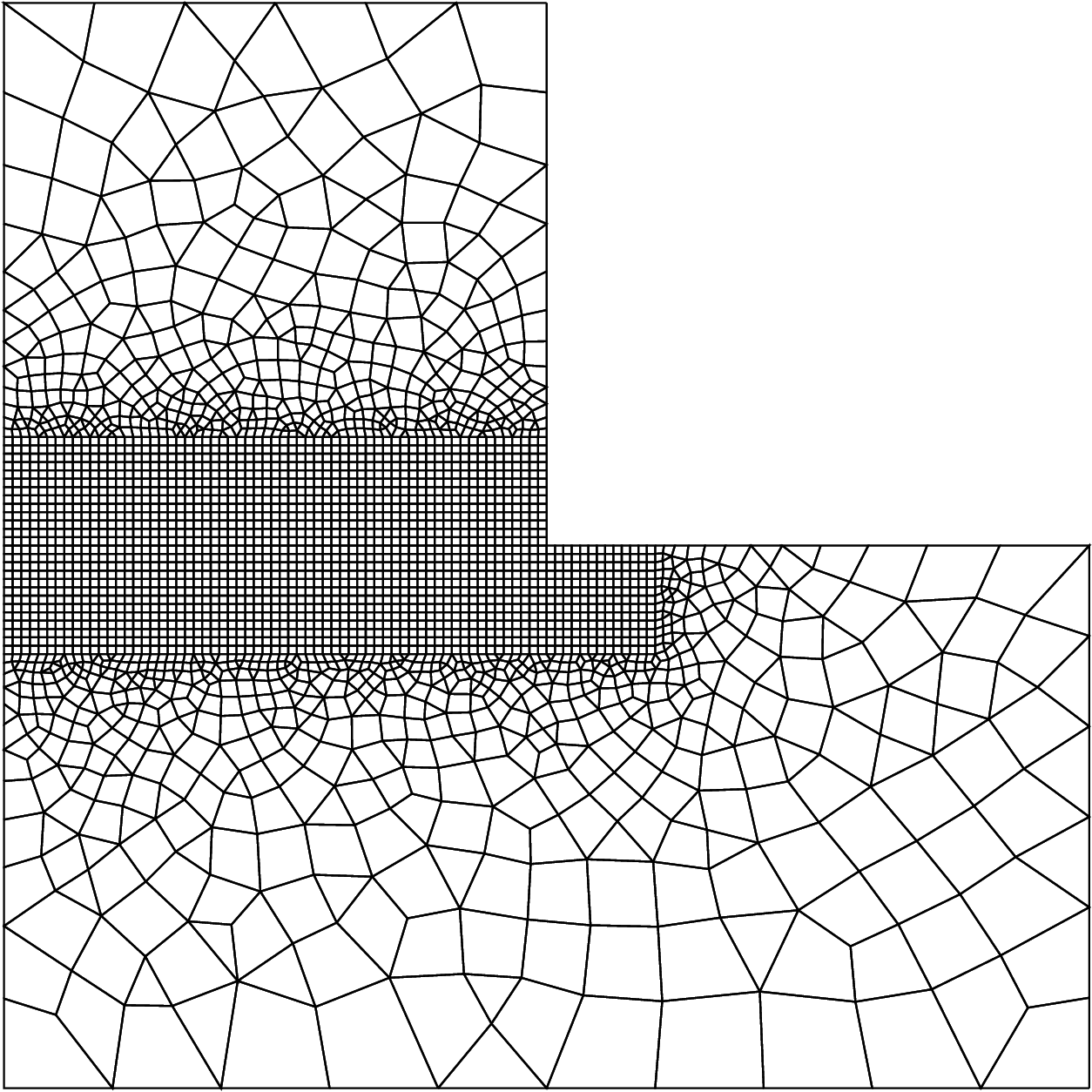}\vspace{2.8em}}
	\caption{Numerical analysis of an L-shaped specimen: (a) mechanical system and (b) finite element triangulation}
	\label{fig:LShape_Sketch}
\end{figure}

Next, the L-shaped specimen depicted in Fig.~\ref{fig:LShape_Sketch} is numerically analyzed.
~%
In contrast to the CT-specimen, this boundary value problem is characterized by a curved crack which starts at the singular inner corner. The system's description as well as the employed finite-element mesh is shown in Fig.~\ref{fig:LShape_Sketch}. Again, the mesh is refined where the crack is expected to grow. However, it is not aligned with the crack's geometry. By doing so, a possible mesh-bias can be reduced. The mesh consists of $2893$ quadrilateral and triangular elements. Plane strain conditions are assumed and loading is applied by prescribing displacement $\bar u(t_j)=t_j \frac{u_{\text{max}}}{8.658}$ with $T=8.658$, $\rho=0.08658$ and $u_{\text{max}}=0.8\,\text{mm}$. The phase-field length-scale parameter is set to $l=10$ mm throughout all computations.

\begin{figure}[htbp]
	\centering
	\subcaptionbox{\label{fig:Lshape_FUcurves_lp}}
	{\begin{psfrags}%
			\psfrag{s1}[tc][tc][1]{\color[rgb]{0,0,0}\setlength{\tabcolsep}{0pt}\begin{tabular}{c}$\bar u$ [mm]\end{tabular}}%
			\psfrag{s2}[bc][bc][1]{\color[rgb]{0,0,0}\setlength{\tabcolsep}{0pt}\begin{tabular}{c}$F$ [N] \end{tabular}}%
			\psfrag{s000001}[cl][cl][0.8]{\color[rgb]{0,0,0}\setlength{\tabcolsep}{0pt}\begin{tabular}{c}AM \end{tabular}}%
			\psfrag{s000002}[cl][cl][0.8]{\color[rgb]{0,0,0}\setlength{\tabcolsep}{0pt}\begin{tabular}{c}$\rho=0.5$ \end{tabular}}%
			\psfrag{s000003}[cl][cl][0.8]{\color[rgb]{0,0,0}\setlength{\tabcolsep}{0pt}\begin{tabular}{c}$\rho=0.05$ \end{tabular}}%
			\psfrag{s000004}[cl][cl][0.8]{\color[rgb]{0,0,0}\setlength{\tabcolsep}{0pt}\begin{tabular}{c}$\rho=0.01$ \end{tabular}}%
			\psfrag{s000005}[cl][cl][0.8]{\color[rgb]{0,0,0}\setlength{\tabcolsep}{0pt}\begin{tabular}{c}$\rho=0.007$ \end{tabular}}%
			\psfrag{s000006}[cl][cl][0.8]{\color[rgb]{0,0,0}\setlength{\tabcolsep}{0pt}\begin{tabular}{c}$\rho=0.005$ \end{tabular}}%
			\psfrag{x1}[tm][tm][1]{\color[rgb]{0,0,0}\setlength{\tabcolsep}{0pt}\begin{tabular}{c}$0.0$\end{tabular}}%
			\psfrag{x2}[tm][tm][1]{\color[rgb]{0,0,0}\setlength{\tabcolsep}{0pt}\begin{tabular}{c}$0.2$\end{tabular}}%
			\psfrag{x3}[tm][tm][1]{\color[rgb]{0,0,0}\setlength{\tabcolsep}{0pt}\begin{tabular}{c}$0.4$\end{tabular}}%
			\psfrag{x4}[tm][tm][1]{\color[rgb]{0,0,0}\setlength{\tabcolsep}{0pt}\begin{tabular}{c}$0.6$\end{tabular}}%
			\psfrag{x5}[tm][tm][1]{\color[rgb]{0,0,0}\setlength{\tabcolsep}{0pt}\begin{tabular}{c}$0.8$\end{tabular}}%
			\psfrag{y1}[cr][cr][1]{\color[rgb]{0,0,0}\setlength{\tabcolsep}{0pt}\begin{tabular}{c}$0$\end{tabular}}%
			\psfrag{y2}[cr][cr][1]{\color[rgb]{0,0,0}\setlength{\tabcolsep}{0pt}\begin{tabular}{c}$100$\end{tabular}}%
			\psfrag{y3}[cr][cr][1]{\color[rgb]{0,0,0}\setlength{\tabcolsep}{0pt}\begin{tabular}{c}$200$\end{tabular}}%
			\psfrag{y4}[cr][cr][1]{\color[rgb]{0,0,0}\setlength{\tabcolsep}{0pt}\begin{tabular}{c}$300$\end{tabular}}%
			\psfrag{y5}[cr][cr][1]{\color[rgb]{0,0,0}\setlength{\tabcolsep}{0pt}\begin{tabular}{c}$400$\end{tabular}}%
			\includegraphics[scale=1]{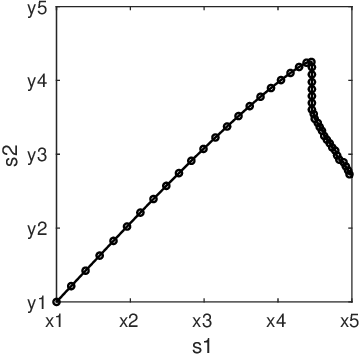}
	\end{psfrags}}
	\subcaptionbox{\label{fig:Lshape_timesteps_lp}}
	{\begin{psfrags}%
			\psfrag{s1}[tc][tc][1]{\color[rgb]{0,0,0}\setlength{\tabcolsep}{0pt}\begin{tabular}{c}time-step $j$\end{tabular}}%
			\psfrag{s2}[bc][bc][1]{\color[rgb]{0,0,0}\setlength{\tabcolsep}{0pt}\begin{tabular}{c}$t_j$/$T$ \end{tabular}}%
			\psfrag{s0001}[cl][cl][0.8]{\color[rgb]{0,0,0}\setlength{\tabcolsep}{0pt}\begin{tabular}{c}AM \end{tabular}}%
			\psfrag{s0002}[cl][cl][0.8]{\color[rgb]{0,0,0}\setlength{\tabcolsep}{0pt}\begin{tabular}{c}$p=4$ \end{tabular}}%
			\psfrag{s0003}[cl][cl][0.8]{\color[rgb]{0,0,0}\setlength{\tabcolsep}{0pt}\begin{tabular}{c}$p=6$ \end{tabular}}%
			\psfrag{s0004}[cl][cl][0.8]{\color[rgb]{0,0,0}\setlength{\tabcolsep}{0pt}\begin{tabular}{c}$p=8$ \end{tabular}}%
			\psfrag{s0005}[cl][cl][0.8]{\color[rgb]{0,0,0}\setlength{\tabcolsep}{0pt}\begin{tabular}{c}$p=10$ \end{tabular}}%
			\psfrag{x1}[tm][tm][1]{\color[rgb]{0,0,0}\setlength{\tabcolsep}{0pt}\begin{tabular}{c}$1$\end{tabular}}%
			\psfrag{x2}[tm][tm][1]{\color[rgb]{0,0,0}\setlength{\tabcolsep}{0pt}\begin{tabular}{c}$50$\end{tabular}}%
			\psfrag{x3}[tm][tm][1]{\color[rgb]{0,0,0}\setlength{\tabcolsep}{0pt}\begin{tabular}{c}$100$\end{tabular}}%
			\psfrag{x4}[tm][tm][1]{\color[rgb]{0,0,0}\setlength{\tabcolsep}{0pt}\begin{tabular}{c}$150$\end{tabular}}%
			\psfrag{x5}[tm][tm][1]{\color[rgb]{0,0,0}\setlength{\tabcolsep}{0pt}\begin{tabular}{c}$200$\end{tabular}}%
			\psfrag{y01}[cr][cr][1]{\color[rgb]{0,0,0}\setlength{\tabcolsep}{0pt}\begin{tabular}{c}$0$\end{tabular}}%
			\psfrag{y02}[cr][cr][1]{\color[rgb]{0,0,0}\setlength{\tabcolsep}{0pt}\begin{tabular}{c}$0.25$\end{tabular}}%
			\psfrag{y03}[cr][cr][1]{\color[rgb]{0,0,0}\setlength{\tabcolsep}{0pt}\begin{tabular}{c}$0.50$\end{tabular}}%
			\psfrag{y04}[cr][cr][1]{\color[rgb]{0,0,0}\setlength{\tabcolsep}{0pt}\begin{tabular}{c}$0.75$\end{tabular}}%
			\psfrag{y05}[cr][cr][1]{\color[rgb]{0,0,0}\setlength{\tabcolsep}{0pt}\begin{tabular}{c}$1.00$\end{tabular}}%
			\includegraphics[scale=1]{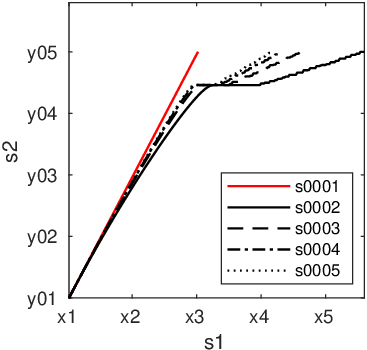}
	\end{psfrags}}
	\caption{Numerical analysis of an L-shaped specimen: (a) force-displacement diagram for $L_4$-norm and (b) evolution of the physical time as a function in terms of the iteration number for different $L_p$-norms \label{fig:Lshape_lpexperiment}}
\end{figure}

The results obtained from the time-adaptive E\&M-algorithm are presented in Fig.~\ref{fig:Lshape_lpexperiment} for numerous $L_p$-norms. For the sake of comparison, results corresponding to standard alternate minimization are also shown.
Since all algorithms predict the same structural response, only the diagram corresponding to the $L_4$-norm is shown in Fig.~\ref{fig:Lshape_lpexperiment}(a). In line with the time adaptivity of the E\&M-algorithm, the time step size is reduced slightly before, during and after brutal crack growth. This is certainly also the case for other $L_p$-norms, see Fig.~\ref{fig:Lshape_lpexperiment}(b). Furthermore and like the computations of the CT-specimen, the choice of the norm influences the time step size. However, the inverse trend can be seen here: the smaller the $p$, the larger the number of load increments, cf. Fig.~\ref{fig:Lshape_lpexperiment}(b). As a consequence, finding the numerically most efficient norm is challenging.

\begin{figure}[htbp]
	\centering
	\subcaptionbox{$t_{1}$=0}
	{\includegraphics[scale=0.14]{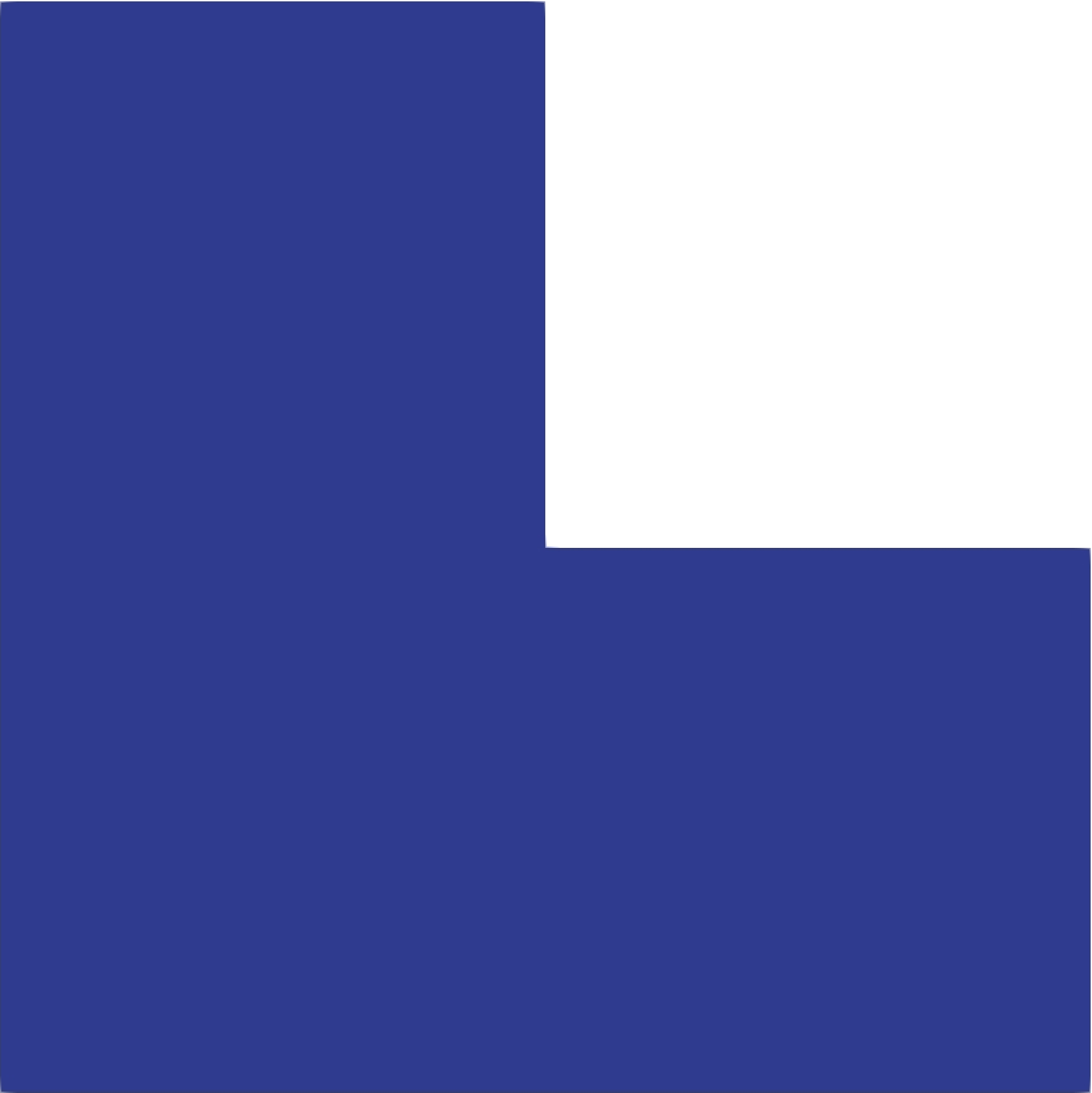}}
	\subcaptionbox{$t_{113}=7.487$}
	{\includegraphics[scale=0.14]{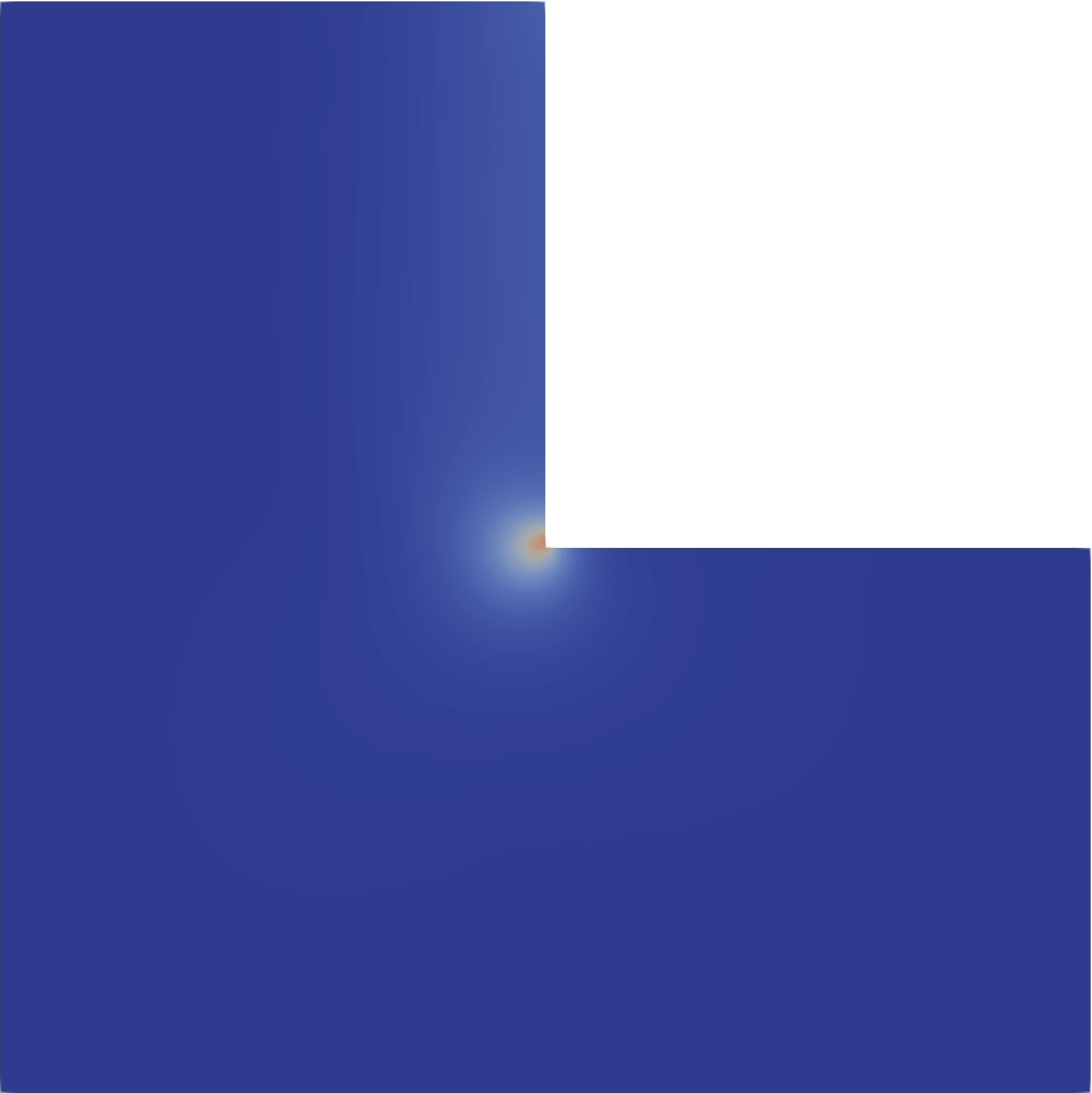}}
	\subcaptionbox{$t_{131}=7.487$}
	{\includegraphics[scale=0.14]{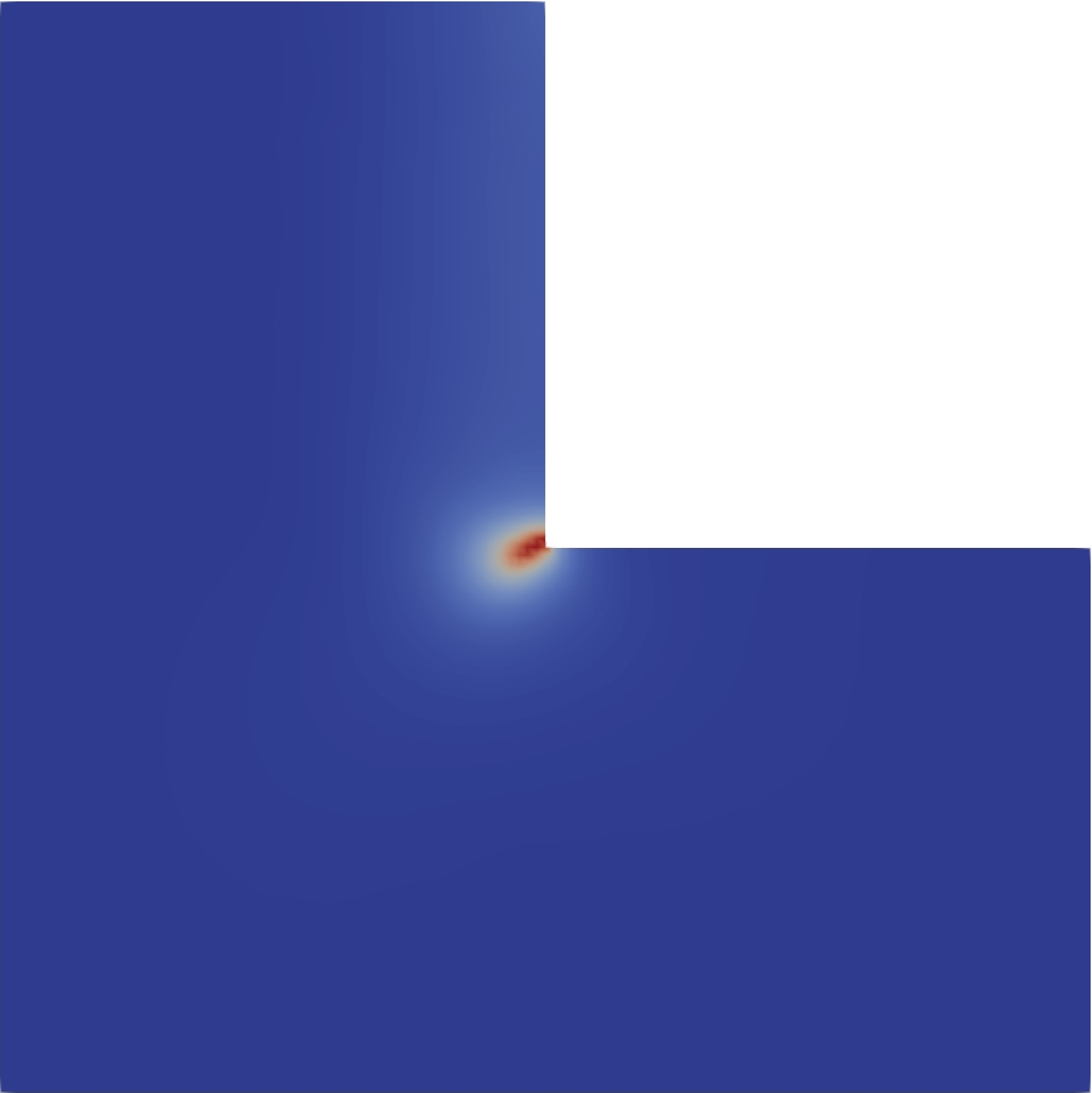}}
	\subcaptionbox{$t_{149}=7.487$}
	{\includegraphics[scale=0.14]{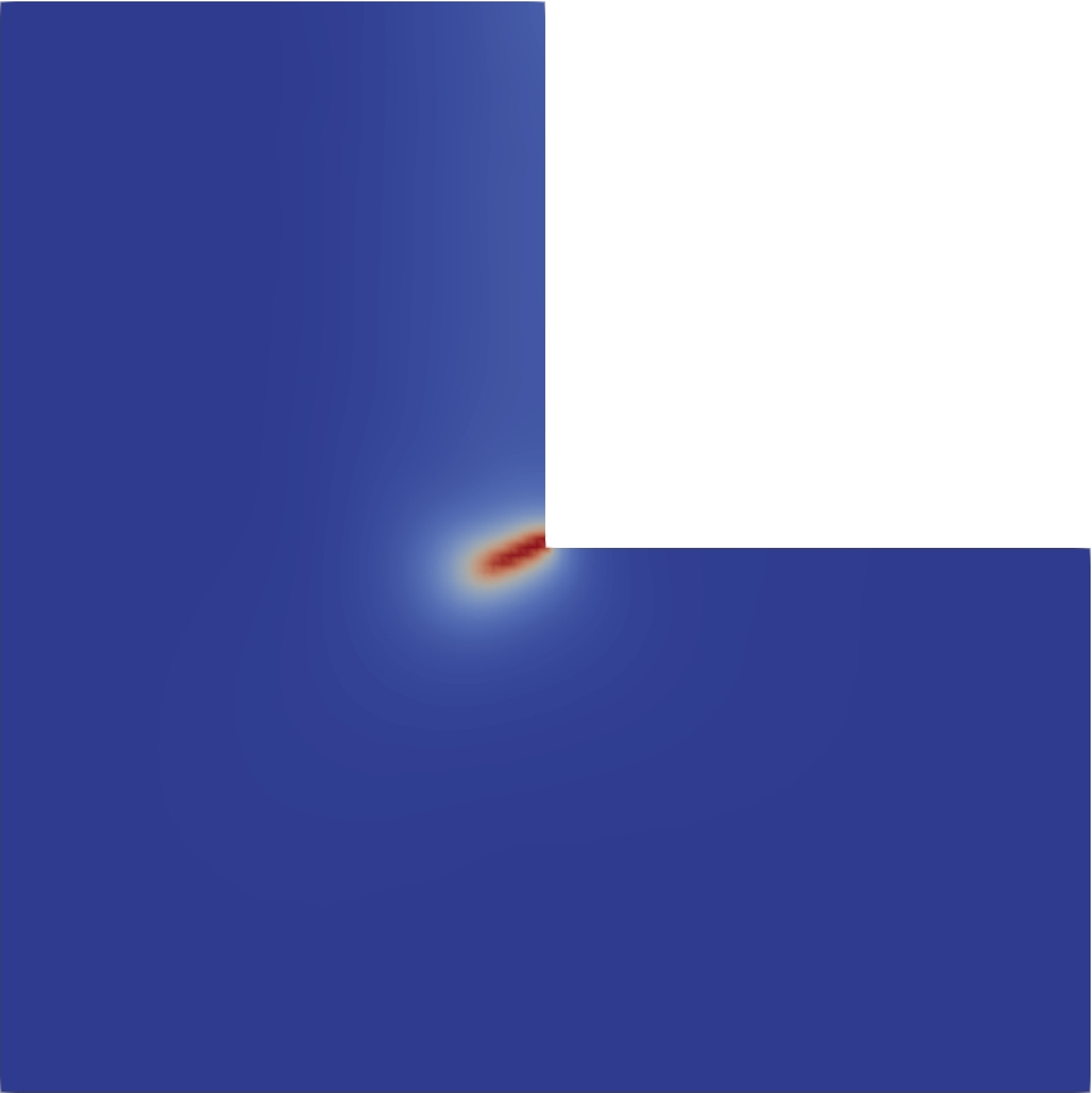}}
	\subcaptionbox{$t_{169}=7.783$}
	{\includegraphics[scale=0.14]{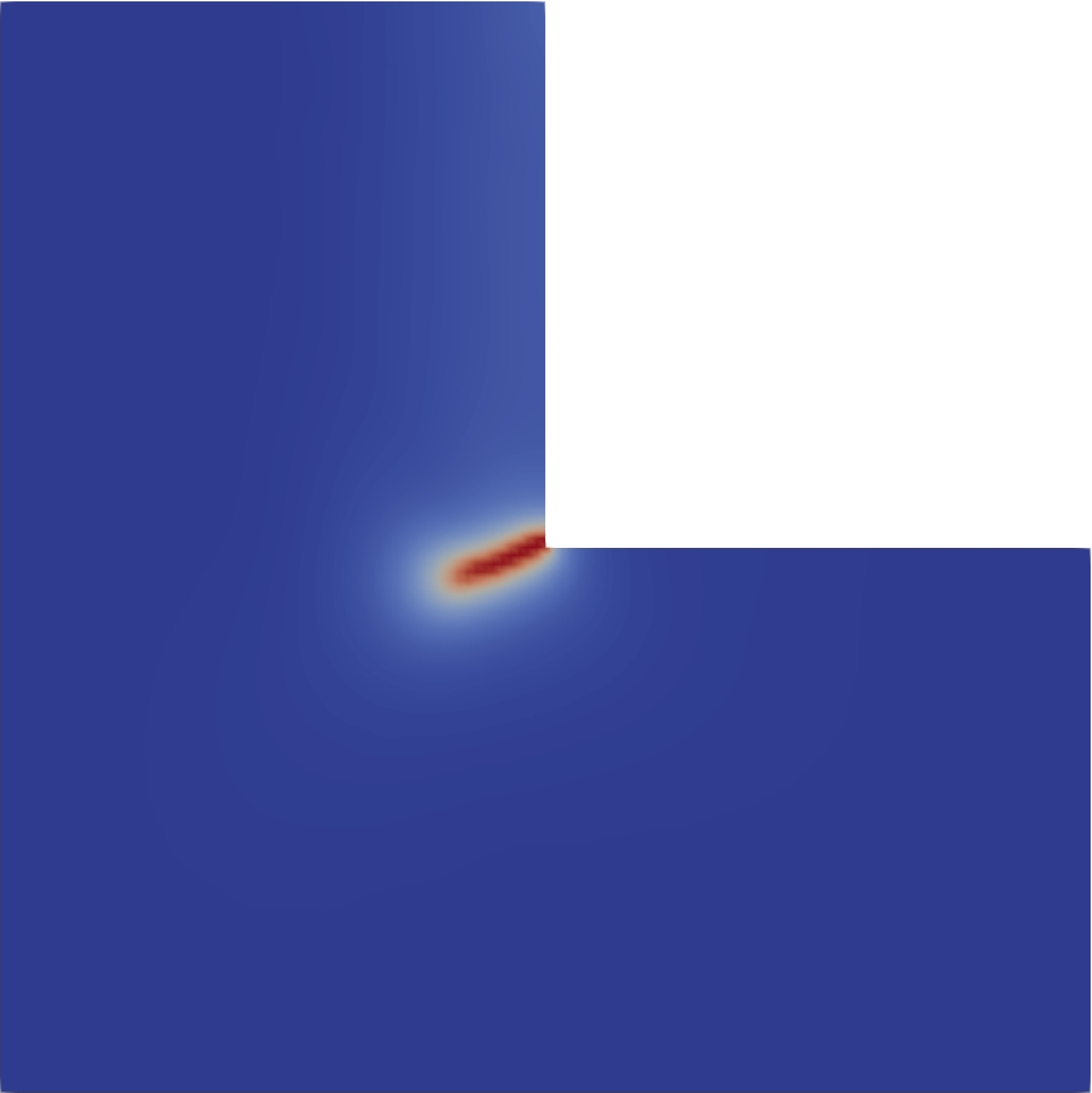}}
	\subcaptionbox{$t_{189}=8.082$}
	{\includegraphics[scale=0.14]{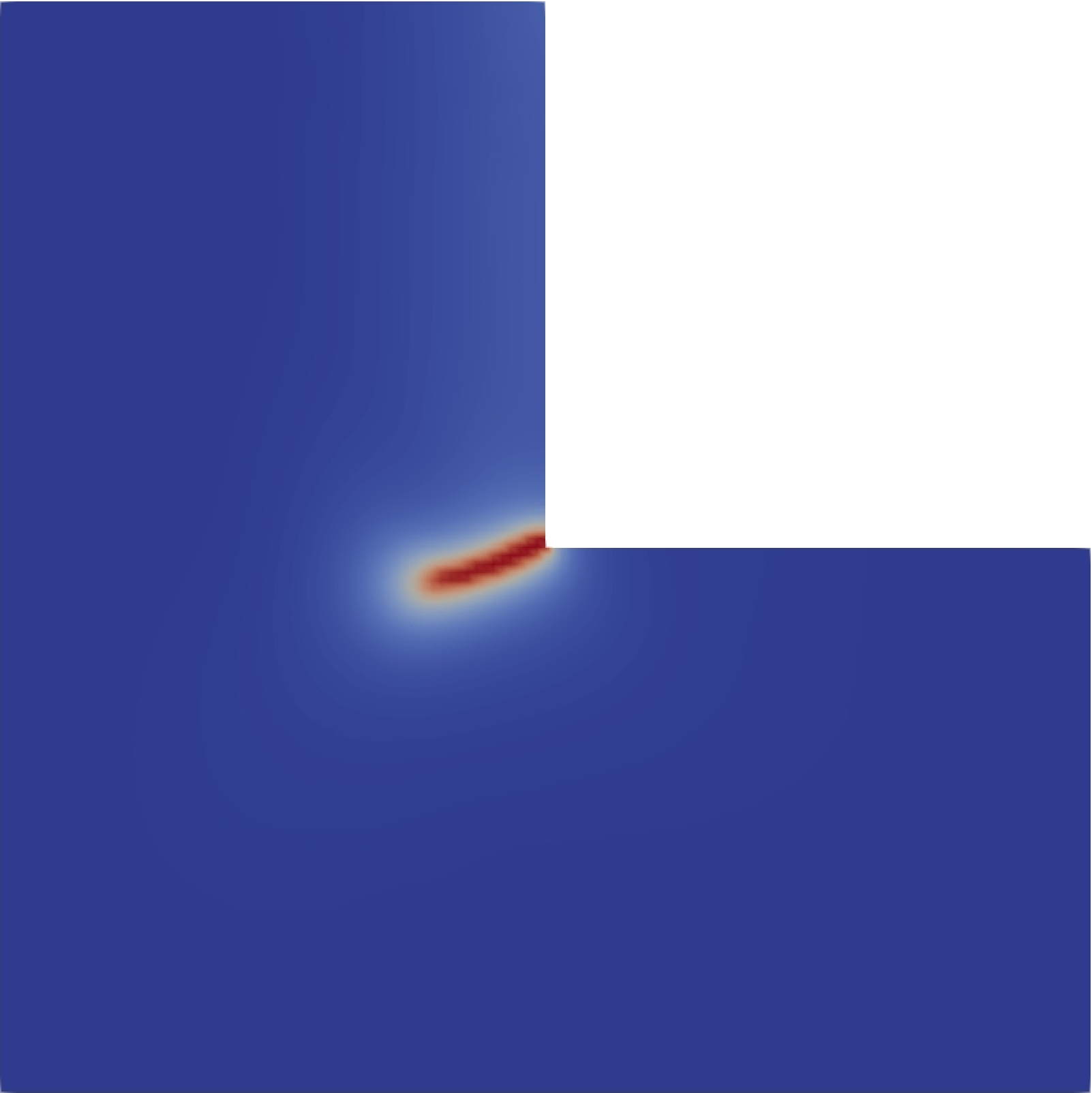}}
	\subcaptionbox{$t_{209}=8.342$}
	{\includegraphics[scale=0.14]{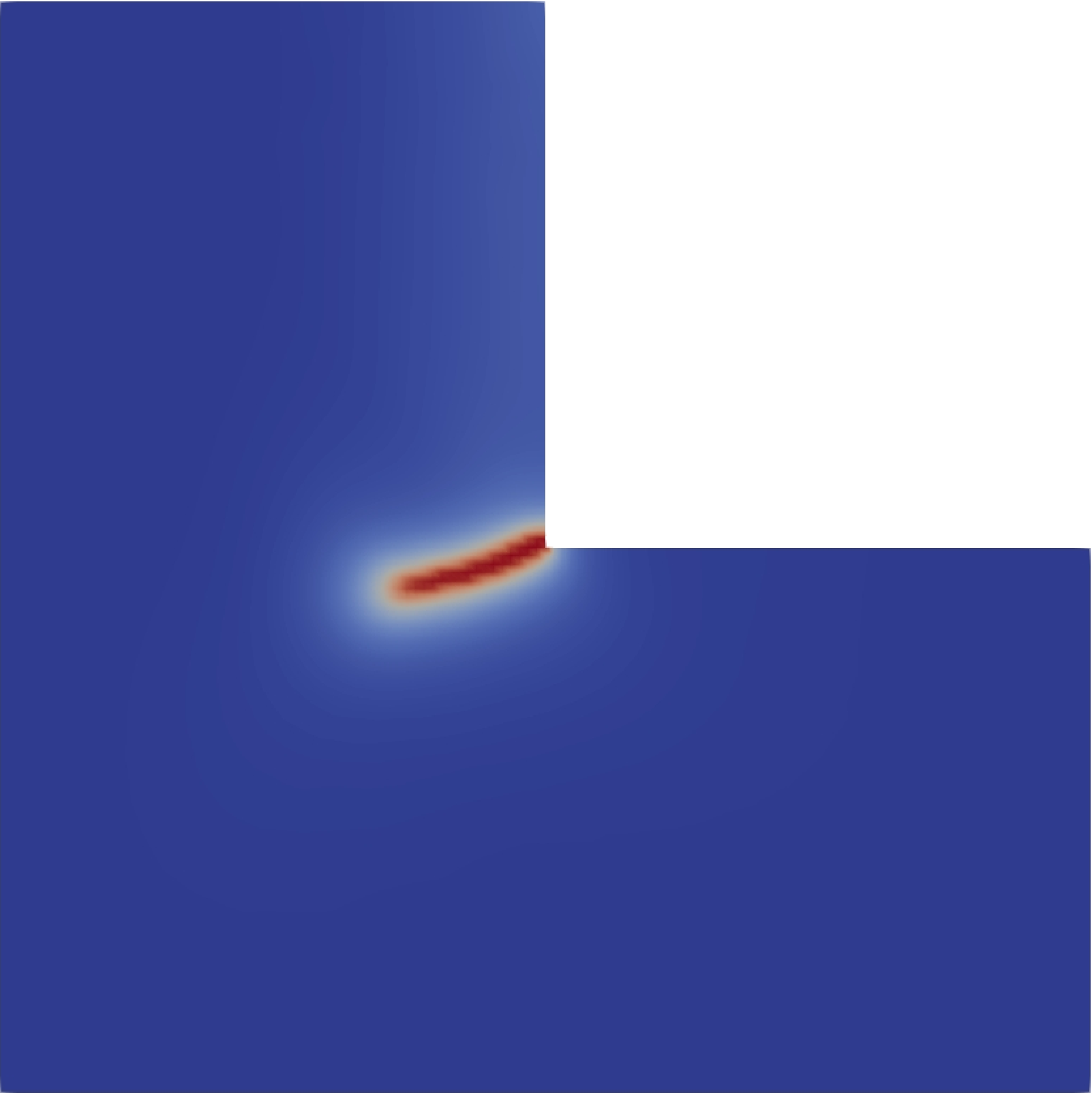}}
	\subcaptionbox{$t_{230}=8.658$}
	{\includegraphics[scale=0.14]{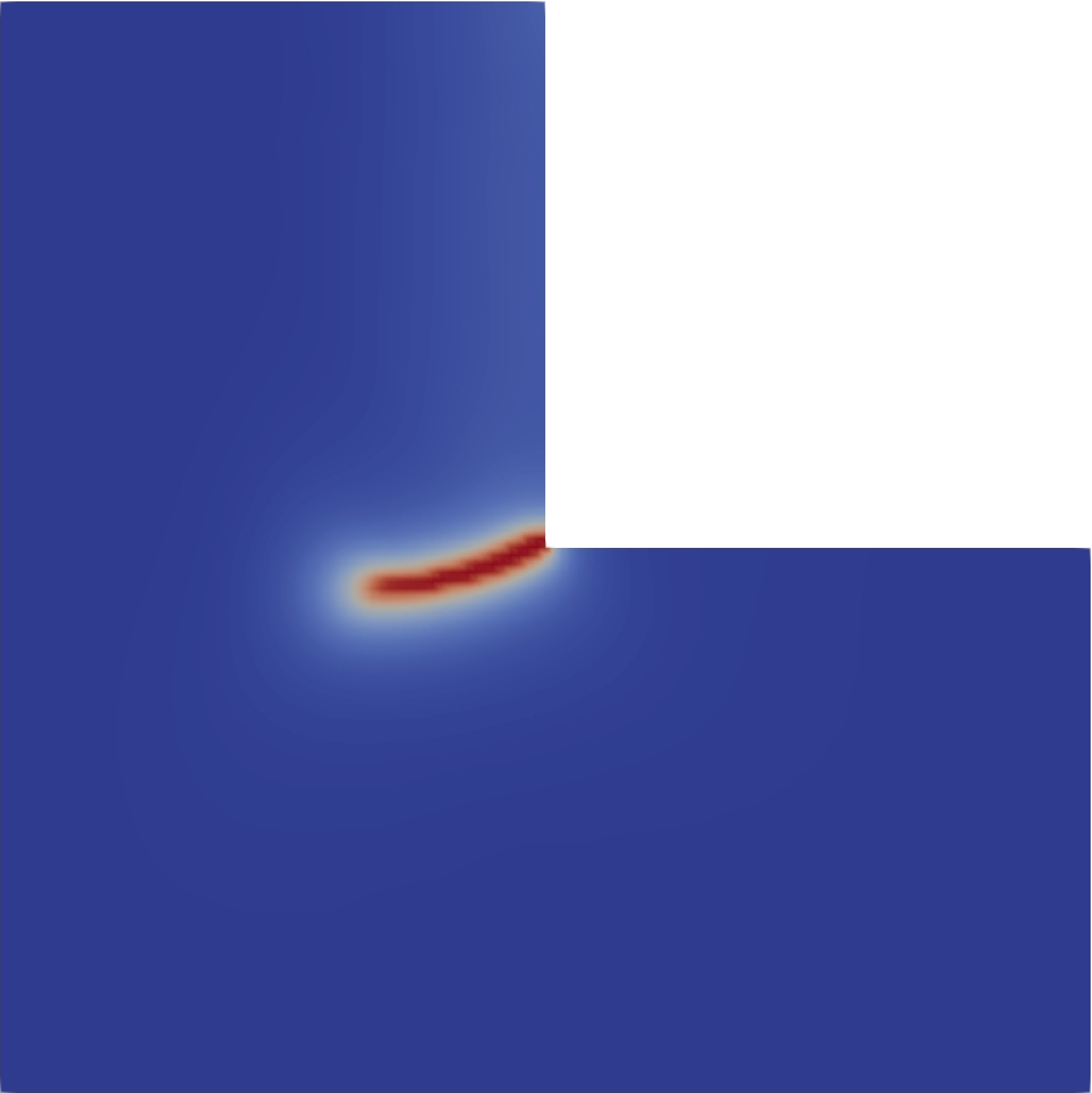}}
	\caption{Numerical analysis of an L-shaped specimen: Distribution of the phase-field variable $z$ computed from the time adaptive scheme combined with alternate minimization for $\rho=0.08658$ and $p=4$. Brutal crack-growth occurs at $\bar u=0.692$ mm \label{fig:LShape_Plots}}
\end{figure}

The evolution of the phase-field variable $z$ is summarized in Fig.\ref{fig:LShape_Plots}.
It can be seen that the crack starts at the inner corner where a stress singularity is present (analytical solution). Subsequently, the crack evolves in a curved manner to the left boundary. According to Fig.~\ref{fig:LShape_Plots} brutal crack growth occurs at $\bar{u}=0.692$ mm. This can also be observed in the load-displacement diagram, cf. Fig.~\ref{fig:Lshape_lpexperiment}(a).